\documentclass[structabstract]{aa}  
\usepackage{calrsfs}
\usepackage{enumerate}
\usepackage{footmisc}
\usepackage{graphicx}
\graphicspath{{graphics/}{../graphics/}}   
\usepackage{longtable, pdflscape}
\usepackage{natbib}
\bibpunct{(}{)}{;}{a}{}{,} 
\usepackage{txfonts, color}
\usepackage{soul}
\usepackage[switch]{lineno}
\usepackage{xfrac} 

\usepackage[draft=False,colorlinks=true,linkcolor=black,citecolor=blue,urlcolor=blue,linktoc=all]{hyperref}

\makeatletter
\def\LongtableFooter{%
  \multicolumn{\LT@cols}{r}{\framebox[1.1\width]{\textbf{continued on the following page}}}\\}
\makeatother

\newcommand{\cdbox}[1]{%
  {\color{blue}%
    \dbox{\color{black}#1}}%
}
\usepackage{dashbox,framed,color,ocg-p}
\newcommand{\ToggleLayer}[2]{%
  \leavevmode
  \pdfstartlink user {
    /Subtype /Link
    /Border [0 0 0]%
    /A <<
      /S/JavaScript
      /JS (
         var aOCGs = this.getOCGs(), Layer;
         var Layers = "#1".split(","), Active = -1, i, l;
         for (l=0; l<Layers.length; l++) {
           Layer = Layers[l];
           for (i=0; aOCGs && i<aOCGs.length; i++) {
             if (aOCGs[i].state && aOCGs[i].name == Layer) {
               Active = l;
               aOCGs[i].state = false;
             }
           }
           if (Active >= 0) break;
         }
         if (Active == -1) {
           for (l=0; l<Layers.length; l++) {
             if (Layers[l] == "") Active = l;
           }
         }
         Active = Active + 1;
         if (Active == Layers.length) Active = 0;
         Layer = Layers[Active];
         for (i=0; aOCGs && i<aOCGs.length; i++) {
           if (aOCGs[i].name == Layer) aOCGs[i].state = true;
         }
      )
    >>
  }#2%
  \pdfendlink
}

\DeclareMathAlphabet{\pazocal}{OMS}{zplm}{m}{n}

\newcommand{\tba}[1]{\textbf{\textcolor{blue}{To be added: #1}}}

\newcommand{\co}[2]{^{#1}\mathrm{C}^{#2}\mathrm{O}}
\newcommand{\transition}[2]{\,({#1}\text{--}{#2})}
\newcommand{\chired}{\chi^{2}_{\mathrm{red}}}

\newcommand{\beamefficiency}{\eta_{\mathrm{MB}}}

\newcommand{\flag}[2]{\pazocal{F}_{\mathrm{#1}}^{\mathrm{#2}}}

\newcommand{\Mach}{\pazocal{M}_{\sigma_{\mathrm{v}_{\mathrm{los}}}\mathrm{,\,3D}}}

\newcommand{\Min}{\mathrm{min}}
\newcommand{\Ncomp}{N_{\mathrm{comp}}}

\newcommand{\Nmed}{N_{\mathrm{med}}}
\newcommand{\Njump}{N_{\mathrm{jump}}}
\newcommand{\rms}{\sigma_{\mathrm{rms}}}
\newcommand{\rmsTa}{\sigma(T_{\mathrm{A}}^{*})}
\newcommand{\sect}{Sect.$\,$}
\newcommand{\fig}{Fig.$\,$}
\newcommand{\significance}[1]{\pazocal{S}_{\mathrm{#1}}}
\newcommand{\opacity}[1]{\tau_{0}^{#1}}
\newcommand{\Tb}{T_{\mathrm{MB}}}
\newcommand{\Ta}{T_{\mathrm{A}}^{*}}
\newcommand{\wco}{W_{\mathrm{CO}}}
\newcommand{\Tex}{T_{\mathrm{ex}}}
\newcommand{\veldisp}{\sigma_{v}}
\newcommand{\vlsr}{v_{\mathrm{LSR}}}

\newcommand{\kms}{km$\,$s$^{-1}$}
\newcommand{\Kkms}{K$\,$km$\,$s$^{-1}$}
\newcommand{\msun}{M$_{\odot}$}

\defcitealias{Deharveng1988}{D88}
\newcommand\gausspy{\textsc{GaussPy}}
\newcommand\gausspyplus{\textsc{GaussPy+}}
\newcommand\scousepy{\textsc{ScousePy}}

\begin{document}
    \title{Autonomous Gaussian decomposition of the Galactic Ring Survey} 
    \subtitle{I. Global statistics and properties of the \textsuperscript{13}CO emission data}

    \author{M. Riener
        \inst{1, }{\thanks{Member of the International Max-Planck Research School for Astronomy and Cosmic Physics at the University of Heidelberg (IMPRS-HD), Germany}}
    \and
    J. Kainulainen
        \inst{2}
    \and
    H. Beuther
        \inst{1}
    \and
    J. D. Henshaw
        \inst{1}
    \and
    J. H. Orkisz
        \inst{2}
    \and
    Y. Wang
        \inst{1}
    }

    \institute{Max-Planck Institute for Astronomy, K\"onigstuhl 17, 69117 Heidelberg, Germany
    \and
    Chalmers University of Technology, Department of Space, Earth and Environment, SE-412 93 Gothenburg, Sweden
    }

    \date{Received ..., 2019; accepted ..., 2019}

 
  \abstract
  {
The analysis of large molecular line surveys of the Galactic plane is essential for our understanding of the gas kinematics on Galactic scales, in particular its link with the formation and evolution of dense structures in the interstellar medium.
An approximation of the emission peaks with Gaussian functions allows for an efficient and straightforward extraction of useful physical information contained in the shape and Doppler-shifted frequency of the emission lines contained in these enormous data sets.
In this work we present an overview and first results of a Gaussian decomposition of the entire Galactic Ring Survey (GRS) $\co{13}{}\transition{1}{0}$ data that consists of about 2.3 million spectra.
We performed the decomposition with the fully automated \gausspyplus\ algorithm and fitted about 4.6 million Gaussian components to the GRS spectra.
These decomposition results enable novel and unexplored ways to interpret and study the gas velocity structure. 
We discuss the statistics of the fit components and relations between the fitted intensities, velocity centroids, and velocity dispersions.
We find that the magnitude of the velocity dispersion values increase toward the inner Galaxy and around the Galactic midplane, which we speculate is partly due to the influence of the Galactic bar and regions with higher non-thermal motions located in the midplane, respectively.
We also use our decomposition results to infer global properties of the gas emission and find that the number of fit components used per spectrum is indicative for the amount of structure along the line of sight. 
We find that the emission lines from regions located on the far side of the Galaxy show increased velocity dispersion values, likely due to beam averaging effects.
We demonstrate how this trend has the potential to aid in characterising Galactic structure by disentangling emission that is belonging to the nearby Aquila Rift molecular cloud from emission that is more likely associated with the Perseus and Outer spiral arms.
With this work we also make our entire decomposition results available.
}


   \maketitle
%



\section{Introduction}

The study of the velocity structure of molecular gas is of vital importance for understanding the origin and evolution of structures in the interstellar medium (ISM).
Carbon monoxyde (CO), which is the most abundant molecule in the ISM after molecular hydrogen (H$_{2}$), has multiple transitions that are excited at the low temperatures prevalent in the molecular part of the ISM and can be observed in the radio and sub-mm part of the electromagnetic spectrum.
Thus CO is often used as a tracer of H$_{2}$ \citep[for a review see][]{Bolatto2013review}, whose observation at the typical temperatures ($\sim\,10\ \text{--}\ 30\,$K) of molecular clouds and clumps are challenging or effectively impossible.
Moreover, the CO emission lines contain useful velocity information: the Doppler shift of the centroid frequency of the line compared to the rest frequency in the local standard of rest (LSR) comes from the bulk motion of the gas, mostly due to Galactic dynamics; and the width of the line beyond the natural line broadening is caused by thermal (or Doppler) and non-thermal contributions, with the line shape usually well described by a Gaussian shape. 

Given the relative ease of observation and the plethora of information they encode, the rotational transitions of the most abundant isotopologues of CO -- $^{12}$CO, $^{13}$CO and C$^{18}$O -- have thus for a long time been prime targets for large mapping surveys of the Milky Way's disk.
Examples of Galactic plane surveys from the last two decades that are already published or currently in preparation are: the Galactic Ring Survey \citep[GRS;][]{Jackson2006}; the CO (3-2) High-resolution Survey of the Galactic Plane \citep[COHRS;][]{Dempsey2013}; the Mopra Southern Galactic Plane CO Survey \citep[MGPS;][]{Burton2013-mgps, Braiding2018Mopra}; the Three-mm Ultimate Mopra Milky Way Survey \citep[ThrUMMS,][]{Barnes2015}; the $^{13}$CO/C$^{18}$O (J=3-2) Heterodyne Inner Milky Way Plane Survey \citep[CHIMPS;][]{Rigby2016}; the FOREST unbiased Galactic plane imaging survey with the Nobeyama 45 m telescope \citep[FUGIN;][]{Umemoto2017}; the Structure, Excitation, and Dynamics of the Inner Galactic Interstellar Medium survey \citep[SEDIGISM,][]{Schuller2017}; and the Milky Way Imaging Scroll Painting survey \citep[MWISP,][]{Su2019}. 
See also Fig.~2 in \citet{Heyer2015review} for an exhaustive overview of large CO surveys of the Galactic plane up until 2015.

These data sets have provided and continue to provide invaluable information about the CO content and kinematics for individual case studies or samples of molecular clouds \citep[e.g.][]{Heyer2015review, MivilleDeschenes2017}, infrared dark clouds (IRDCs) \citep[e.g.][]{Simon2006irdcs, Kainulainen2013_irdcs, Barnes2018-irdcs, Zhou2019-irdc}, molecular clumps \citep[e.g.][]{Urquhart2018-atlasgal} or filaments \citep[e.g.][]{Zucker2018-gmfs, Zhang2019-gmfs} covered by these surveys.
While for many works the average kinematic properties of their objects have been of prime interest, some studies have further analysed the detailed velocity structure of their objects by exploiting the full spatial and spectral resolution of one or more of these surveys \citep[e.g.][]{Barnes2018-irdcs}.

There are also studies that used the entire data set of one of the Galactic plane surveys; these have so far mostly focussed on either the global properties of the gas emission at larger scales to obtain information about the Galactic structure \citep[e.g.][]{Dame2001, Nakanishi2006, Rigby2016, Roman-Duval2016-grs} or on segmenting the observed gas emission into molecular clouds, clumps, or filaments \citep[e.g.][]{Solomon1987, Rathborne2009, Rice2016-clouds, MivilleDeschenes2017, Colombo2019}, to infer useful average physical properties of the gas emission on the scales of these objects and trying to put them into the context of their location within our Galaxy \citep[e.g.][]{RomanDuval2010gmcs, Heyer2015review, MivilleDeschenes2017}.

What is currently still missing is a study of the detailed velocity structure of the molecular gas on Galactic scales, with a spatial resolution sufficient to resolve the inner structure of molecular clouds ($\lesssim 1\arcmin$) and spectral resolution sufficient to resolve the thermal linewidth of the cold molecular gas ($\sim 0.2$~\kms).
Such a study requires analysing the entire data set of one of the more recent large Galactic plane surveys in its native spatial and spectral resolution. 
Presently, we do not know what this velocity structure looks like on Galactic scales and its study could reveal systematic trends linked to physical processes, for instance the effects spiral arms have on the kinematics of the gas in the context of star formation.

However, a study of the detailed velocity structure of any of these Galactic plane surveys requires a reduction of the complexity of these data sets.
One possible approach is the decomposition of the emission lines of each spectrum into individual (velocity) components.
In case of CO emission of Galactic plane surveys, these velocity components can be associated with different structures along the line of sight.
This association is non-trivial and complicated due to multiple physical and observational effects, such as velocity crowding (mixing of emission along the line of sight), velocity gradients, optical depth, and difficulties in transforming $\vlsr$ velocities to physical distances.
The main problem in the decomposition itself is how to correctly identify the (sometimes blended) emission lines, so that we can determine the shape of the individual lines and subsequently can infer the correct physical properties producing these line shapes.

\citet{Riener2019} recently presented the Gaussian decomposition package \gausspyplus\, which was designed for the fully automated and efficient decomposition of large data sets of emission lines, such as Galactic plane surveys, and fits multi-Gaussian models to the spectra.
In \citet{Riener2019}, \gausspyplus\ was already applied on a challenging test field of the GRS data set, for which it performed well in terms of recovered flux and spatial coherence of the fit results.

In this work, we present decomposition results with \gausspyplus\ for the entire GRS data set.
We mostly focus on the global properties of the gas emission and discuss distributions of and correlations between the fit parameters. 
We provide a characterisation of the complexity of gas emission along the line of sight and try to combine it with information about the dust emission from the Herschel infrared Galactic Plane Survey \citep[Hi-GAL;][]{Molinari2016-higal}.
We also present a way in which the simplification of the data set via the Gaussian decomposition may allow to narrow down the location of the emitting gas within our Galaxy. 
In a forthcoming work (Riener et al., in prep.), we aim to present distance estimates to the Gaussian fit components, which will open up discussion about the Galactic distribution of the gas emission and variations with Galactocentric distance of the kinematic properties of the gas, and will foster synergies with other ISM tracers, such as the upcoming large-scale dust extinction map from the PROMISE project (Kainulainen et al., in prep.).


\section{Observational data and decomposition method}
\label{sec:data}

\subsection{Galactic Ring Survey}
\label{sec:grs}
In this work we use data from the Boston University–Five College Radio Astronomy Observatory GRS (\citealt{Jackson2006}). 
This survey covers a longitude range of $\ell = 18^\circ - 55.7^\circ$ and a latitude range of $\left|b\right| < 1.1^\circ$ with an angular resolution of $46^{\prime\prime}$ and a pixel sampling of $22^{\prime\prime}$. 
The velocity coverage of the survey is $-5$ to $135$~\kms\ for $\ell \leq 40^\circ$ and $-5$ to $85$~\kms\ for $\ell > 40^\circ$.
We also used the additional limited data from $l = 14^\circ - 18^\circ$ that does not cover the full latitude range of the rest of the data; the velocity range of this additional data is again $-5$ to $135$~\kms.
The GRS data set has a velocity resolution of $0.21$~\kms.
We used \textsc{Swarp}\footnote{\url{http://www.astromatic.net/software/swarp}\label{foot:swarp}} \citep{Bertin2002} to combine the original data cubes of the GRS\footnote{\url{https://www.bu.edu/galacticring/new_data.html}\label{foot:grs}} into a single mosaicked cube (see App.~\ref{app:grs-preparation}). 
The values in the GRS data set are given in terms of antenna temperatures ($\Ta$) that we converted to main beam temperatures ($\Tb$) by dividing by the main beam efficiency\footref{foot:grs} of $\beamefficiency = 0.48$.

\subsection{Hi-GAL}
\label{sec:ppmap}

We use maps of mean dust temperatures and H$_{2}$ column densities from \citet{Marsh2017-ppmap}\footnote{\url{http://www.astro.cardiff.ac.uk/research/ViaLactea/}} that are based on dust emission observations from Hi-GAL \citep{Molinari2016-higal}.
The maps from \citet{Marsh2017-ppmap} combine continuum data spanning a wavelength range of $70-500$~$\mu$m at a spatial resolution of 12$^{\prime\prime}$.
We used \textsc{Swarp}\footref{foot:grs} to combine 21 PPMAP fields overlapping with the GRS coverage.
We spatially smoothed the mosaicked maps of mean dust temperatures and H$_{2}$ column densities to the GRS resolution and regridded the smoothed maps so that the pixels aligned with the GRS mosaic.

\subsection{The \gausspyplus\ algorithm}
\label{sec:gauss_decomp}

For the decomposition of the GRS data set we used \gausspyplus\footnote{\url{https://ascl.net/1907.020}} \citep{Riener2019}, a fully automated Gaussian decomposition package that is based on \gausspy\footnote{\url{https://ascl.net/1907.019}} \citep{Lindner2015}.
The \gausspy\ algorithm uses derivative spectroscopy to automatically decide on the number of fit components and the initial guesses for their parameters.
The \gausspyplus\ package enhances the performance of \gausspy\ by introducing an improved fitting routine that tries to refit decomposition results that failed in-built and optional user-selected quality criteria.
Moreover, \gausspyplus\ includes spatially coherent refitting routines that aim to improve spatial consistency of the decomposition results.
In addition, \gausspyplus\ offers many convenience functions and automated preparatory steps, such as a reliable noise estimation, signal identification and masking of instrumental artefacts. 
For further details about the \gausspyplus\ algorithm, such as an in-depth explanation of its functionality and discussion about its performance for decompositions of synthetic spectra and a GRS test field, see \citet{Riener2019}.

Our main aim in this work is to present a homogeneous decomposition and analysis of the GRS data set in its native spatial and spectral resolution.
We used the same parameter settings in \gausspyplus\ throughout the entire survey region to guarantee a comparable analysis of the data set. 
This means that we did not finetune our settings to individual regions and the decomposition results with \gausspyplus\ thus might show differences in their performance, given the non-uniform noise coverage of the GRS that can show significant variation (see \fig\ref{fig:noise_map}).
However, the variations in the noise values between different regions of the survey means that the signal-to-noise (S/N) based thresholds of \gausspyplus\ could extract more signal peaks in the spectra that showed reduced noise values.
Some of the emission peaks may also have a non-Gaussian shape due to, for example, line blending or optical depth effects and will thus likely not be well fit by \gausspyplus\ (see App.~\ref{sec:optical_depth} for a discussion about the impact of optical depth on our fitting results). 


\section{Gaussian decomposition of the GRS data}
\label{sec:decomp_results}

In this section, we present results of the Gaussian decomposition with \gausspyplus\ for the entire GRS data set.
The appendix contains further, more technical discussions about the decomposition.
Details about the data preparation, parameter settings, and decomposition runs can be found in App.~\ref{app:grs-decomp}; we also present quality assurance metrics for the fit results in App.~\ref{sec:quality}.
Furthermore, we give a detailed discussion about the effects of optical depth on our decomposition in App.~\ref{sec:optical_depth}.
We find that issues due to optical depth only affect the densest regions in the GRS and should not be a problem for the vast majority of the decomposition results presented in this work.

The whole GRS data set contains in total $2\,283\,920$~spectra.
Of these, we excluded $1188$~spectra ($0.05\%$) that showed extremely high $\rmsTa$ values above $0.75$~K (see \sect\ref{sec:noise-values}).
In the data preparation step (see App.~\ref{app:grs-decomp}), \gausspyplus\ identified signal peaks in $75.3\%$ of all spectra; $\sim 96\%$ of these were fitted with one or multiple components in the decomposition.

The final decomposition contains $4\,648\,985$ fitted Gaussian components. The best fit solutions include fit components with S/N ratios as low as 1.5 \citep[it is beneficial to allow fit components with amplitudes below S/N < 3, cf.,][]{Riener2019}.
We highlight the $\sim 75\%$ of the fit components with S/N ratios $> 3$ in our discussion of the fit parameter statistics in Sects.$\,$\ref{sec:ncomps} \text{--} \ref{sec:velocity-dispersion}, since the components with lower S/N ratios can already be severely affected by the noise.
Depending on what the decomposition results are used for, it may thus be beneficial to only select more reliable fit components with S/N ratios $> 3$ or amplitude values above a specific $\Tb$ threshold.

\subsection{Catalogue description}
\label{sec:catalogue}

With this work, we also make a catalogue of all our decomposition results for the GRS available\footnote{\tba{Link to CDS or webpage}}. 
In this section we describe the entries of the catalogue that include quality flags that can be useful in identifying fit solutions that might have problems or are inconsistent with neighbouring decomposition results. 

\begin{table*}
    \caption{Decomposition results.}
    \centering
    \scriptsize
    \renewcommand{\arraystretch}{1.2}
    \setlength{\tabcolsep}{4.5pt}
\begin{tabular}{ccccccccccccccccccccc}
\hline\hline
$x_{\text{pos}}$ & $y_{\text{pos}}$ & $\ell$ & $b$ & T$_{\text{B}}$ & $\Delta\text{T}_{\text{B}}$ & v$_{\text{LSR}}$ & $\Delta\text{v}_{\text{LSR}}$ & $\sigma_{\text{los}}$ & $\Delta\sigma_{\text{los}}$ & $\sigma_{\text{rms}}$ & $p$ & AICc & $\chired$ & $\Ncomp$ & $N_{\text{med}}$ & $N_{\text{jump}}$ & F$_{1}$ & F$_{2}$ & F$_{3}$ & F$_{4}$ \\
 & & {[$\degr$]} & [$\degr$] & [K] & [K] & [\kms] & [\kms] & [\kms] & [\kms] & [K] & [$\%$] &  &  &  &  &  &  &  &  &  \\
(1) & (2) & (3) & (4) & (5) & (6) & (7) & (8) & (9) & (10) & (11) & (12) & (13) & (14) & (15) & (16) & (17) & (18) & (19) & (20) & (21) \\
\hline

286 & 2 & 54.241 & -1.088 & 2.26 & 0.39 & 24.81 & 0.04 & 0.21 & 0.04 & 0.41 & 5.9 & -419.4 & 1.21 & 2 & 1 & 0 & 0 & 0 & 0 & 0 \\
286 & 2 & 54.241 & -1.088 & 1.52 & 0.25 & 45.19 & 0.10 & 0.50 & 0.10 & 0.41 & 5.9 & -419.4 & 1.21 & 2 & 1 & 0 & 0 & 0 & 0 & 0 \\
287 & 2 & 54.235 & -1.088 & 1.51 & 0.24 & 24.85 & 0.08 & 0.46 & 0.08 & 0.38 & 1.1 & -465.7 & 0.96 & 2 & 1 & 0 & 0 & 0 & 0 & 0 \\
287 & 2 & 54.235 & -1.088 & 1.43 & 0.26 & 44.96 & 0.08 & 0.39 & 0.08 & 0.38 & 1.1 & -465.7 & 0.96 & 2 & 1 & 0 & 0 & 0 & 0 & 0 \\
288 & 2 & 54.229 & -1.088 & 2.09 & 0.34 & 24.88 & 0.04 & 0.20 & 0.04 & 0.36 & 0.8 & -367.3 & 1.09 & 1 & 1 & 0 & 0 & 0 & 0 & 0 \\
290 & 2 & 54.217 & -1.088 & 1.32 & 0.26 & 24.84 & 0.13 & 0.60 & 0.13 & 0.47 & 19.2 & -337.7 & 1.00 & 1 & 0 & 0 & 0 & 0 & 0 & 0 \\
291 & 2 & 54.210 & -1.088 & 2.28 & 0.36 & 24.76 & 0.04 & 0.20 & 0.04 & 0.37 & 30.5 & -356.9 & 0.97 & 1 & 1 & 0 & 0 & 0 & 0 & 0 \\
292 & 2 & 54.204 & -1.088 & 2.16 & 0.34 & 24.84 & 0.06 & 0.34 & 0.06 & 0.48 & 47.5 & -326.4 & 0.88 & 1 & 1 & 0 & 0 & 0 & 0 & 0 \\
294 & 2 & 54.192 & -1.088 & 1.04 & 0.20 & 24.79 & 0.15 & 0.69 & 0.15 & 0.38 & 55.3 & -369.4 & 1.15 & 1 & 1 & 0 & 0 & 0 & 0 & 0 \\
295 & 2 & 54.186 & -1.088 & 2.26 & 0.26 & 24.86 & 0.06 & 0.43 & 0.06 & 0.41 & 8.2 & -351.1 & 1.17 & 1 & 1 & 0 & 0 & 0 & 0 & 0 \\
296 & 2 & 54.180 & -1.088 & 1.84 & 0.24 & 24.72 & 0.08 & 0.54 & 0.08 & 0.42 & 10.6 & -450.0 & 1.02 & 1 & 1 & 0 & 0 & 0 & 0 & 0 \\

    \hline
    \end{tabular}
    \label{tbl:table_decomp}
    \tablefoot{ 
    \scriptsize 
    (This table is available in its entirety on this webpage: \tba{Link to CDS or webpage}. A portion is shown here for guidance regarding its form and content.)
    }
\end{table*}

We show a subset of the decomposition results in Table~\ref{tbl:table_decomp}.
Each row corresponds to a single Gaussian fit component; a spectrum fitted with eight Gaussian components will have eight rows in the table.

Columns~(1) and (2) show the pixel position of the spectrum in the mosaicked cube of the GRS with the corresponding Galactic coordinate values given in columns~(3) and (4).
The next columns list the parameters and associated errors of the Gaussian fit parameters: peak main beam brightness temperature or amplitude value $\Tb$ (5) and its error $\Delta\Tb$ (6); mean position $\vlsr$ (7) and its error $\Delta\vlsr$ (8); and velocity dispersion $\veldisp$ (9) and its error $\Delta\veldisp$ (10).
Column~(11) gives the root-mean-square value of the estimated noise of the spectrum $\rms$ given in $\Tb$ values.
The remaining columns list quality metrics that can indicate problems with the fit solutions, such as inconsistencies with neighbouring fit results.
Column~(12) gives the resulting $p$-value of a normality test for normally distributed residual values (see App.~\ref{sec:goodness-of-fit} for more details); column~(13) shows the value of the corrected Akaike information criterion \citep[AICc;][]{Akaike1973} for the best fit solution; and column~(14) gives the reduced chi-squared ($\chired$) values (see App.~\ref{sec:goodness-of-fit} for more details).
The next three columns can be used to identify fit solutions whose number of fit components is inconsistent with neighbouring fit solutions.
Columns~(15), (16), and (17) give the number of fit components $\Ncomp$, the weighted median number of fit components $N_{\text{med}}$ determined from the fit solutions of the direct neighbours, and the number of component jumps $N_{\text{jump}}$ $>2$ toward directly neighbouring fit solutions (see Sect.~3.2.2 in \citealt{Riener2019} for more information on the last two parameters). 
Finally, the last four columns indicate whether the fit component was flagged with one of the remaining optional quality flagging criteria described in Sect.~3.2.2 of \citealt{Riener2019}, with "1" indicating that the component was flagged with the respective criterion.
The flagging criteria are as follows: F$_{1}$ (18) indicates whether the fit component was blended with another component; F$_{2}$ (19) indicates whether the fit component caused a negative feature in the residual; F$_{3}$ (20) indicates whether the fit component was flagged as broad; and F$_{4}$ (21) indicates whether the fit component was part of a region in the spectrum, which was flagged in Phase~2 of the spatially coherent refitting routine in \gausspyplus, which aims at identifying inconsistencies between the centroid values of the fit components of neighbouring fit solutions (see Sect.~3.3.2 in \citealt{Riener2019} for more details).

The criteria for the quality flags are very strict in the default settings of \gausspyplus. 
To catch the majority of potential problems with the fit solutions, these criteria were designed to be biased towards producing false positives rather than false negatives.
It is thus likely that a significant fraction of flagged components, in particular those flagged as blended (F$_{1}$) or broad (F$_{3}$), will be good fits.

\subsection{Flux recovery}
\label{sec:recovered_flux}

\begin{figure*}
    \centering
    \includegraphics[width=\textwidth]{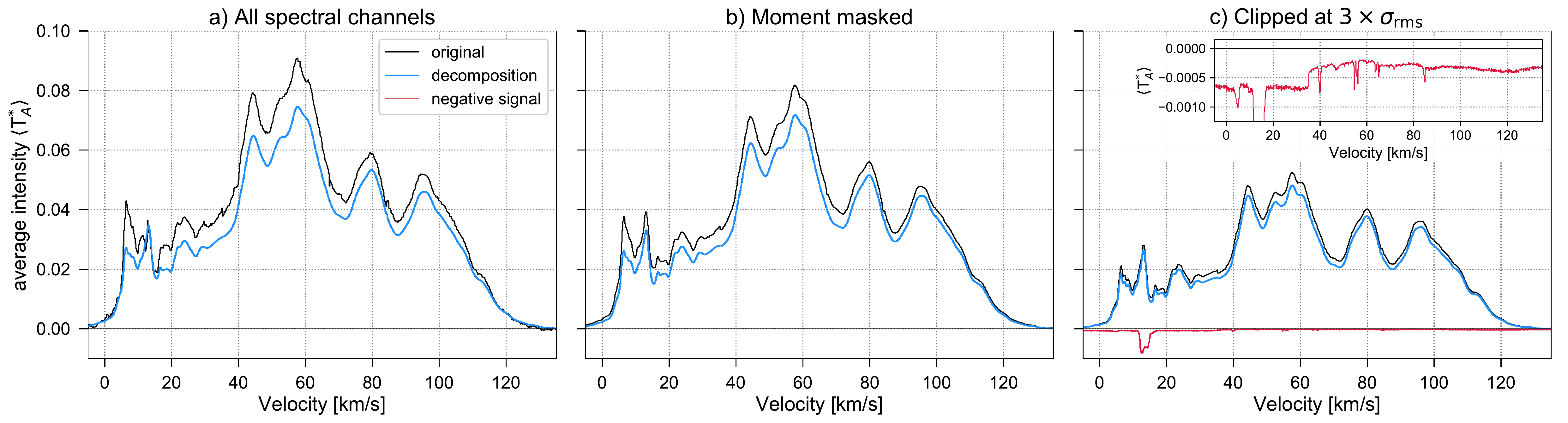}
    \caption{Average spectra of the full GRS data set (grey lines) and our final Gaussian decomposition results (blue lines).
    The three panels show different ways the average spectra were created: in \emph{a)} we use all voxels, in \emph{b)} we use only the voxels retained from moment masking, and in \emph{c)} we use only voxels whose intensity value is above a S/N threshold of 3.
    The red line in \emph{panel~c} shows an average of high negative values throughout the GRS data set that was obtained by using only voxels with a value below $-3\times\text{S/N}$.
    The inset in the right panel shows a zoom-in of the red line to better display the individual smaller negative peaks.
    See \sect\ref{sec:recovered_flux} for more details.}
    \label{fig:average_spectra}
\end{figure*}

\begin{figure*}[t]
    \centering
    \includegraphics[width=0.9\hsize]{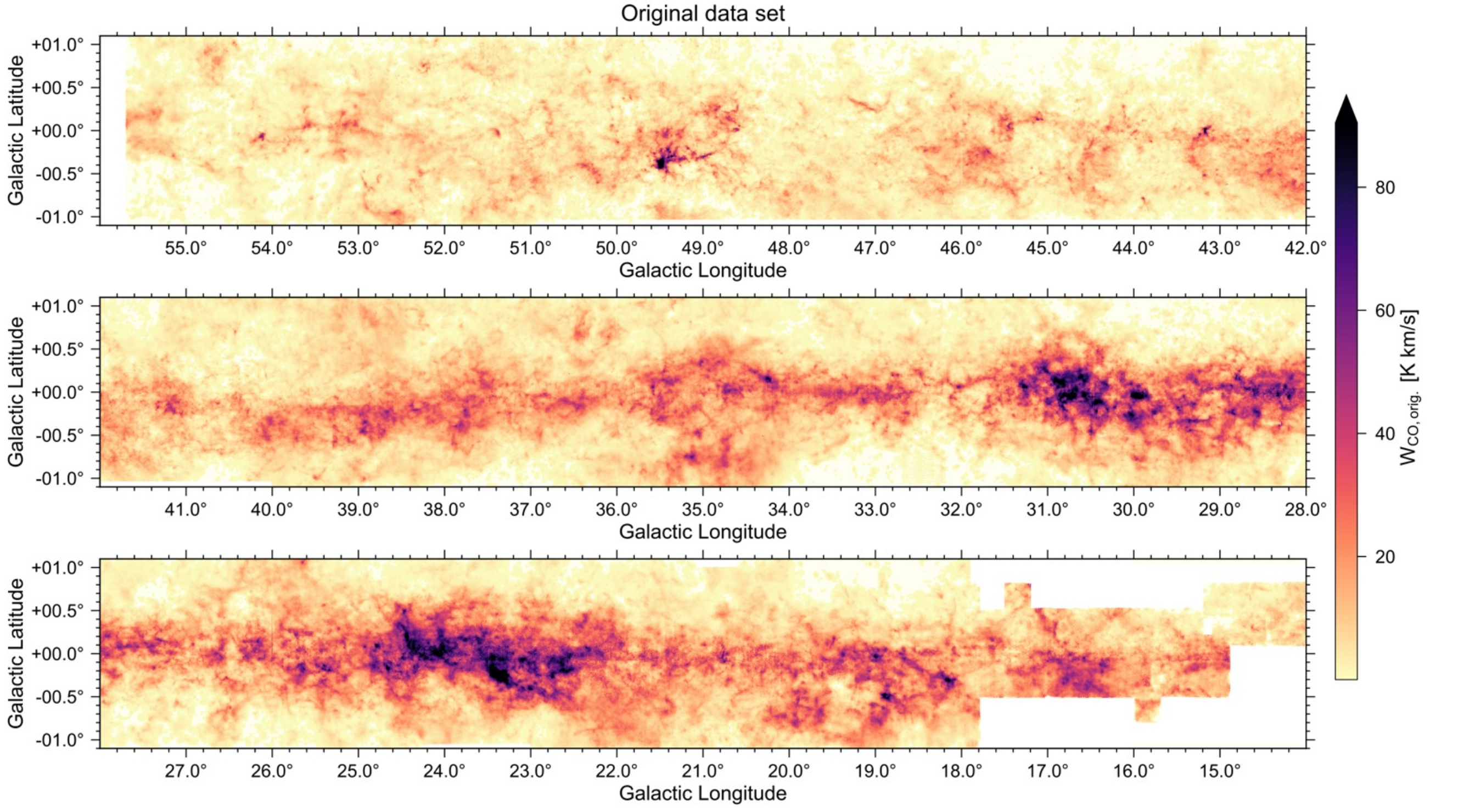}%
    \hspace{-0.9\hsize}%
    \begin{ocg}{fig:mom0_original_off}{fig:mom0_original_off}{0}%
    \end{ocg}%
    \begin{ocg}{fig:mom0_original_on}{fig:mom0_original_on}{1}%
    \includegraphics[width=0.9\hsize]{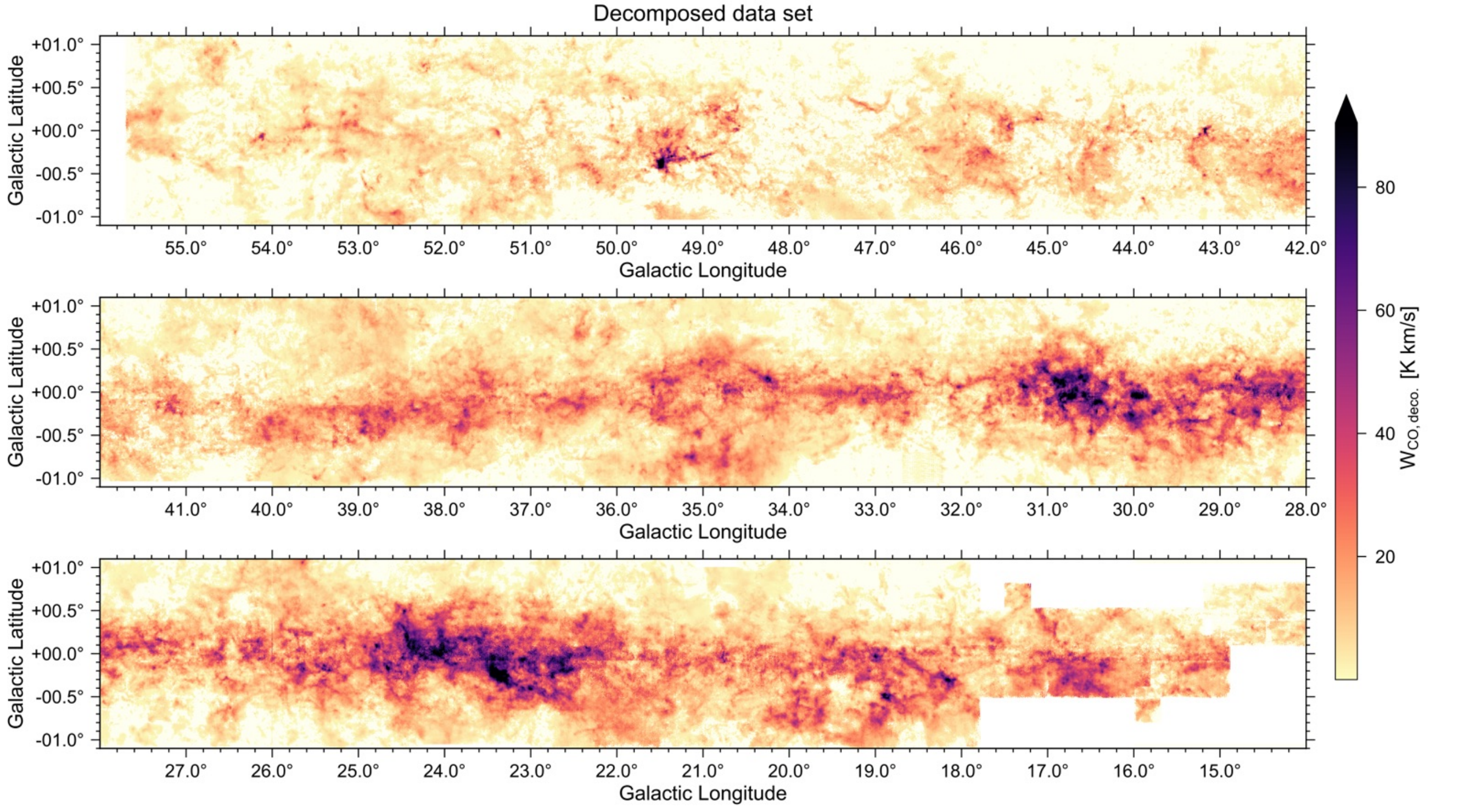}%
    \end{ocg}\\%
    \includegraphics[width=0.9\hsize]{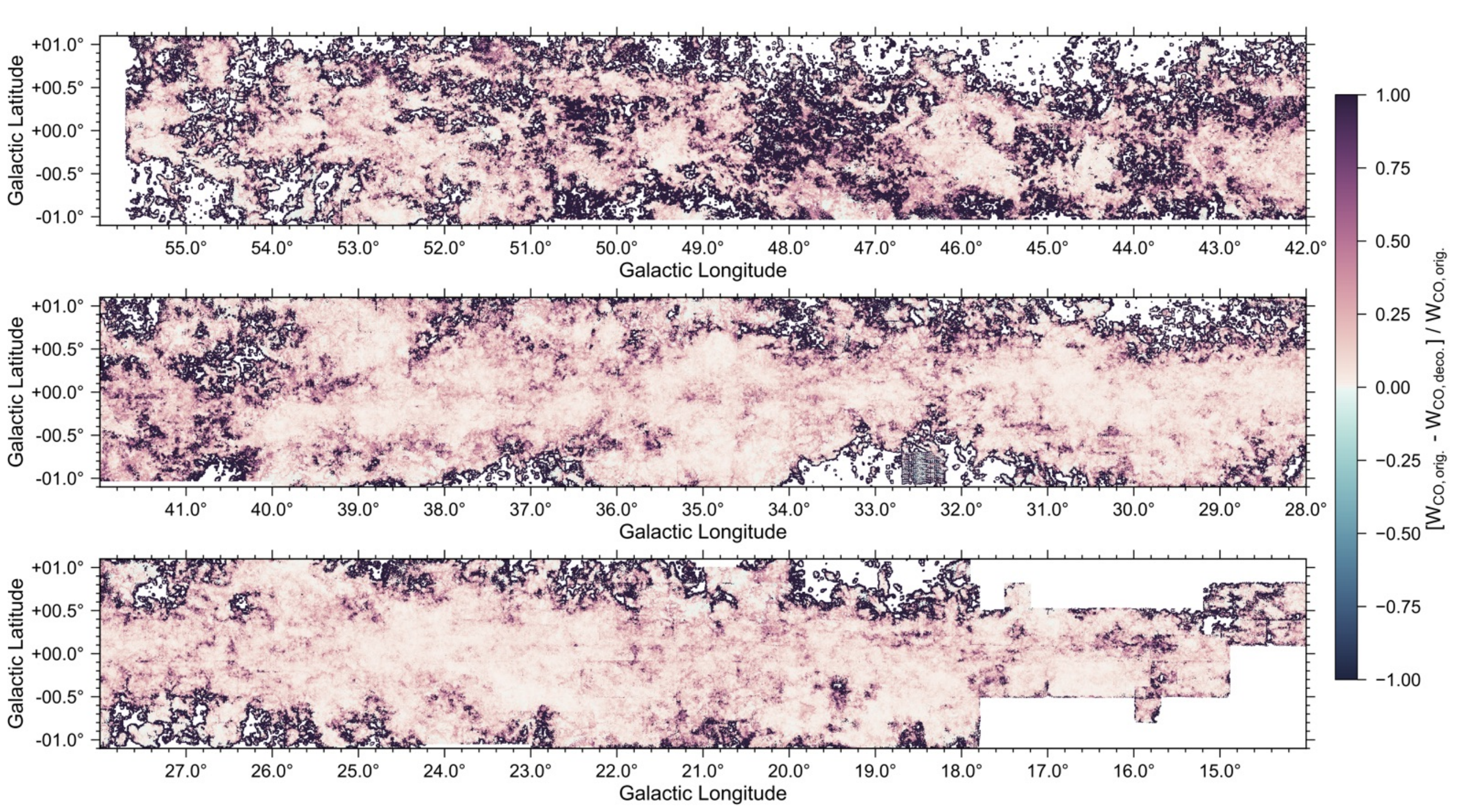}
    \caption{\emph{Top}: Zeroth moment map of the GRS data set recreated from the Gaussian fit components and integrated over the entire velocity velocity range ($-5 \lesssim \vlsr \lesssim 135$~\kms). 
    When displayed in Adobe Acrobat, it is possible to switch to the map of the  \ToggleLayer{fig:mom0_original_off,fig:mom0_original_on}{\protect\cdbox{original data set}}.
    \emph{Bottom}: Map of the normalised residual values. }
    \label{fig:zero_mom}
\end{figure*}

\begin{figure*}
    \centering
    \includegraphics[width=0.9\hsize]{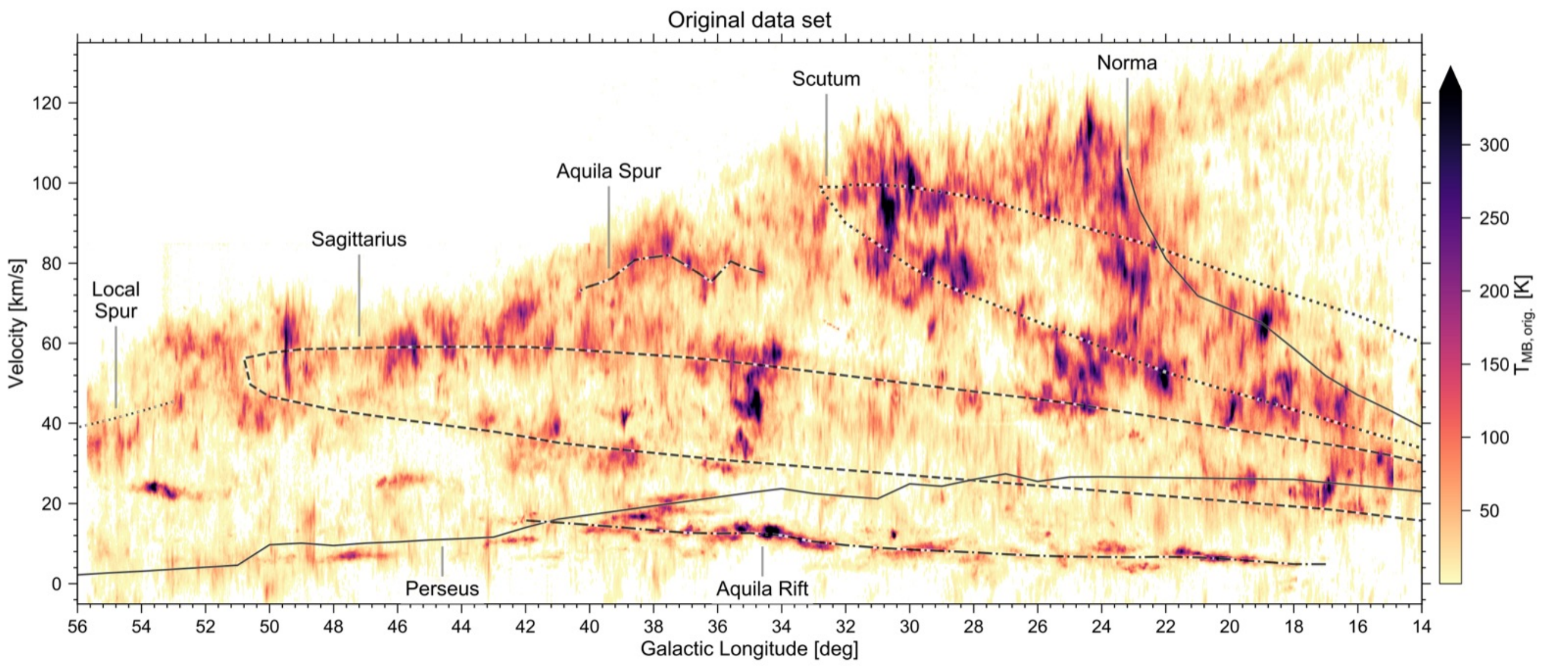}%
    \hspace{-0.9\hsize}%
    \begin{ocg}{fig:pv_original_off}{fig:pv_original_off}{0}%
    \end{ocg}%
    \begin{ocg}{fig:pv_original_on}{fig:pv_original_on}{1}%
    \includegraphics[width=0.9\hsize]{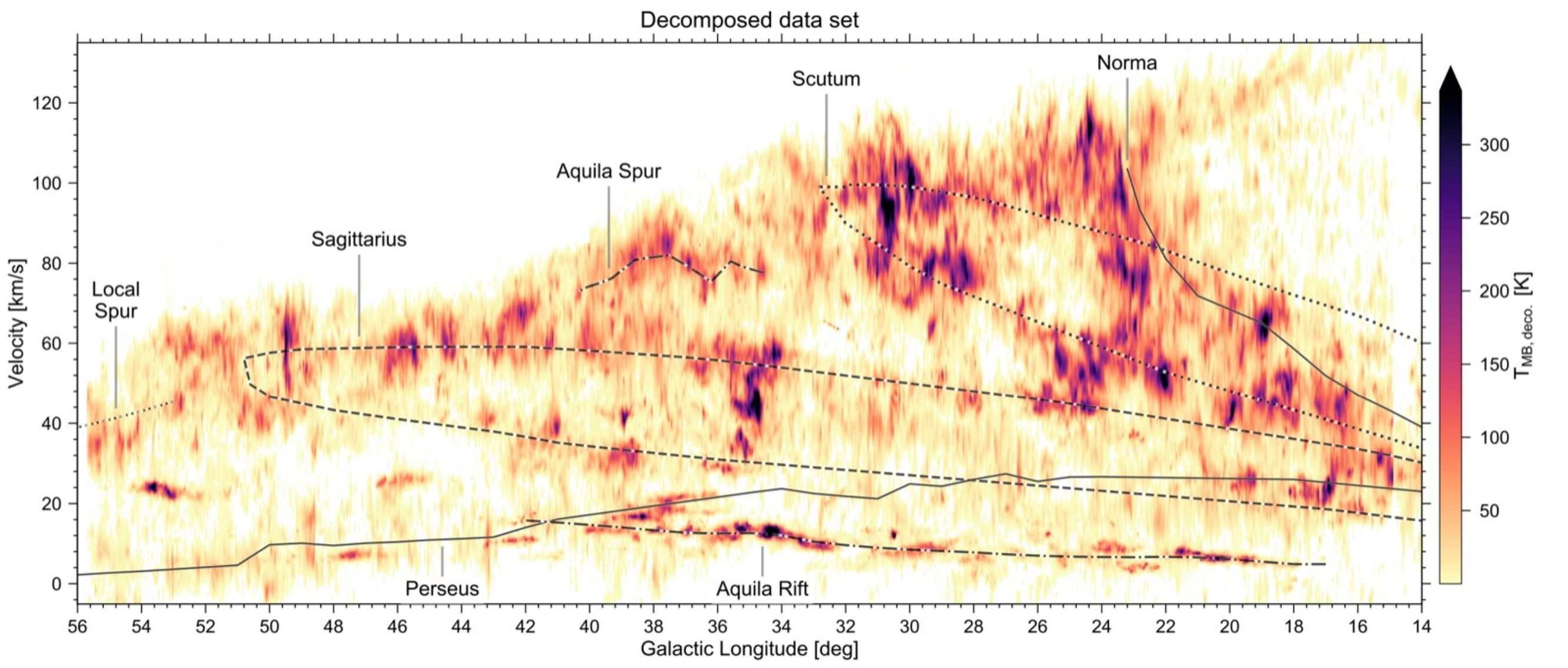}%
    \end{ocg}\\%
    \includegraphics[width=0.9\textwidth]{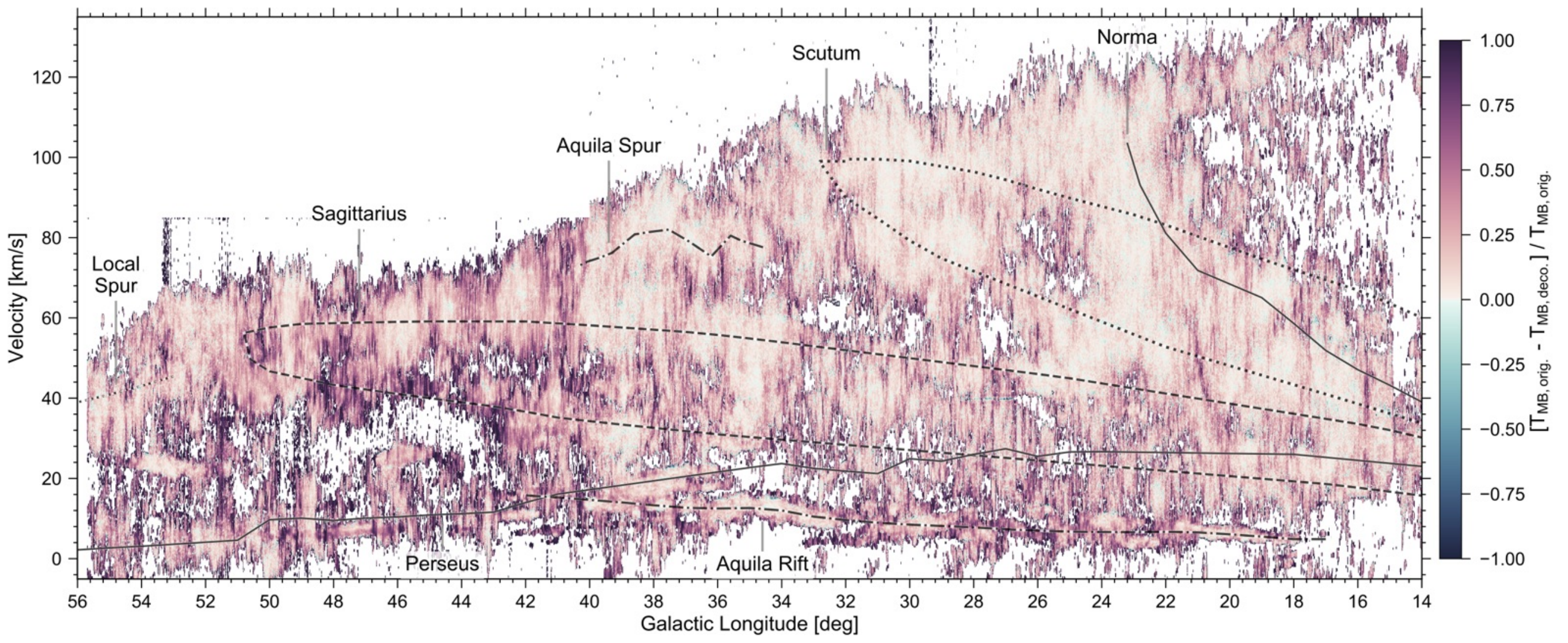}
    \caption{\emph{Top:} PV map of the decomposed GRS data set.
    When displayed in Adobe Acrobat, it is possible to switch to the map of the  \ToggleLayer{fig:pv_original_off,fig:pv_original_on}{\protect\cdbox{original data set}}.
    The emission was integrated over the full Galactic latitude range ($\pm 1.1\degr$). 
    \emph{Bottom}: Map of the normalised residual values.
    Overplotted on both panels are positions of spiral arms, spurs, and local Galactic features from \citet{Reid2016}.}
    \label{fig:grs-mosaic_pv+spiral_arms}
\end{figure*}

In this section we discuss the quality of our decomposition results in terms of recovered flux.
We start by comparing the average spectrum of the original GRS data set to that of the decomposition. 
Figure~\ref{fig:average_spectra} shows this comparison for spectra obtained using three different averaging methods.
In \fig\ref{fig:average_spectra}\emph{a}, we use all available voxels; in \fig\ref{fig:average_spectra}\emph{b}, we use the moment masking technique as outlined in \citet{Dame2011}; and in \fig\ref{fig:average_spectra}\emph{c}, we only use voxels that have a value above $3\times\rms$.
By comparing the average spectra of the original data and the decomposition, we can estimate the recovered flux in our decomposition, which is $84.3\%, 87.5\%, \text{and}~92.1\%$ for the three respective cases.
The moment masking technique provides the most accurate comparison, as it recovers most of the emission while still being relatively unaffected by noise contributions \citep{Riener2019}.

We can partly understand the missing flux by the fact that the moment masking technique recovers flux that is buried deep within the noise and will thus likely not be recovered in our decomposition.
The clipped average spectra in \fig\ref{fig:average_spectra}\emph{c} neglect all of the smaller intensity contributions below a S/N threshold of $3$. 
As expected, the fraction of recovered flux in our decomposition is higher for these high S/N regions. 
A comparison with \fig\ref{fig:average_spectra}\emph{b} shows that about a third of the total flux is coming from voxels with amplitude values below a S/N threshold of 3. 

Next, we discuss the average spectrum of negative values to show that the GRS data set contains significant negative spectral features.
The red line in \fig\ref{fig:average_spectra}\emph{c} shows an average spectrum of all voxels from the original data set with values smaller than $-3\times\rms$.
The inset in \fig\ref{fig:average_spectra}\emph{c} shows a zoomed-in version of the average spectrum of negative values. 
Most values of this average negative spectrum are due to random noise fluctuations that exceed a threshold of $-3\times\rms$ and cause the general offset (the jump occurring at $\vlsr \sim 35$~\kms\ is due to a change in the observing mode that resulted in lower noise values; see \citealt{Jackson2006}).
However, on top of this general offset are also many individual negative spikes located at specific $\vlsr$ values.
The most prominent negative peak is situated at a velocity range of $\vlsr = 11-17$~\kms~and reaches its lowest value at $\vlsr \sim 12.5$~\kms.
This artefact is due to the presence of $\co{13}{}$ emission in one of the "off" positions used in the subtraction of sky emission \citep{Jackson2006}.
The negative spikes at this position affect a significant number of spectra.
We suspect that the remaining smaller negative spikes are likely also due to contaminating $\co{13}{}$ emission in other "off" positions used in the sky subtraction.

In the default settings of \gausspyplus\ all negative peaks smaller than $-5\times\rms$ are automatically masked out for the decomposition, which led to the identification and masking out of negative spectral features for about $1.6\%$ of all spectra from the GRS data set.
The vast majority of identified negative spikes are located in the region with $32.8\degr \leq \ell \leq 38.1\degr$ and $-0.7\degr \leq b \leq 1.1\degr$.
Another region containing many negative spikes is located at $32.2\degr \leq \ell \leq 32.7\degr$ and $-1.1\degr \leq b \leq -0.71\degr$.
Not excluding these negative spikes leads to incorrect comparisons if all spectral channels are summed up, since the average spectrum from the original data set is substantially reduced at the spectral range where the largest negative peaks are located (cf. \fig\ref{fig:average_spectra}\emph{a}).

We now shift our comparison to the examination of the spatial distribution of the integrated intensity, that means the zeroth moment maps of the data and decomposition results (\fig\ref{fig:zero_mom}).
These maps were obtained with the moment masking technique outlined in \citet{Dame2011}, that means for the decomposition results we used the same unmasked spectral channels that were also used for the original data set.
The similarity of the two maps gives us already a qualitative confirmation that the decomposition manages to reproduce the data set well.
We provide a normalised residual map to quantify this similarity better (\emph{bottom} part of \fig\ref{fig:zero_mom}).
Positive and negative values can indicate spectra for which emission features were left unfit and spectra for which the final fit solution might fit a single component over multiple emission peaks, respectively.
The empty spaces correspond to unobserved regions and regions where the moment masking technique identified no signal.
The positions of high positive values in the normalised residual map are predominantly associated with diffuse emission in the original data set.
This diffuse emission was likely too buried within the noise to be identified in the decomposition.
More quantitatively, the $19.4\%$ of the spectra with normalised residual values of $1$ are responsible for $21\%$ of the residual emission but only account for $2.6\%$ of the total flux in the zeroth moment map of the original data set.
For $4.1\%$ of the spectra the value in the zeroth moment map of the decomposed data set is higher than for the original data set; this could indicate that noise was fitted or that a single component was incorrectly fitted over multiple signal peaks.
However, since for these spectra the emission from the decomposed data set is on average only higher by about $1\%$ than the emission in the original data set, we conclude that this is only a minor issue.

Features of high values in the normalised residual map can also be traced back to elevated noise levels (cf. \fig\ref{fig:noise_map}).
This correlation is expected, given that we use the same S/N thresholds over the entire survey region and the noise properties do vary significantly throughout the map.
We try to quantify this correlation by splitting the data set in two parts, using the median of the noise distribution ($\sigma(\Tb)=0.25$~K) as the threshold.
We find that the $50\%$ of spectra with noise values $> 0.25$~K contribute $62.8\%$ to the residual, confirming that the regions with higher noise are indeed correlated with higher residual values.

Next, we discuss the recovered flux in terms of the position-velocity (PV) map of the GRS (\fig\ref{fig:grs-mosaic_pv+spiral_arms}).
We obtained these PV maps by integrating the moment masked data over the full Galactic latitude range.
The positions of spiral arms and spurs from \citet{Reid2016} are also shown.
The GRS covers both the near and far sides of the Scutum and Sagittarius spiral arms, which is why they make a turn in the PV map with the near sides having the lower velocity values.

The similarity of the PV maps of the original and decomposed data sets is another reassurance that our fits managed to reproduce all main features of the GRS data set, which is also confirmed in the normalised residuals (\emph{bottom} part of \fig\ref{fig:grs-mosaic_pv+spiral_arms}).
More quantitatively, the $12.5\%$ of the data points in the normalised residual map with positive values of 1 -- so the positions where none of the features of the original PV map could be recreated in the decomposed version --  are responsible for $17.9\%$ of the residual flux but are associated with positions in the PV map that account for only $3.5\%$ of the emission.
These percentages again confirm that positions with the highest normalised residual values are correlated with weak or diffuse emission in the original data set that was difficult to decompose.


\section{Statistics of the Gaussian components}
\label{sec:stats}

In this section, we present the distributions of the Gaussian fit parameters, namely the amplitude or intensity $\Tb$, velocity dispersion $\veldisp$, and mean position $\vlsr$. We also examine the relationships between the fit parameters and discuss some general properties of the GRS emission line data.

\subsection{Number of fitted components}
\label{sec:ncomps}

\begin{figure*}
  \centering
  \includegraphics[width=0.9\textwidth]{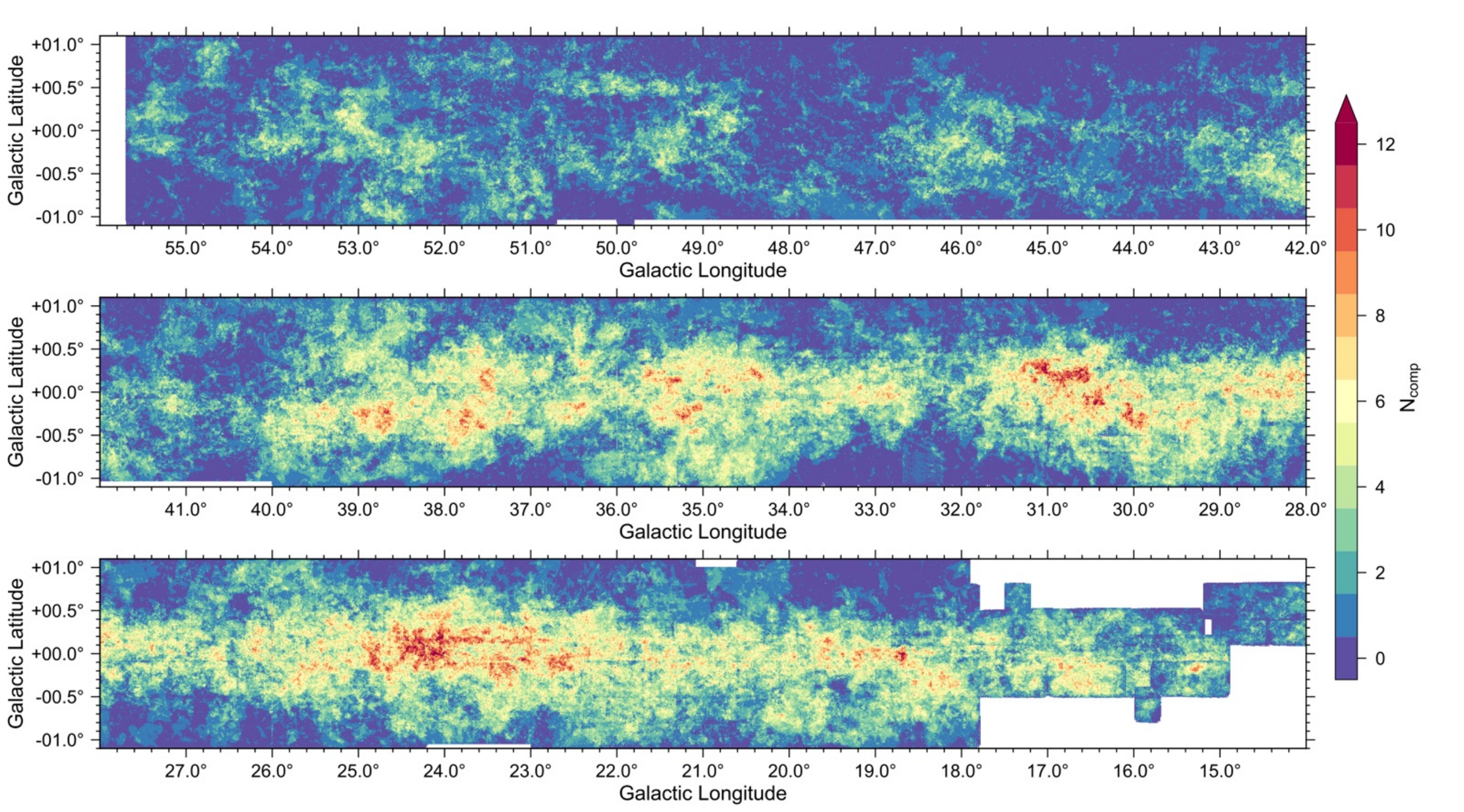}
  \caption{Map of the number of Gaussian fit components per spectrum.}
  \label{fig:component_map}
\end{figure*}

The number of fitted Gaussian components per spectrum ($\Ncomp$) is an interesting quantity because it is a measure of complexity of the CO emission along the line of sight.
For most of the GRS, we can assume that emission peaks in a spectrum that are well separated in $\vlsr$ will be associated with different Galactic orbits and will thus originate from different structures along the line of sight.
Figure~\ref{fig:component_map}, which shows the $\Ncomp$ values for the entire GRS coverage, is then a proxy for how many structures there are along the line of sight.
Especially near the Galactic midplane, several Gaussian components are required to fit the spectra.

\begin{figure}
    \centering
    \includegraphics[width=\columnwidth]{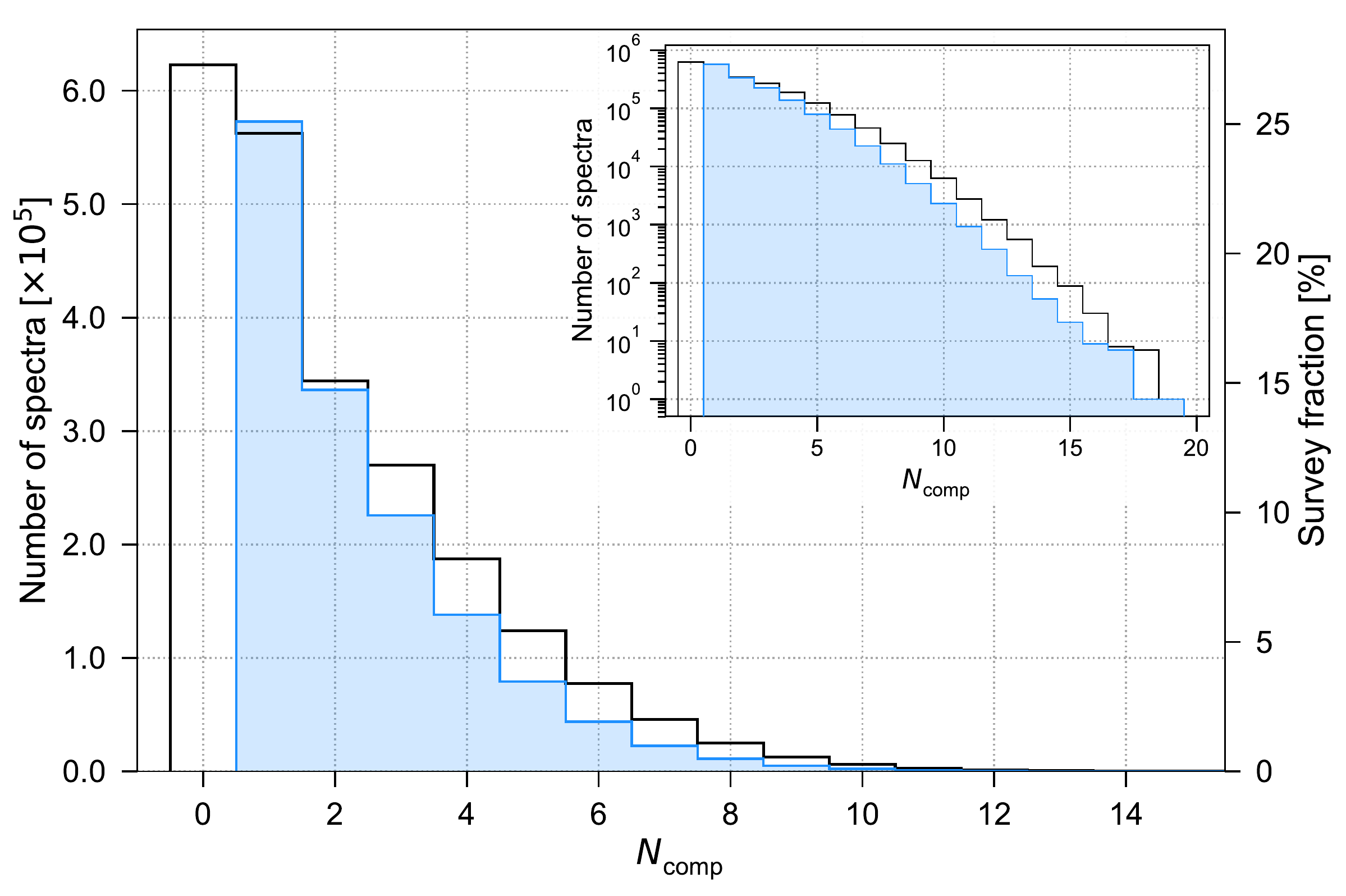}
    \caption{Histogram of the number of fitted Gaussian components per spectrum across the whole GRS survey for the full decomposition results (black line) and the fit components with S/N ratio > 3 (blue).
    The inset shows the same distribution on a logarithmic scale.}
    \label{fig:histogram_ncomps}
\end{figure}

We show a histogram of the number of fitted Gaussian components for the entire decomposition in \fig\ref{fig:histogram_ncomps}. 
For about $27.3\%$ of the spectra, we could fit no Gaussian components and for about $24.7\%$ of the spectra, only one Gaussian component was fitted.
The percentage of GRS spectra that have at least one or more fit components ($72.7\%$) is close to the percentage of spectra for which \gausspyplus\ identified signal peaks ($75.3\%$).
If we assume all signal peaks identified by \gausspyplus\ are correct, we get that for $2.6\%$ of the spectra from the GRS data set we do not fit valid signal peaks in our decomposition.
Most of these unfit signal peaks likely did not satisfy the minimum requirements for fit components in the \gausspyplus\ decomposition run.

We considered whether the $\sim 0.1\%$ of GRS spectra with best-fit solutions that use a high number of components ($\Ncomp > 10$) are indeed connected with very complex lines of sight or if they result from artefacts or problems in the decomposition.
Most of the spectra that were fitted with a large number of components do occur in groups near the mid-plane where complex spectra are expected (cf. \fig\ref{fig:component_map}).
We can use the information contained in Table~\ref{tbl:table_decomp} to gauge to first order whether these complex fit solutions are significantly different from their neighbours. 
A small difference between $\Ncomp$ and $\Nmed$, the weighted median number of components calculated from the immediate neighbours, and the number of component jumps $\Njump$ is a good first indication that the fit solution is similar to its neighbours.
In the default settings of \gausspyplus\ spectra get flagged if $\Delta N_{\text{max}} = \vert\Ncomp - \Nmed\vert > 1$ or $\Njump > 1$.
With these criteria $\sim 71\%$ of the $4884$ fit solutions using more than 10 components get flagged. 
Relaxing the criteria to $\Delta N_{\text{max}} > 2$ and $\Njump > 2$, to factor in uncertainties in the neighbouring fit solutions, and requiring that both of these criteria are satisfied reduces the percentage of flagged fit solutions to $\sim 31\%$.
Based on this analysis, we conclude that for about a third, but possibly the majority, of the spectra with high number of components ($\Ncomp > 10$) the fit solutions could be (partly) inconsistent with their neighbouring fit solutions.

There are multiple possible explanations for these inconsistencies, for example: \emph{i)} spectra that show instrumental artefacts (e.g. regions of the spectrum that fluctuate to very high and low values causing the decomposition algorithm to fit many of the high positive peaks), \emph{ii)} spectra for which the estimated noise value was too low, causing the decomposition algorithm to mistake noise peaks for signal peaks with high S/N, \emph{iii)} spectra that show leftover continuum emission that was fit with many individual components, or \emph{iv)} spectra that contained emission features deviating from a Gaussian shape that could not be fit well with Gaussian components.
We will return to the question of complexity along the line of sight in \sect\ref{sec:complexity}, where we will compare the number of fit components with the integrated CO emission and molecular gas surface densities derived from dust emission.

\subsection{Intensity values}
\label{sec:statistics}

\begin{figure}
    \centering
    \includegraphics[width=\columnwidth]{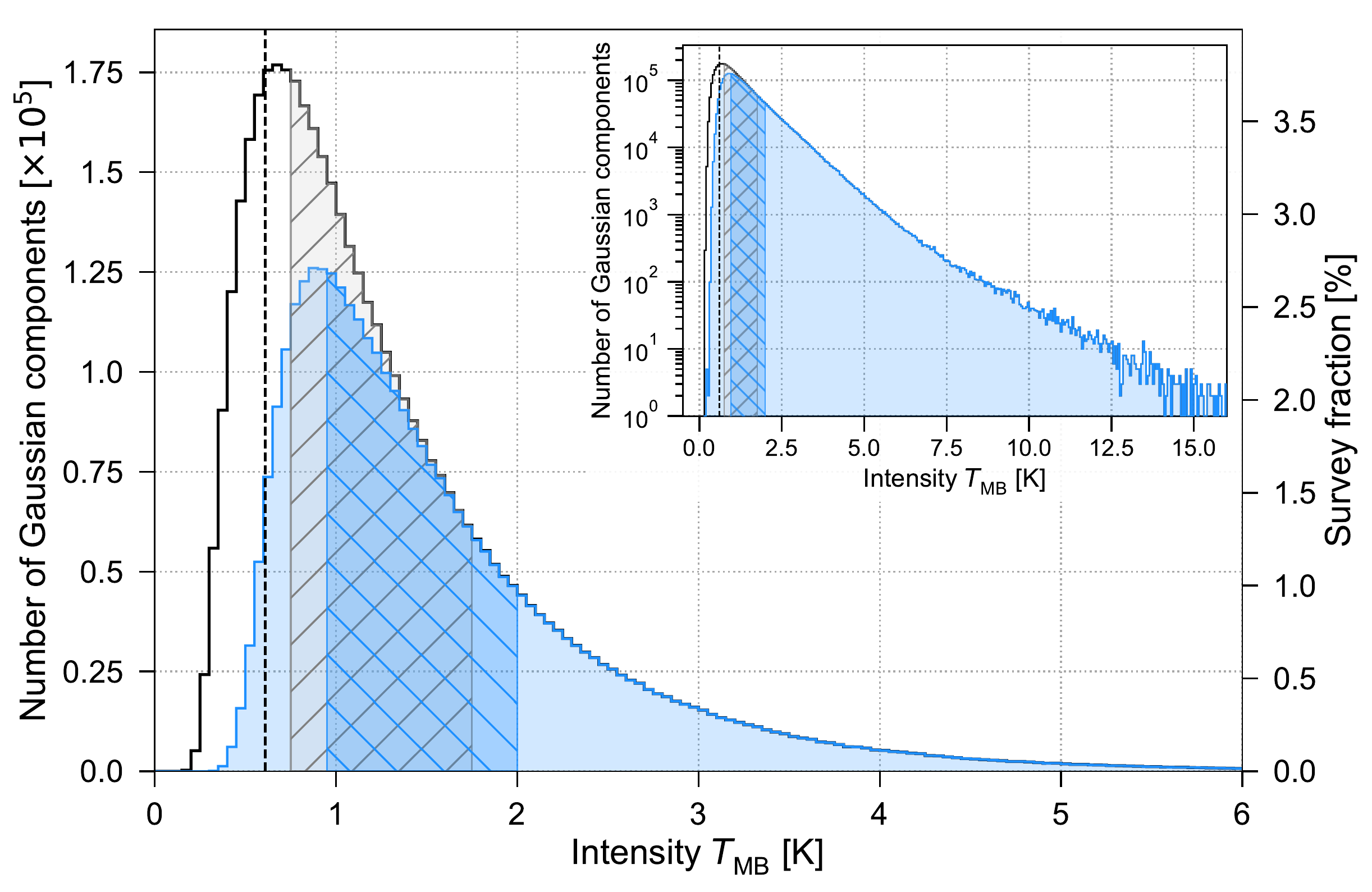}
    \caption{Histogram of intensity values for all fit components (black line) and fit components with S/N ratios $> 3$ (blue).
    The inset shows the same distribution on a logarithmic scale.
    The dotted vertical line shows the $3\times\text{S/N}$ limit for the peak value of the noise distribution (shown in \fig\ref{fig:histogram_noise}).
    The hatched areas mark the IQRs of the two distributions.
    The bin width is 0.05~K.}
    \label{fig:histogram_intensity}
\end{figure}

The distribution of the fitted amplitude values for the entire GRS data set (\fig\ref{fig:histogram_intensity}) peaks at about $\Tb = 0.68$~K and the interquartile range (IQR) is $0.71 - 1.71$~K.
For the subset of components with S/N > 3, the distribution peaks at a value of about $0.88$~K and the IQR is $0.95 - 1.98$~K. 
The dotted vertical line at $\Tb = 0.61$~K shows the typical sensitivity limit of $3\times\text{S/N}$ based on the peak value from the noise distribution shown in \fig\ref{fig:histogram_noise} ($\sigma(\Tb) = 0.2$~K), which is very close to the peak value of the distribution of intensity values.
The drop in the intensity distribution below the indicated sensitivity limit is thus likely not a physical effect but a result of the noise properties preventing the extraction of weaker signal peaks. 
When viewed logarithmically, the distribution shows an almost linear decrease between $\Tb \sim 1 - 6\,$K, after which it flattens.
We checked the spatial distribution of the fit components with high amplitude values of $\Tb > 6$~K.
The vast majority of these components form separate connected structures on scales of individual molecular clouds, with $\Tb$ values $\gtrsim 10$~K concentrated at their centres.
We thus conclude that most of the components with high $\Tb$ are not due to instrumental artefacts, but come from high column density regions.

\subsection{Centroid velocity values}
\label{sec:centroids}

\begin{figure}
    \centering
    \includegraphics[width=\columnwidth]{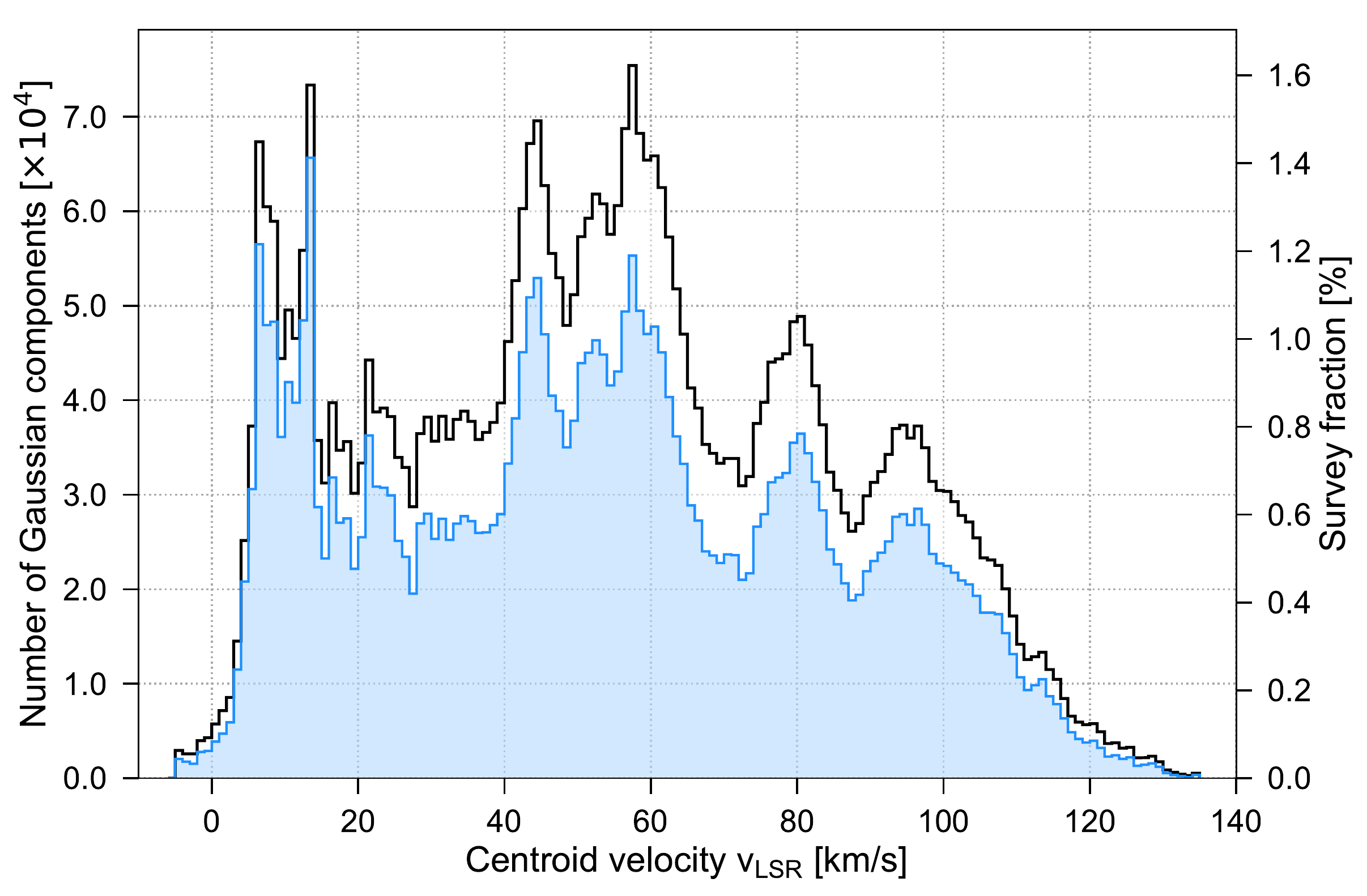}
    \caption{Histogram of centroid velocity values for all fit components (black) and fit components with S/N ratios $> 3$ (blue).
    The bin width is $1$~\kms.}
    \label{fig:histogram_means}
\end{figure}

The distribution of the $\vlsr$ positions of the Gaussian fit parameters (\fig\ref{fig:histogram_means}) shows that we do fit components across the entire velocity range ($-5~ \text{to}~135$~\kms) spanned by the GRS.
Some of the peaks in the distribution can be attributed to inferred positions of spiral arms, spurs, or local Galactic features (cf. the PV map in \fig\ref{fig:grs-mosaic_pv+spiral_arms}), for example the Aquila Rift cloud at centroid velocities of about $10$~\kms\ and the far and near portion of the Sagittarius and Scutum spiral arm, respectively, at around $60$~\kms. 
As expected, the shape of the distribution is very similar to the average spectra from \fig\ref{fig:average_spectra}.
A comparison between those two figures shows that even though the number of fit components with $\vlsr$ values of $0 - 20$~\kms\ is high, their average intensity values are much lower than for the $\vlsr$ range between $\sim 40 - 70$~\kms.
This is likely due to a larger contribution of diffuse, faint emission from local gas at low $\vlsr$ values.
For regions close to the sun that are spatially well resolved, we can have only diffuse emission in the beam, which causes comparatively weak emission lines.
At larger distances, where much larger physical areas are covered in the beam, this diffuse emission will likely be diluted and merged with stronger emission peaks, so that diffuse and strong emission is detected simultaneously in the beam.
Moreover, the moderate spatial resolution of the GRS can cause stronger emission lines for molecular clouds observed at larger distances (if the beam filling factor is still approximately unity).

\subsection{Velocity dispersion values}
\label{sec:velocity-dispersion}

\begin{figure}
    \centering
    \includegraphics[width=\columnwidth]{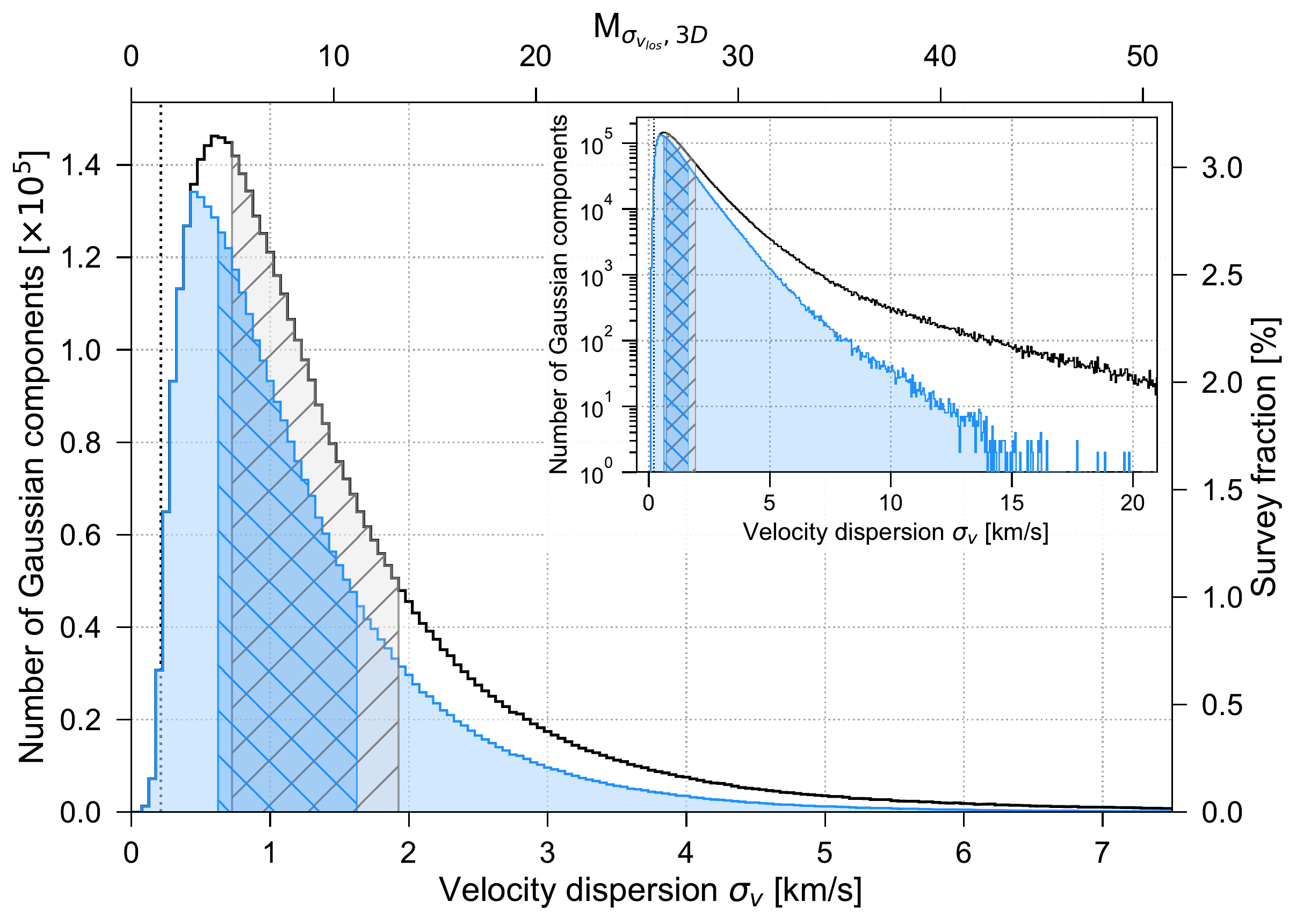}
    \caption{Histogram of velocity dispersion values for all fit components (black) and fit components with S/N ratios $> 3$ (blue).
    The upper abscissa indicates corresponding upper limits for turbulent Mach number values.
    The inset shows the same distribution on a logarithmic scale.
    The dotted vertical line indicates the velocity resolution of $0.21$~\kms.
    The hatched areas mark the IQRs of the two distributions.
    The bin width is $0.05$~\kms.}
    \label{fig:histogram_veldisp}
\end{figure}

The distribution of the $\veldisp$ values of the Gaussian fit components (\fig\ref{fig:histogram_veldisp}) is interesting, as it allows us to already estimate to first order upper limits for the turbulent Mach number associated with the emission lines.

The peak for the distribution of $\veldisp$ values for all fit components is at $\veldisp \sim 0.6$~\kms\ and the IQR is $0.68 < \veldisp < 1.89$~\kms. 
For the subset of fit components with S/N ratios > 3, the peak is shifted toward a lower value of $0.45$~\kms\ and the IQR is $0.59 - 1.58$~\kms. 
About $0.6\%$ of all fitted components have a velocity dispersion value below the resolution limit.
In the default settings of \gausspyplus\, the minimum allowed value for the full width at half maximum (FWHM) is set to the width of a single spectral channel, which yields $\veldisp$ values below the resolution limit.

When plotted logarithmically, the distribution has a linear dropoff from about $1 - 4$~\kms, after which it flattens and has a shallower decline; it also shows that most of the fit components with very broad FWHM values have S/N ratios $< 3$.
It is likely that most of these broad components were fit over multiple low S/N peaks that could not be correctly deblended.
A visual inspection of some of the GRS spectra showed that a small fraction also suffers from what seems to be an incorrect or insufficient baseline subtraction, which could lead to broad features with low S/N ratios in the spectrum.

We can get upper limits for the turbulent Mach number $\Mach$ by assuming that all non-thermal contributions to the velocity dispersion $\veldisp$ are due to turbulence:

\begin{equation}
	\pazocal{M}_{\sigma_{v_{los}}, 3\text{D}} \approx \sqrt{3} \dfrac{\sigma_{v_{\text{turb},\,1\text{D}}}}{c_{\mathrm{S}}} = \sqrt{3}\left[\left(\dfrac{\veldisp}{c_{\mathrm{S}}}\right)^{2} - \left(\dfrac{\bar{\mu}_{p}}{\mu_{obs}}\right)\right]^{1/2},
    \label{eq:mach}
\end{equation}

\noindent where $\veldisp$ is the velocity dispersion along the line of sight, $c_{\mathrm{S}}$ is the isothermal sound speed, $\bar{\mu}_{p}$ is the mean molecular mass ($\bar{\mu}_{p} = 2.33$~amu), and $\mu_{obs}$ is the molecular mass of the observed molecule ($29$~amu for $^{13}$CO).
We estimate the isothermal sound speed $c_{\mathrm{S}}$ with

\begin{equation}
	c_{\mathrm{S}} = \sqrt{\dfrac{k_{\mathrm{B}}T_{\mathrm{k}}}{\bar{\mu}_{p}m_{\mathrm{H}}}},
\label{eq:soundspeed}
\end{equation}

\noindent where $T_{\mathrm{k}}$ is the kinetic temperature of the gas, $k_{\mathrm{B}}$ is the Boltzmann constant, and $m_{\mathrm{H}}$ is the mass of atomic hydrogen.
For $T_{\mathrm{k}}$ we assume a uniform value of $18$~K throughout the survey, which corresponds to the average mean line of sight dust temperature for the HiGAL data overlapping with the GRS as estimated by \citet{Marsh2017-ppmap}.

The upper abscissa in \fig\ref{fig:histogram_veldisp} shows the resulting Mach number axis. The peak of the distribution corresponds to a Mach number of about $4$.
About $36.7\%$ of all fit components and $28.6\%$ of the fit components with S/N ratio $>3$ are associated with turbulent Mach number values $> 10$.
About $3\%$ of the fit components have $\veldisp$ values $> 5$~\kms, resulting in associated high Mach numbers that are greater than $34$.
To put these large $\veldisp$ values into perspective, we can compare them to typical line width values found on cloud scales.
For their catalogue of molecular clouds in the GRS, \citet{Rathborne2009} found average and maximum values for the line width of $3.6$ and $9.8$~\kms, which translate to $\veldisp$ values of $1.5$ and $4.2$~\kms, respectively.
Given these values, it seems unlikely that the fit components with large $\veldisp$ values trace regions of extreme turbulence; it seems more likely they are: \emph{i)} due to non-random ordered motion, for example velocity gradients along the line of sight; \emph{ii)} incorrect fits of multiple signal peaks with a single component; \emph{iii)} fits of artefacts in the spectrum, introduced for example by insufficient or incorrect baseline subtraction; \emph{iv)} associated with warmer gas temperatures (that increase the thermal linewidth and decrease the non-thermal contribution).

\subsection{Relationships between the parameters}
\label{sec:relation}

\begin{figure}
    \centering
    \includegraphics[width=\columnwidth]{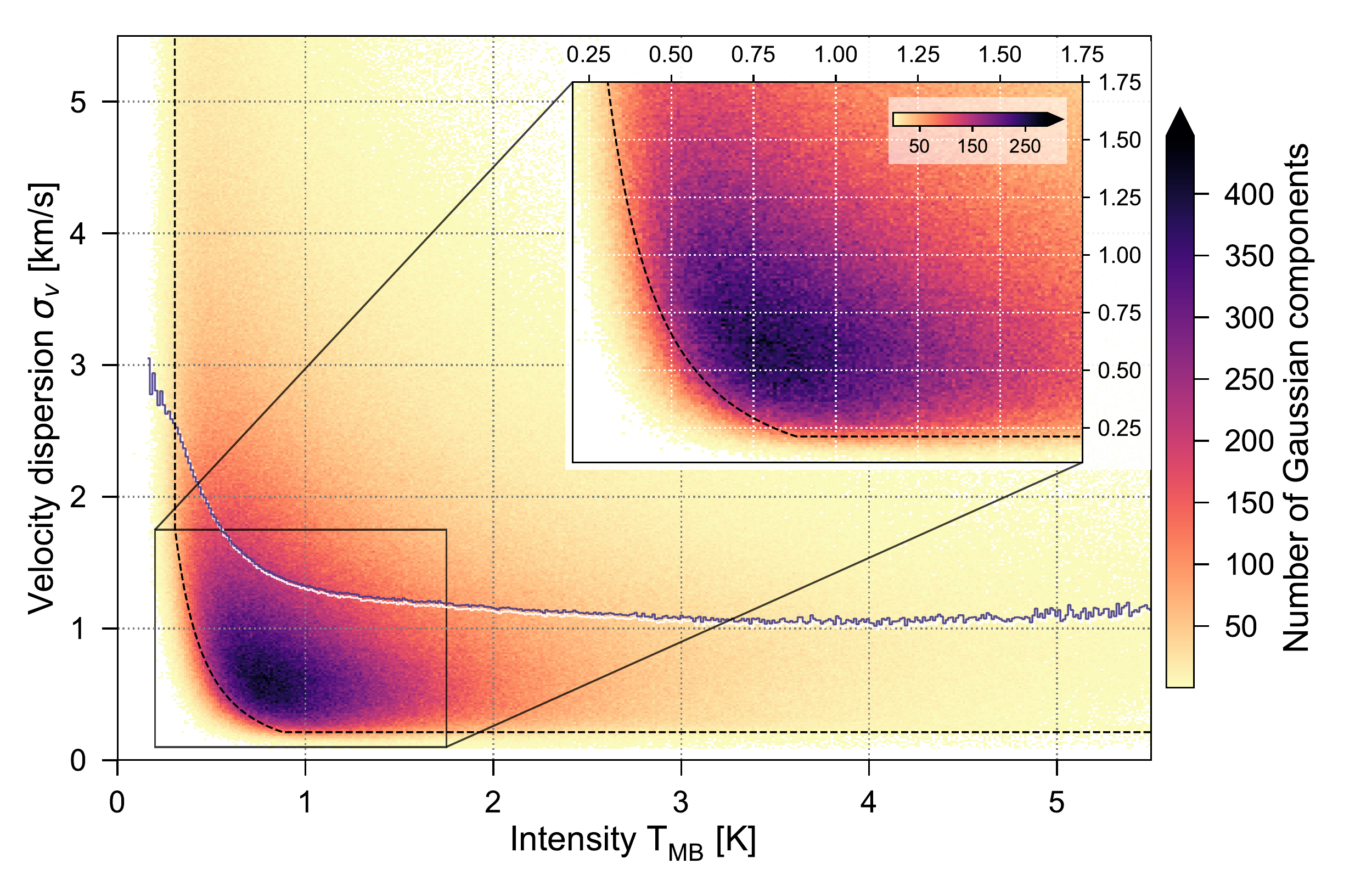}
    \caption{2D histogram of intensity and velocity dispersion values for all fit components.
    The dashed line shows the lower limit for a chosen significance value of $\significance{{\Min}} = 5$ and a a velocity dispersion value corresponding to the spectral resolution limit of the survey ($0.21$~\kms).
    The blue line shows the mean $\veldisp$ value per intensity bin.
    The number of bins in either direction is 400.
    The inset is a zoom-in of the most dense region of the distribution and uses a finer grid with 175 bins in either direction.}
    \label{fig:histogram_2d_intensity-veldisp}
\end{figure}

\begin{figure}
    \centering
    \includegraphics[width=\columnwidth]{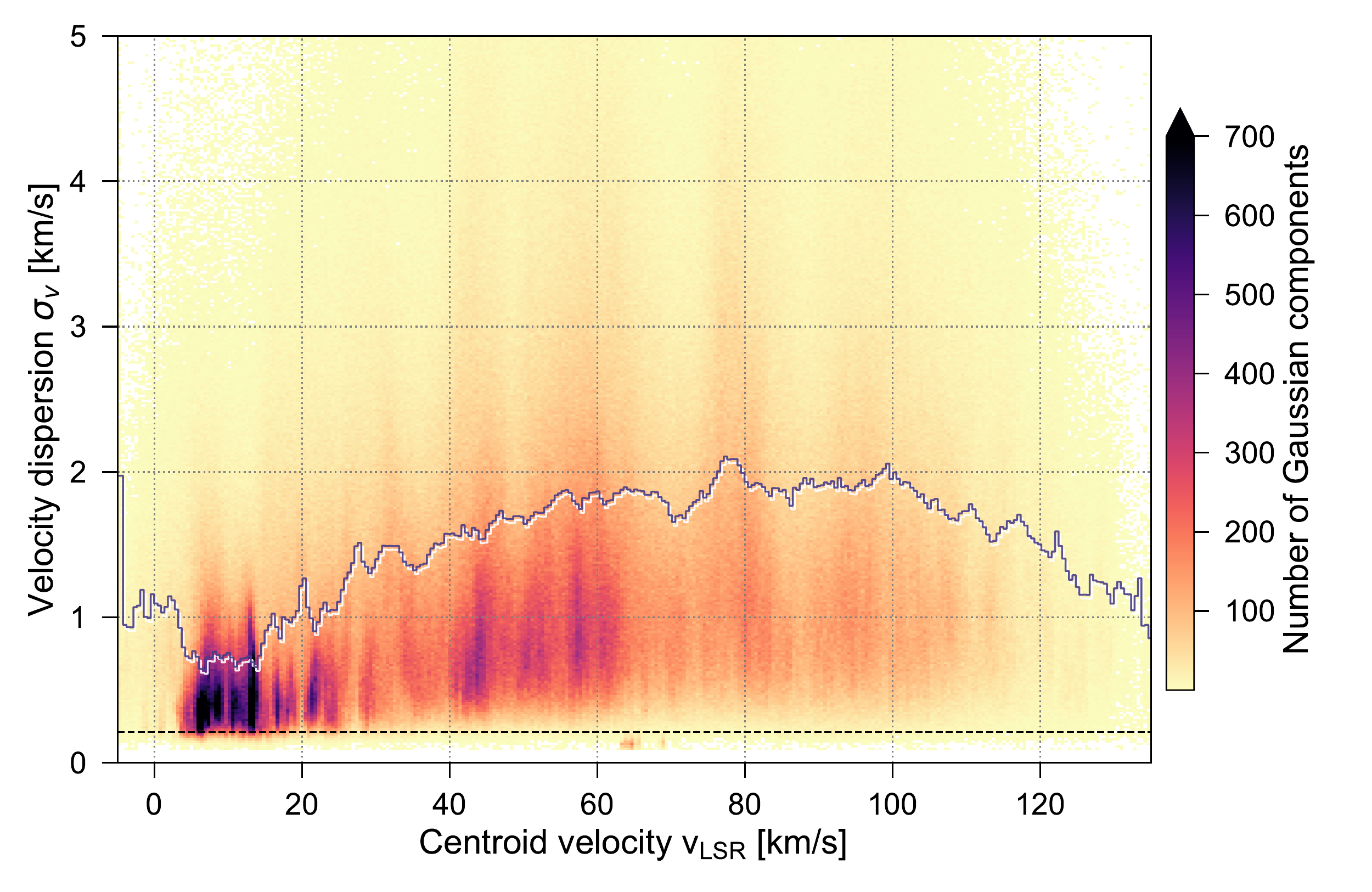}
    \caption{2D histogram of centroid velocity and velocity dispersion values for all fit components.
    The dashed horizontal line indicates the velocity resolution of $0.21$~\kms.
    The blue line shows the mean $\veldisp$ value per centroid velocity bin.
    The number of bins in either direction is $300$.
    The data points with very low $\veldisp$ values located at $\sim 63 < \vlsr\ < 66$~\kms\ are due to an instrumental artefact.}
    \label{fig:histogram_2d_centroid_vel-vel_disp}
\end{figure}

\begin{figure}
    \centering
    \includegraphics[width=\columnwidth]{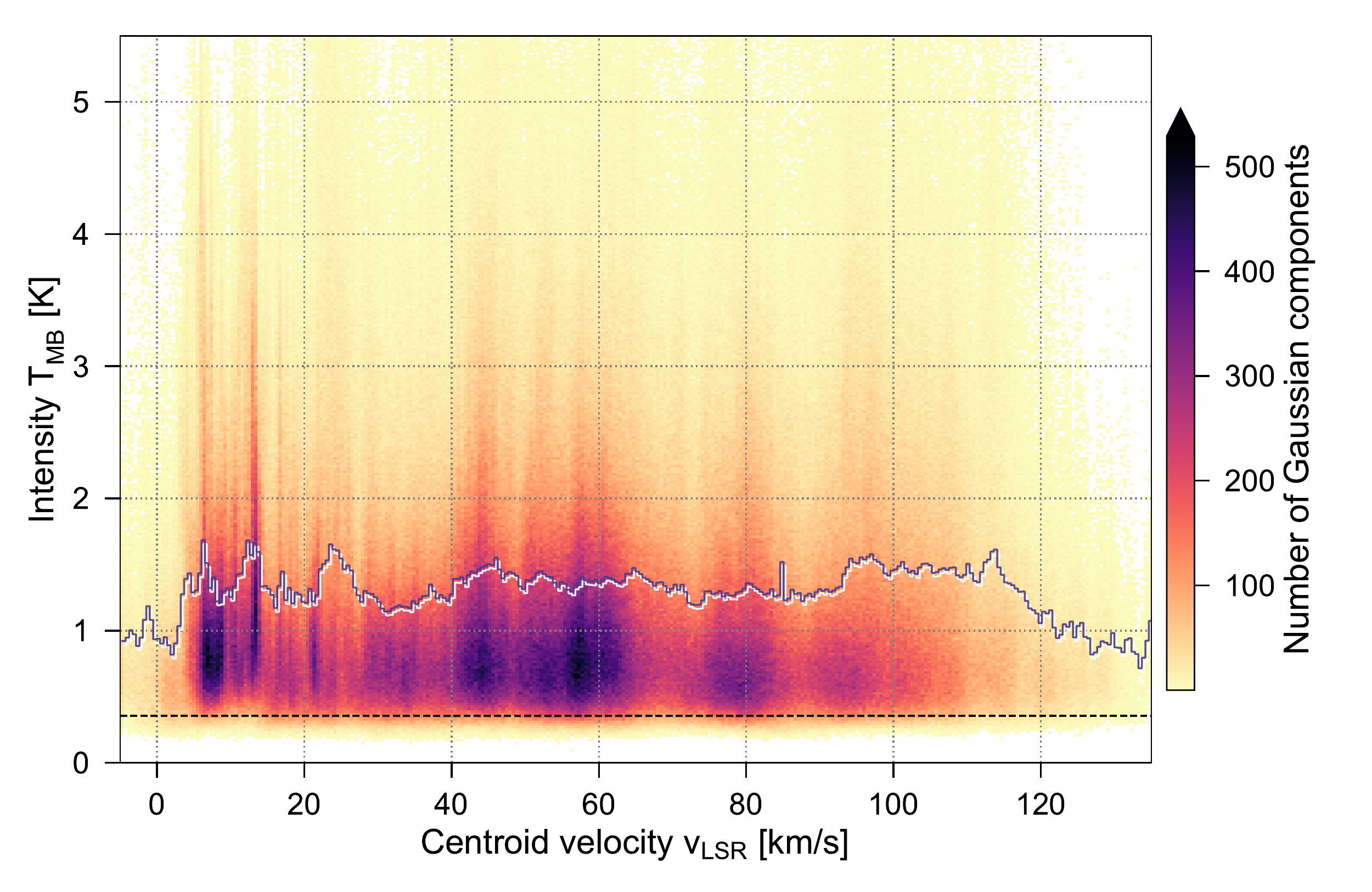}
    \caption{2D histogram of centroid velocity and intensity values for all fit components.
    The dashed horizontal line indicates a S/N limit of $3$ for $\sigma(\Tb)=0.12$~K, which corresponds to the $0.1^{st}$ percentile of the noise distribution shown in \fig\ref{fig:histogram_noise}.
    The blue line shows the mean intensity value per centroid velocity bin.
    The number of bins in either direction is $300$.}
    \label{fig:histogram_2d_centroid_vel-intensity}
\end{figure}

In this section we describe the relationships between the fit parameters using two-dimensional (2D) histograms. 
This will allow us to better characterise our population of fit components, for example, the typical shape of the fitted lines.
We will also look at trends of $\Tb$ and $\veldisp$ values with $\vlsr$ position, which we can use to infer to first order properties of the gas emission at different Galactic distances.

Figure.~\ref{fig:histogram_2d_intensity-veldisp} shows the 2D distribution of the intensity and velocity dispersion values of all Gaussian fit components. 
The majority of components show both moderate intensity and velocity dispersion values ($0.5 \lesssim \Tb \lesssim 1.5$~K, $0.25 \lesssim \veldisp \lesssim 1.5$~\kms).
Beyond this concentration of most data points, the distribution shows a bifurcation, with the components having either high intensity and small velocity dispersion values or low intensity and high velocity dispersion values.
That components with high $\veldisp$ values are predominantly connected with low $\Tb$ values is an indication that these are likely fits to artefacts in the spectrum, as discussed in \sect\ref{sec:velocity-dispersion}.

The absence of components with both low intensity and low velocity dispersion values is due to a selection effect in our decomposition.
We only retain Gaussian components above a chosen significance value $\significance{\Min}$ (see Sect.~3.2.1 in \citealt{Riener2019}), which excludes components with low intensity and low velocity dispersion values, because they are indistinguishable from individual random noise peaks.
In \fig\ref{fig:histogram_2d_intensity-veldisp}, we indicate the expected lower limit for a significance limit of $\significance{\Min} = 5$, the spectral resolution limit of the GRS survey ($0.21$~\kms), and a typical noise value of $\rmsTa = 0.098$~K (corresponding to the peak of the noise distribution shown in \fig\ref{fig:histogram_noise}).
This lower limit corresponds very well to the shape of the distribution, with velocity dispersion values being clearly limited by the spectral resolution.
Thus, \fig\ref{fig:histogram_2d_intensity-veldisp} also serves as a good indicator of the sensitivity limit of our decomposition.

Only about $4\%$ of all fit components have both high values of $\Tb$ ($> 2$~K) and $\veldisp$ ($> 2$~\kms).
Moreover, the IQR of the $\vlsr$ values of these strong fit components stretches from $57.3\ \text{to}\ 94.5$~\kms\ and is thus clearly shifted toward higher values compared to the IQR of the full distribution ($30.5\ \text{to}\ 78.1$~\kms). 

We can see a similar trend of increasing line widths at higher $\vlsr$ values in the relation between the fitted $\vlsr$ and $\veldisp$ values (\fig\ref{fig:histogram_2d_centroid_vel-vel_disp}).
In contrast, the distribution of fitted intensity values with $\vlsr$ position (\fig\ref{fig:histogram_2d_centroid_vel-intensity}) is on average very constant at a value of about $1.5$~K, which is expected from the distance-independence of the surface brightness.
However, \fig\ref{fig:histogram_2d_centroid_vel-intensity} also shows very bright intensity peaks, in particular from components located at velocities of $\sim 10$~\kms, most of which also have narrow line widths (\fig\ref{fig:histogram_2d_centroid_vel-vel_disp}).
This emission at velocities of about $\sim 10$~\kms\ predominantly originates from the very nearby Aquila Rift cloud and is thus spatially and spectrally well resolved in the GRS observations, which explains the narrow and bright emission peaks.
For regions farther away, beam averaging effects will cause broader emission lines, which will be reflected in the shape of the fit components.
We will discuss these trends further in \sect\ref{sec:confusion}, where we will put them into the context of a size-linewidth relationship.


\section{Global properties of the gas emission}
\label{sec:properties}

In this section we focus on the global properties of our decomposition. We first look at the distribution of the velocity dispersion values with Galactic coordinates and then use the number of fit components per spectrum to gauge the complexity of the gas emission along the line of sight.

\subsection{Distribution of velocity dispersion values as a function of Galactic coordinates}
\label{sec:distribution-veldisp-plane}

\begin{figure}
    \centering
    \includegraphics[width=\columnwidth]{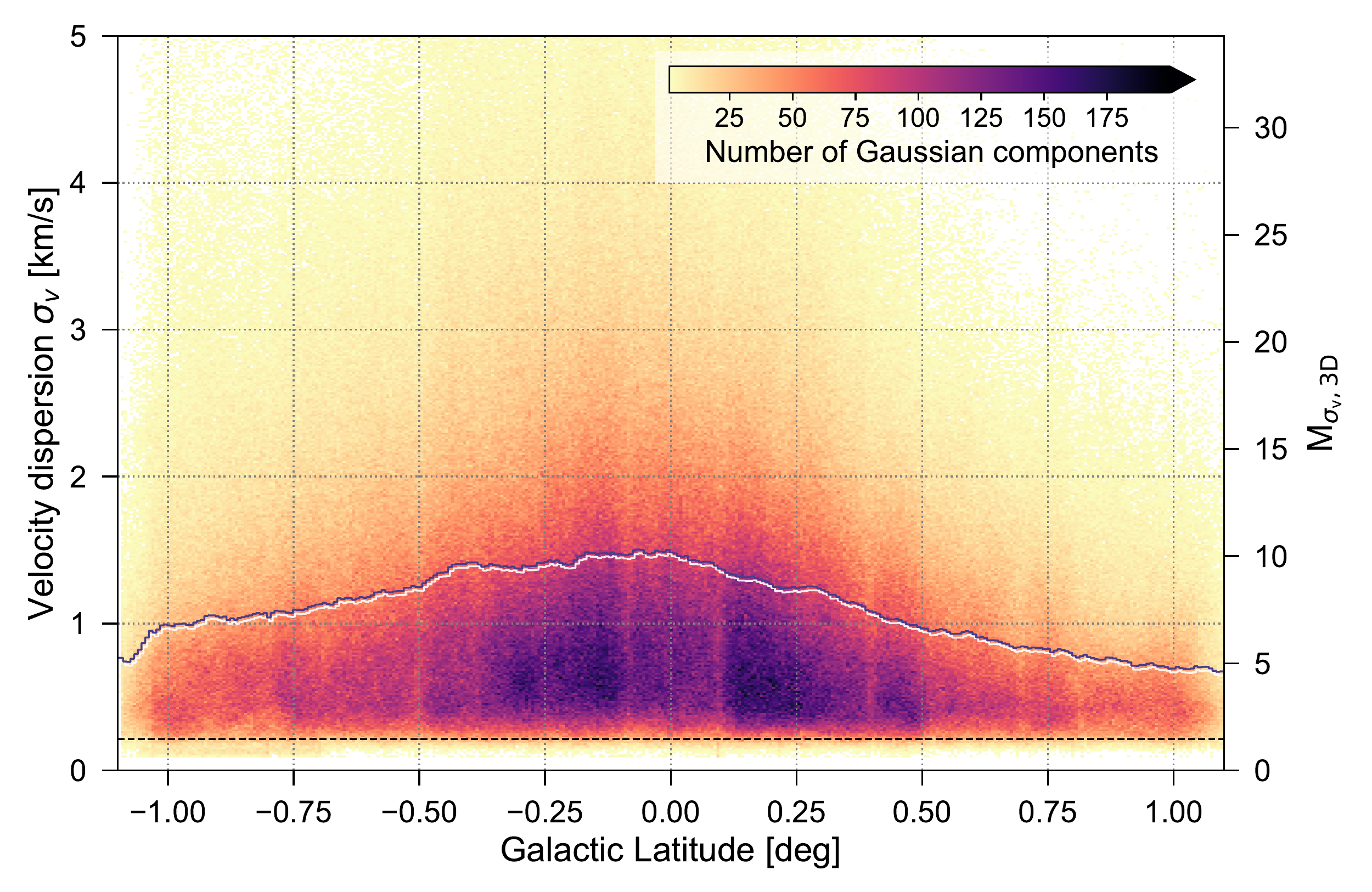}
    \caption{2D histogram of the velocity dispersion values against their Galactic latitude positions for all fit components with S/N $> 3$.
    The dashed horizontal line indicates the velocity resolution of $0.21$~\kms.
    The blue line shows the mean $\veldisp$ value per Galactic latitude bin.
    The number of bins used in either direction is $358$.
    }
    \label{fig:histogram_2d_vel_disp-GLAT}
\end{figure}

\begin{figure*}
    \centering
    \includegraphics[width=\textwidth]{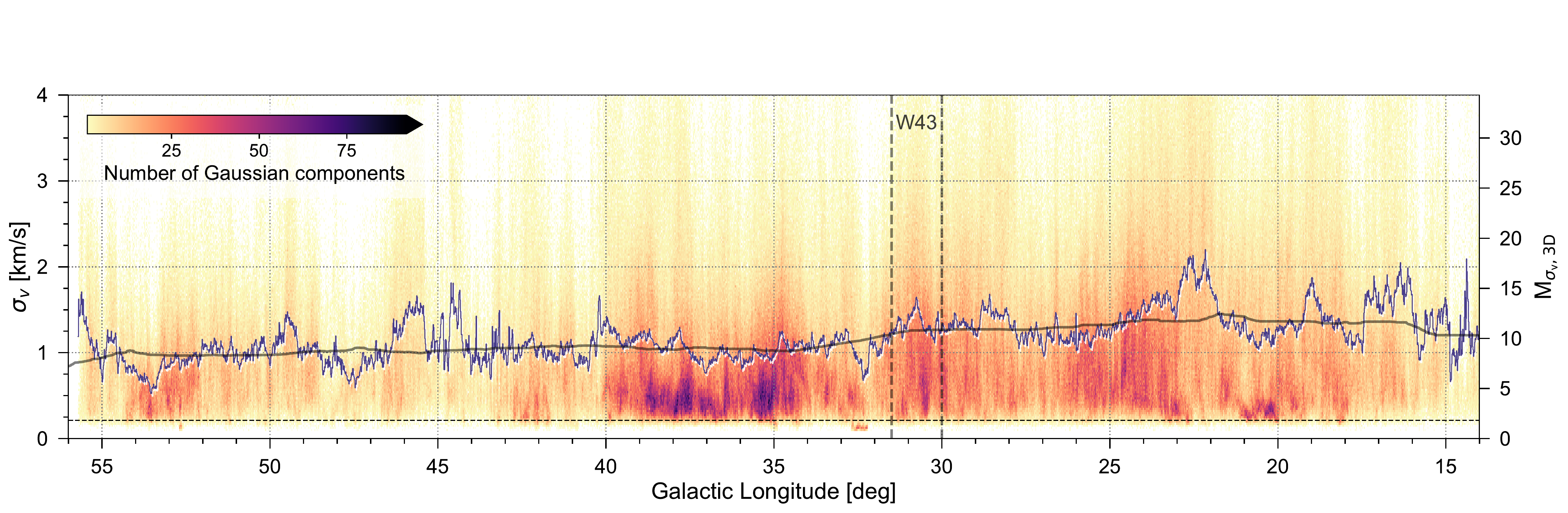}
    \caption{2D histogram of velocity dispersion values against their Galactic longitude positions for all fit components with S/N ratio > 3.
    The dashed horizontal line indicates the velocity resolution of $0.21$~\kms.
    In the horizontal and vertical direction we use $3415$ and $200$ bins, respectively.
    The blue line indicates the mean velocity dispersion value per Galactic longitude bin.
    The gray solid line is a smoothed version of the mean $\veldisp$ curve using a median filter with a kernel of $\sim 3\degr$.
    Dashed vertical lines show the approximate location of the giant HII region complex W43.
    }
    \label{fig:histogram_2d_GLON-vel_disp}
\end{figure*}

Here we examine how the velocity dispersion values of the fit components are distributed as a function of the Galactic coordinates.
We search for global trends and outlying regions, such as regions with above-average non-thermal motion, within the Galactic disk.
We focus on fit components with S/N ratios $> 3$ in this analysis, since the line shape of components with a lower S/N ratio could be significantly impacted by noise.
We again use the $\veldisp$ values to estimate upper limits for the turbulent Mach numbers assuming isothermal gas temperatures of 18~K (see \sect\ref{sec:velocity-dispersion}).

We first focus on the distribution of the $\veldisp$ values as a function of Galactic latitude (\fig\ref{fig:histogram_2d_vel_disp-GLAT}).
The number of high velocity dispersions (and Mach numbers) increase toward the Galactic midplane, which is likely due to the following three effects. 
First, most substantial star-forming regions are concentrated in the midplane \citep[e.g.][]{Beuther2012_Atlasgal}. 
These star-forming regions are associated with warmer molecular gas and plenty of high-velocity dispersion events (such as protostellar jets, outflows, shells, and supernova remnants), which will cause broader emission lines and higher $\veldisp$ values of the associated fit components.
Second, we would expect more confusion along the line of sight toward the midplane, where multiple individual velocity components could be blended together and would thus be observed as a single broad component at the moderate resolution of the GRS.
Third, to maintain a stable disk, the gas pressure of the Galactic disk must balance the imposed pressure created by the weight of the vertical layers of gas and stars.
This pressure will be higher toward the midplane and will accordingly increase the velocity dispersion in the gas.

We also detect a clear asymmetry, with higher mean velocity dispersions at negative Galactic latitude values. 
This could be partly explained by a vertical offset of the Sun from the physical Galactic midplane \citep[e.g.][however, see also \sect\ref{sec:midplane}]{Goodman2014_nessie}.

Next, we look at the distribution of the $\veldisp$ values as a function of Galactic longitude (\fig\ref{fig:histogram_2d_GLON-vel_disp}).
The distribution shows that the number of components with very high $\veldisp$ values increases toward the inner Galaxy.
The mean of the $\veldisp$ distribution shows multiple peaks, some of which are likely associated with large HII region complexes, for example W51 at $\ell \sim 49.5\degr$, and W39 at $\ell \sim 19\degr$.
Apart from these individual peaks and even though the velocity coverage is reduced for $\ell > 40\degr$, the general offset of the mean $\veldisp$ curve is remarkably constant at a value of $\sim 1$~\kms\ from the outermost coverage in Galactic longitude up until $\ell \sim 32\degr$.
For the GRS coverage with $\ell < 32\degr$ the offset of the mean $\veldisp$ curve is shifted to higher values of $\sim 1.2\ \text{to}\ 1.3$~\kms, in agreement with recent studies that also found increased molecular gas velocity dispersion towards the inner part of the Milky Way \citep{MivilleDeschenes2017}.
The giant HII region complex W43 is located at a Galactic longitude range of about $30\degr < \ell < 31.5\degr$ and is close to the near end of the Galactic bar \citep{Zhang2014w43}.
Recent simulations of barred galaxies showed that gas within the central regions dominated by the Galactic bar can reach high gas velocity dispersion values \citep{Khoperskov2018}.
We thus speculate that the increase in the offset of the mean $\veldisp$ curve is at least partly due to increased turbulence introduced by the Galactic bar.
Another reason for the increased non-thermal motions could be that there is more feedback from star formation toward the inner part of the Galaxy, as evidenced for example by the increased HII~region density (see Fig.~3 in \citealt{Anderson2017}) and an increase in the fraction of clumps that show signs of embedded star formation \citep{Ragan2018}.

\subsection{Complexity along the line of sight}
\label{sec:complexity}

\begin{figure*}
    \centering
    \begin{tabular}{cc}
    \includegraphics[width=0.5\textwidth]{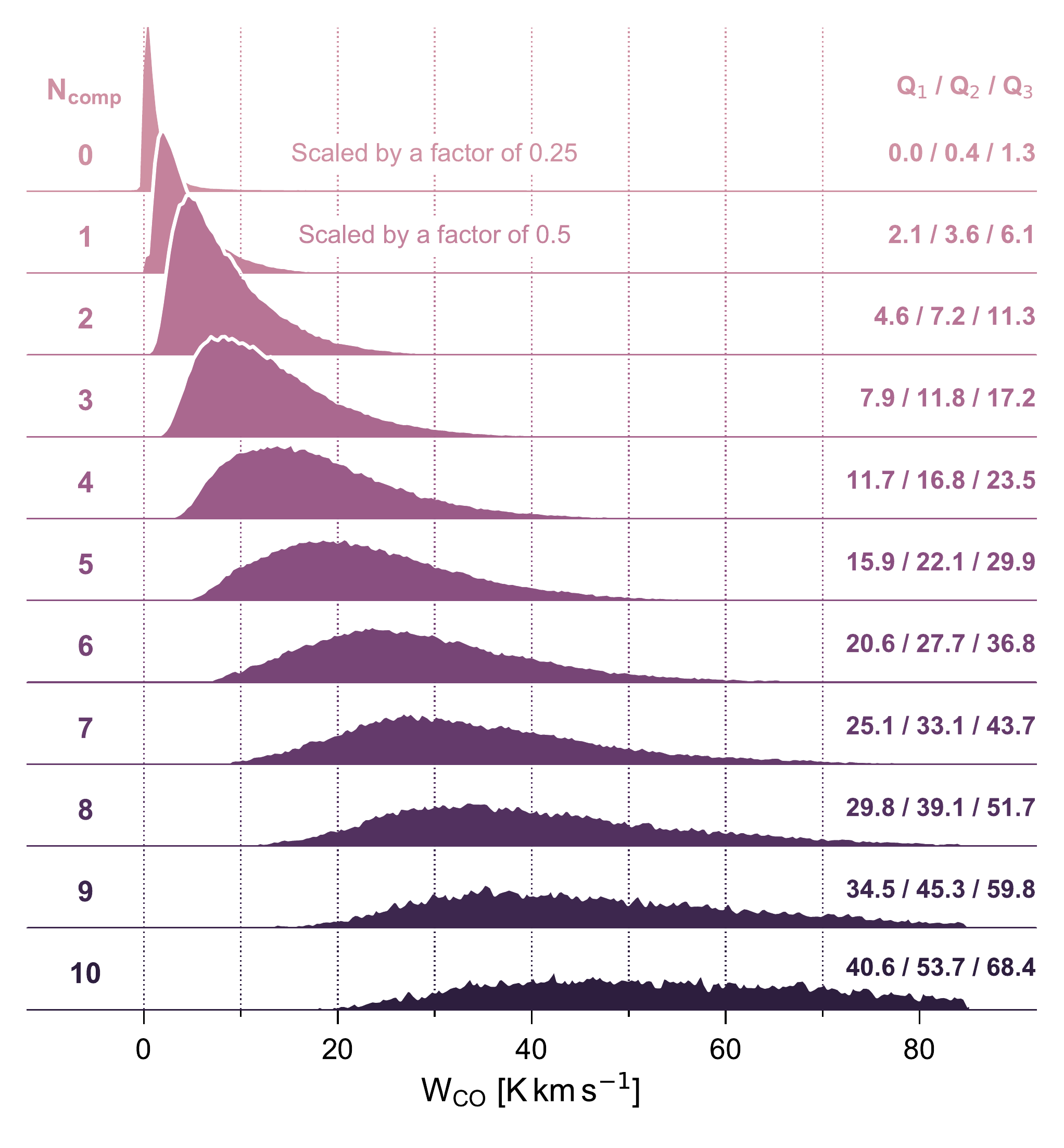}
    \includegraphics[width=0.5\textwidth]{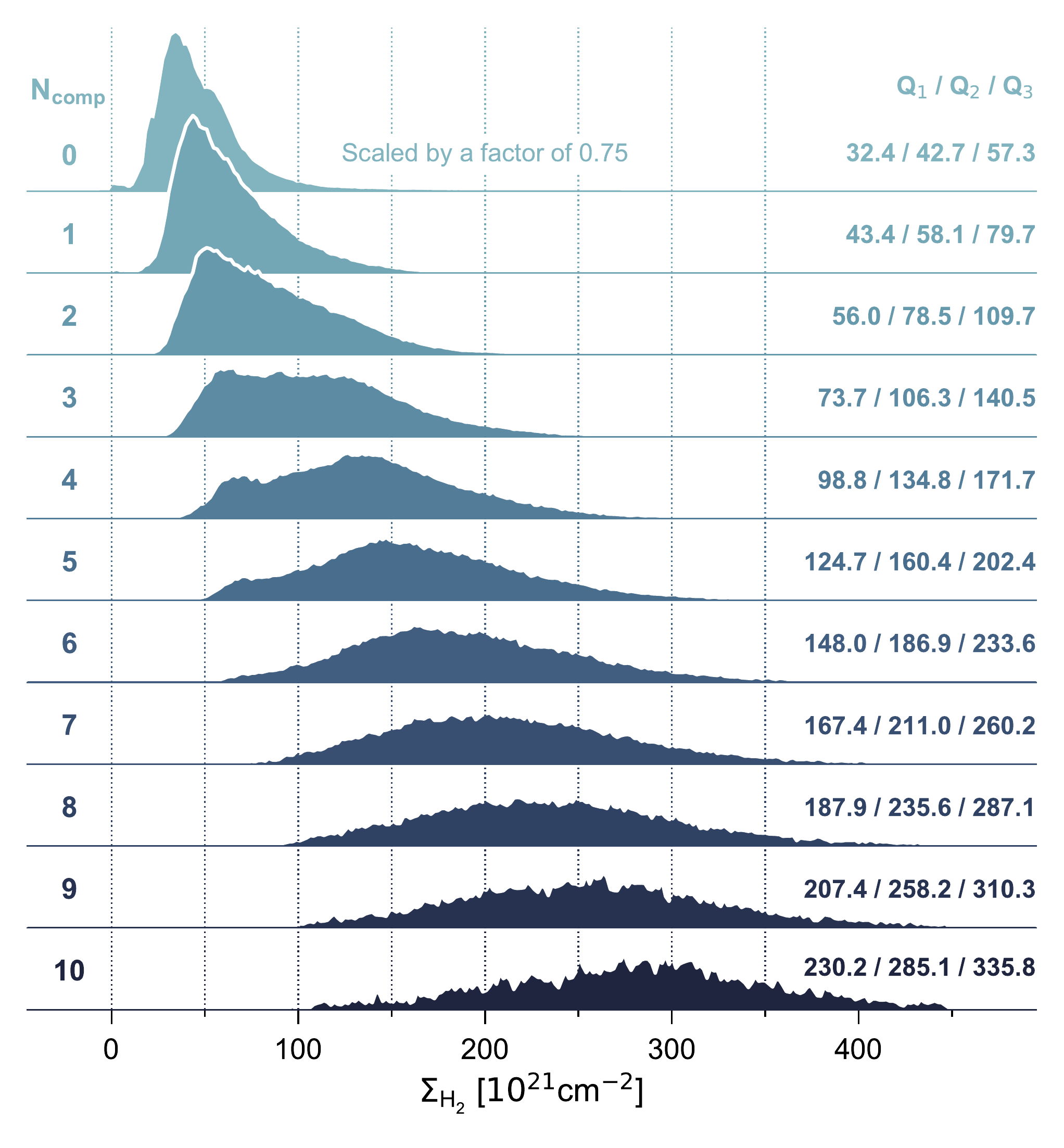} \\
    \end{tabular}
    \caption{Ridge plots showing PDFs of integrated $\co{13}{}$ emission (\emph{left panel}) and H$_{2}$ surface density values inferred from dust emission (\emph{right panel}).
    For the \emph{left panel}, each row shows the PDF of the $\wco$ values of all spectra fitted with $\Ncomp$ components (indicated with the number to the left and the colour-shading of the PDF).
    For the \emph{right panel}, each row shows the PDF of the $\Sigma_{\text{H}_{2}}$ values associated with $\co{13}{}$ spectra fitted with $\Ncomp$ components (indicated with the number to the left and the colour-shading of the PDF). 
    For better visibility three of the PDFs have been scaled in size, which is indicated next to the PDF.
    The values to the right of the PDFs indicate the first quartile (Q$_{1}$), the median (Q$_{2}$), and the third quartile (Q$_{3}$) of the distribution.
    }
    \label{fig:ridgeplots}
\end{figure*}

In this section we discuss the relation between the integrated intensity of $\co{13}{}$ emission, H$_{2}$ column densities inferred from dust emission, and complexity along the line of sight as measured by the number of fit components. 
We aim at determining if the $\co{13}{}$ emission preferentially originates from a few strong components or a larger number of smaller components, possibly spread wide along the spectrum.
We test this by comparing the $\wco$ values of the moment masked GRS data (shown in \fig\ref{fig:zero_mom}) with the number of fit components per spectrum (\fig\ref{fig:component_map}).
We illustrate the results in form of a ridge plot (\emph{left panel} of Fig. \ref{fig:ridgeplots}).
In this plot, each row shows the probability distribution functions (PDFs) of $\wco$ values for all spectra fitted with $\Ncomp$ components.
For better visibility, we chose upper limits of $\wco = 85$~\Kkms\ and $\Ncomp = 10$.
Figure~\ref{fig:ridgeplots} shows that the integrated intensity of $\co{13}{}$ correlates with the number of emission peaks.
The spread of the distributions increases with the number of components, and the distributions overlap in the $\wco$ range.
However, based on \fig\ref{fig:ridgeplots} we can say that, for example, lines of sight with $\wco$ values of $20$~\Kkms\ are predominantly associated with four to six fit components.

A similar analysis can provide insight into interpreting dust continuum emission features in the Galactic plane; for such data, there is otherwise no straightforward way to estimate how complex a line of sight is.
This comparison is interesting, because it immediately shows us how well the gas and dust are mixed.
If the dust column density is independent from the number of $\co{13}{}$ fit components, it could indicate differences in the distribution of dust and $\co{13}{}$ gas.
The \emph{right panel} of \fig\ref{fig:ridgeplots} shows a ridge plot constructed from H$_2$ column density values derived from dust emission (\sect\ref{sec:ppmap}).
We can see the same trend with dust as with $\co{13}{}$: the dust-derived column densities correlate with the number of fit components. To first order, this confirms that the $\co{13}{}$ gas and dust are indeed well mixed and thus largely originate from the same structures along the line of sight. 
Moreover, this analysis indicates that high dust column densities (leading to inferred H$_2$ column density values $> 200 \times 10^{21}$~cm$^{-2}$) arise from a composite of many dust components along the line of sight, rather than a single very dense structure.

Figure~\ref{fig:ridgeplots} also shows that dust emission is present along lines of sight for which no components could be fit in the $\co{13}{}$ spectra.
This implies that the dust emission traces also a more diffuse gas component that shows little to no $\co{13}{}$ emission.
This is in agreement with previous studies that have established that $\co{13}{}$ mostly traces the denser parts of molecular clouds, with surface densities exceeding $\sim 25\,$\msun$\,$pc$^{-2}$ \citep[e.g.,][]{Roman-Duval2016-grs}.


\section{Example application: disentangling emission from the near and far side of the Galaxy}
\label{sec:confusion}

Determining the exact location of molecular gas emission, and hence the Galactic distribution of it, is a major problem for understanding the structure of the Milky Way. 
The location of the emitting gas is usually determined using the kinematic distance method \citep{Clemens1988, Reid2014}, which yields two distance solutions (near and far) for the lines of sight within the solar orbit.
This distance ambiguity can only be resolved with additional information, for example using HI self-absorption \citep[e.g.][]{RomanDuval2009hisa}.
In this section, we discuss how our decomposition results can be useful in solving the distance ambiguity of features at low $\vlsr$ values, specifically between $-5\ \text{to}\ 20$~\kms.
We will use the fact that beam averaging effects for regions at larger distances to the sun will cause broader emission lines and thus higher $\veldisp$ values of the fit components.
The general principle of using the $\veldisp$ value of the fit components to help resolve the kinematic distance ambiguity should also be applicable to other $\vlsr$ values, albeit with less accuracy.
In \sect\ref{sec:kinematics}, we first recall the relationship between the Galactic structure and radial velocities of $-5$ to $20$~\kms.
We will then discuss the expected location and width of the physical midplane for this $\vlsr$ regime (\sect\ref{sec:midplane}) and how we would expect the linewidths to differ between emission coming from nearby and far away regions (\sect\ref{sec:larson}).  
We will then use these three discussions in \sect\ref{sec:nearfar} to argue how our decomposition results can give useful prior information in solving kinematic distance ambiguities.

\subsection{Considerations based on Galactic kinematics}
\label{sec:kinematics}

\begin{figure}
    \centering
    \includegraphics[width=0.676\columnwidth]{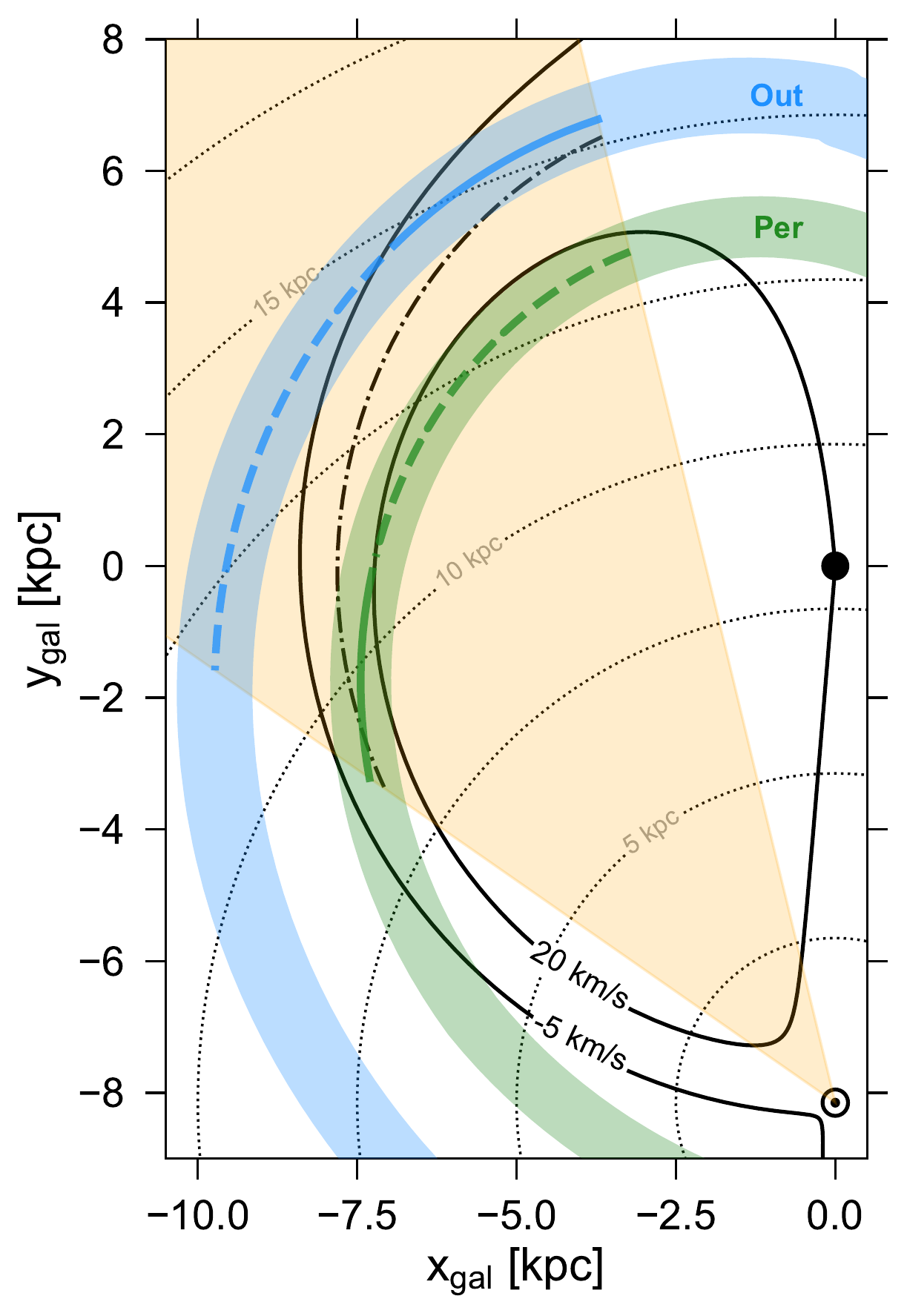}
    \caption{Face-on view of the Galactic plane. 
    The black solid lines show constant projected $\vlsr$ values of $-5$ and $20$~\kms. 
    The orange-shaded area indicates the coverage of the GRS. 
    Inferred positions and estimated widths for the Perseus and Outer spiral arm from \citet{Reid2019} are shown with the shaded green and blue areas, respectively.
    The central positions of the spiral arms are marked with solid and dashed lines depending on whether the majority of the arm is visible in the selected velocity range of $-5 < \vlsr < 20$~\kms.
    The position of the Galactic centre and the sun are indicated with a black dot and the sun symbol, respectively.
    Dotted grey lines indicate distances to the sun in 2.5~kpc intervals.
    The black dash-dotted line shows the distances used to estimate the FWHM extent of the molecular gas disk.
    See Sects.$\,\ref{sec:kinematics}$ and $\ref{sec:midplane}$ for more details.
    }
    \label{fig:mach_vlsr}
\end{figure}

The observed line of sight radial velocity $\vlsr$ of a point along Galactic longitude $\ell$ can be calculated as

\begin{equation}
    \vlsr = R_{0}\, \text{sin}\,\ell \left(\dfrac{\Theta(R_{\text{gal}})}{R_{\text{gal}}} - \dfrac{\Theta_{0}}{R_{0}} \right),
    \label{eq:vlsr}
\end{equation}

\noindent where $R_{\text{gal}}$ is the Galactocentric radius to the point along the line of sight, $\Theta(R_{\text{gal}})$ is the value of the rotation curve for $R_{\text{gal}}$, and $R_{0}$ and $\Theta_{0}$ are the radius of the solar circle and the corresponding rotational speed of that orbit. 
We use values of $R_{0} = 8.15$~kpc and $\Theta_{0} = 236$~\kms\ as estimated by \citet{Reid2019}.
For simplicity, we assume a flat rotation curve (i.e. $\Theta(R_{\text{gal}}) = \text{const.}$) and we do not correct for effects of non-circular motions toward the Galactic center or the direction of Galactic rotation.
We also do not correct for solar peculiar motions used by the telescope.
For a given value of $\vlsr$, Eq.~\ref{eq:vlsr} can be solved for $R_{\text{gal}}$ and rewritten as a function of $\ell$.
We can then use the relation

\begin{equation}
   R_{\text{gal}} = \sqrt{R_{0}^{2} + d_{\odot}^{2} - 2\,R_{0}\,d_{\odot}\,\text{cos}\,\ell}
    \label{eq:rgal}
\end{equation}

\noindent to solve for $d_{\odot}$, the distance to the point along the line of sight associated with the $\vlsr$ value:

\begin{equation}
   d_{\odot} = R_{0}\,d_{\odot}\,\text{cos}\,\ell \pm \sqrt{R_{\text{gal}}^{2} - (R_{0}\,\text{sin}\,\ell)^{2}}.
    \label{eq:kindist}
\end{equation}

\noindent For lines of sight inside the solar orbit, Eq.~\ref{eq:kindist} always yields two distance solutions, corresponding to points on the near and far side of the Galaxy that would be observed as having the same $\vlsr$ value. 

In Figure~\ref{fig:mach_vlsr} we show estimated lines of constant projected $\vlsr$ values of $-5$ and $20$~\kms\ in a face-on view of the first quadrant of the Milky Way.
Given the assumptions listed above, emission peaks observed in GRS with $-5 < \vlsr < 20$~\kms\ are thus expected to come from regions between those two curves of constant $\vlsr$. 
On the near side, this corresponds to the regions in the GRS that are located closest to the sun, with distances mostly $\lesssim 300$~pc.
On the far side, this area overlaps with the inferred locations of the Perseus and Outer spiral arms \citep{Reid2019}, with emission at lower longitude values expected to come from regions that can be located more than 15~kpc away.

\subsection{Expected location and extent of the Galactic disk}
\label{sec:midplane}

Recent studies by \citet{Anderson2019} and \citet{Reid2019} found a good correspondence of the vertical position of the Sun with the IAU definition of the Galactic midplane, which is in contrast to past studies that determined vertical offset positions of z$_{0} = 25 \pm 5$~pc \citep{BlandHawthorn2016}.
Even a larger offset of z$_{0} = 25$~pc would only have strong implications for the most nearby emission from the Galactic midplane, which would be shifted toward negative Galactic latitudes. 
We would nevertheless expect the midplane to be centred at about $b = -0.1\degr\ \text{to}\ 0\degr$ for GRS emission with $\vlsr$ values of $-5$ to $20$~\kms\ originating from the far side of the Galaxy.
However, previous studies of the Milky Way HI and molecular gas disk showed that the first quadrant of the Galactic disk is warped toward positive Galactic latitude values at the larger distances ($\gtrsim 8$~kpc) probed by GRS \citep[e.g.][]{Kalberla2009, Roman-Duval2016-grs, MivilleDeschenes2017}.
Assuming a moderate warp of the Galactic midplane of 50 to 100 pc at a distance of about 10~kpc would translate to positive shifts in Galactic latitude of $\sim 0.3\ \text{to}\ 0.6\degr$.
Factoring in a warp of the Galactic disk, we would thus expect the physical midplane to be shifted toward positive Galactic latitude values. 
The FWHM extent of the molecular gas disk in the outer Galaxy ($R_{\text{gal}} \gtrsim 8.5$~kpc) is on average about 200~pc \citep{Roman-Duval2016-grs}, which at distances of 8 and 15~kpc (the typical distances of GRS emission with $-5 < \vlsr < 20$~\kms\ at the far side of the disk) corresponds to a range in Galactic latitude of about $0.7\degr$ and $0.4\degr$, respectively.

\subsection{Expected velocity dispersion values}
\label{sec:larson}

\begin{table}
    \caption{Expected velocity dispersion values based on a size-linewidth relationship for different physical extents of the beam size at different distances.}
    \centering
    \small
    \renewcommand{\arraystretch}{1.3}
    \begin{tabular}{cccc}
    \hline\hline
    D [kpc] & d$_{\text{beam}}$ [pc] & $\sigma_{\text{exp.}}$ [\kms]\\
    \hline
0.25 & 0.06 pc & 0.2 \\
0.5 & 0.11 pc & 0.2 \\
1.0 & 0.22 pc & 0.3 \\
8.5 & 1.90 pc & 1.0 \\
15.0 & 3.35 pc & 1.3 \\
    \hline
    \end{tabular}
    \label{tbl:veldisp_expected}
\end{table}

\begin{table}
\begin{tabular}{cccc}

\end{tabular}
\end{table}

The physical beam size (d$_{\text{beam}}$) of the GRS varies significantly for distances to gas emission on the near and far side of our Galaxy: d$_{\text{beam}}$ is about 0.06~pc at the distance of 250~pc and 3.35~pc at 15~kpc (Table~\ref{tbl:veldisp_expected}). 
The larger beam for regions farther away can be argued to result in broader linewidths; the widths of emission lines have been shown to exhibit scale dependency \citep{Solomon1987}\footnote{The normalisation factor of 0.7 corrects for the different distance to the Galactic center of R$_{0} = 10$~kpc used by \citet{Solomon1987}.}:

\begin{equation}
\sigma\,(\text{\kms}) = 0.7\cdot \bigg( \frac{L}{1 \,\text{pc}} \bigg)^{0.5}.
\end{equation}

\noindent Table~\ref{tbl:veldisp_expected} shows the linewidths and Mach numbers predicted by this scaling.
For the scales probed by the GRS data at near distances ($\sim 250\ \text{to}\ 500$~pc), the size-linewidth relation predicts narrow velocity dispersions of about $0.2$~\kms.
For distances beyond $8.5$~kpc, the relation predicts velocity dispersions $> 1$~\kms.

\subsection{Information contained in the velocity dispersion values}
\label{sec:nearfar}

\begin{figure*}
    \centering
    \includegraphics[width=0.9\textwidth]{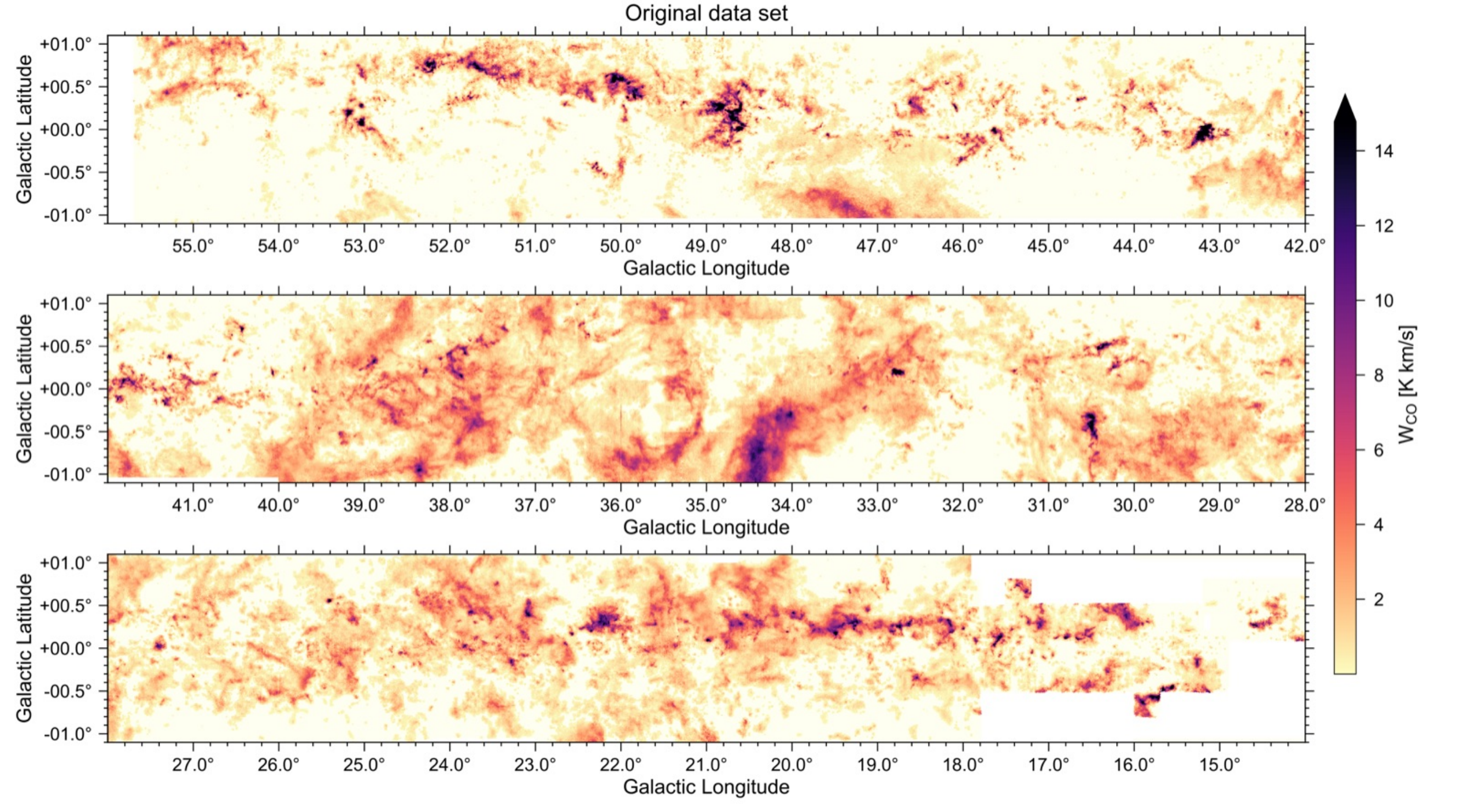}%
    \hspace{-0.9\textwidth}%
    \begin{ocg}{fig:original_off}{fig:original_off}{0}%
    \end{ocg}%
    \begin{ocg}{fig:original_on}{fig:original_on}{1}%
    \includegraphics[width=0.9\textwidth]{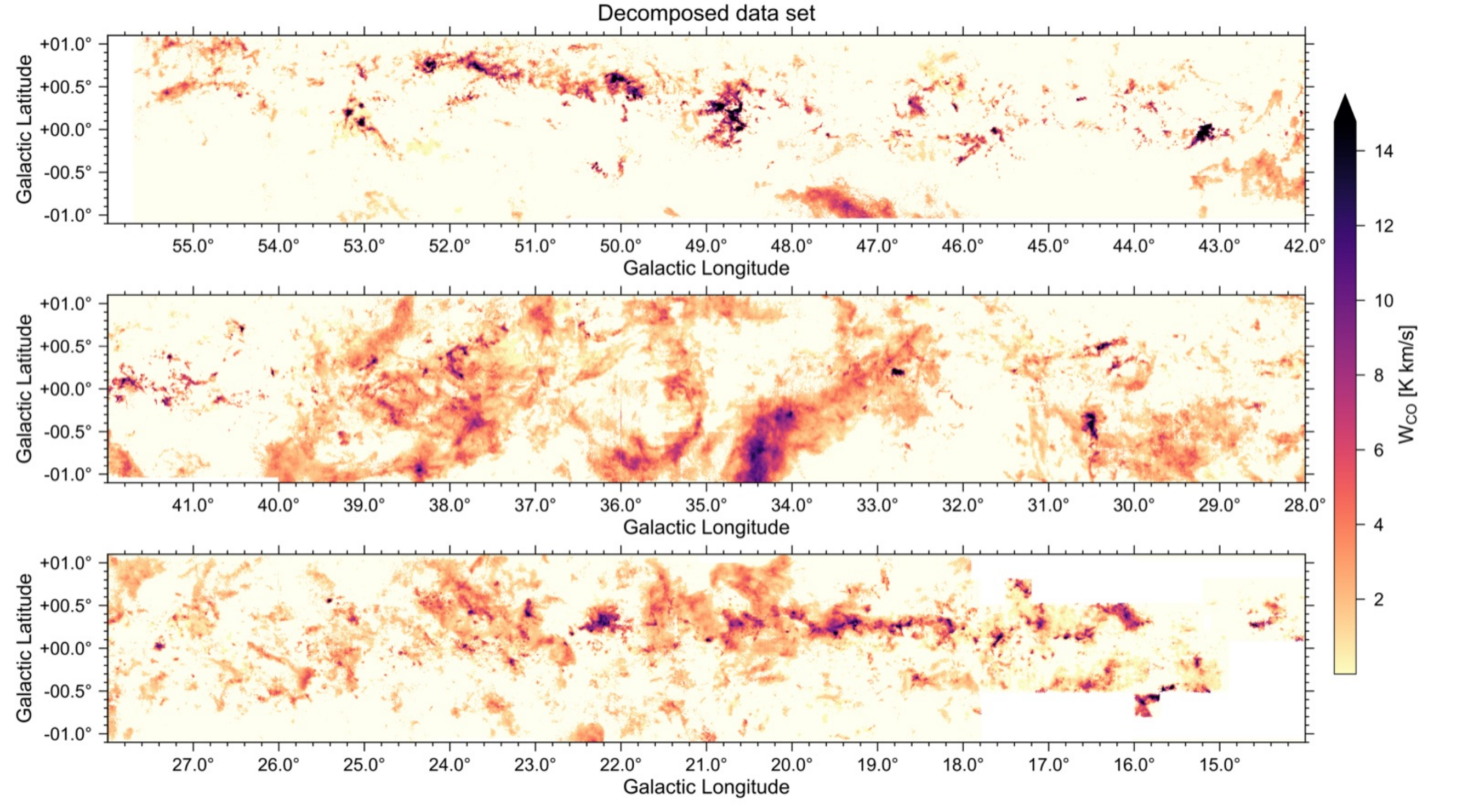}%
    \end{ocg}%
    \hspace{-0.9\textwidth}%
    \begin{ocg}{fig:perseus_off}{fig:perseus_off}{0}%
    \end{ocg}%
    \begin{ocg}{fig:perseus_on}{fig:perseus_on}{1}%
    \includegraphics[width=0.9\textwidth]{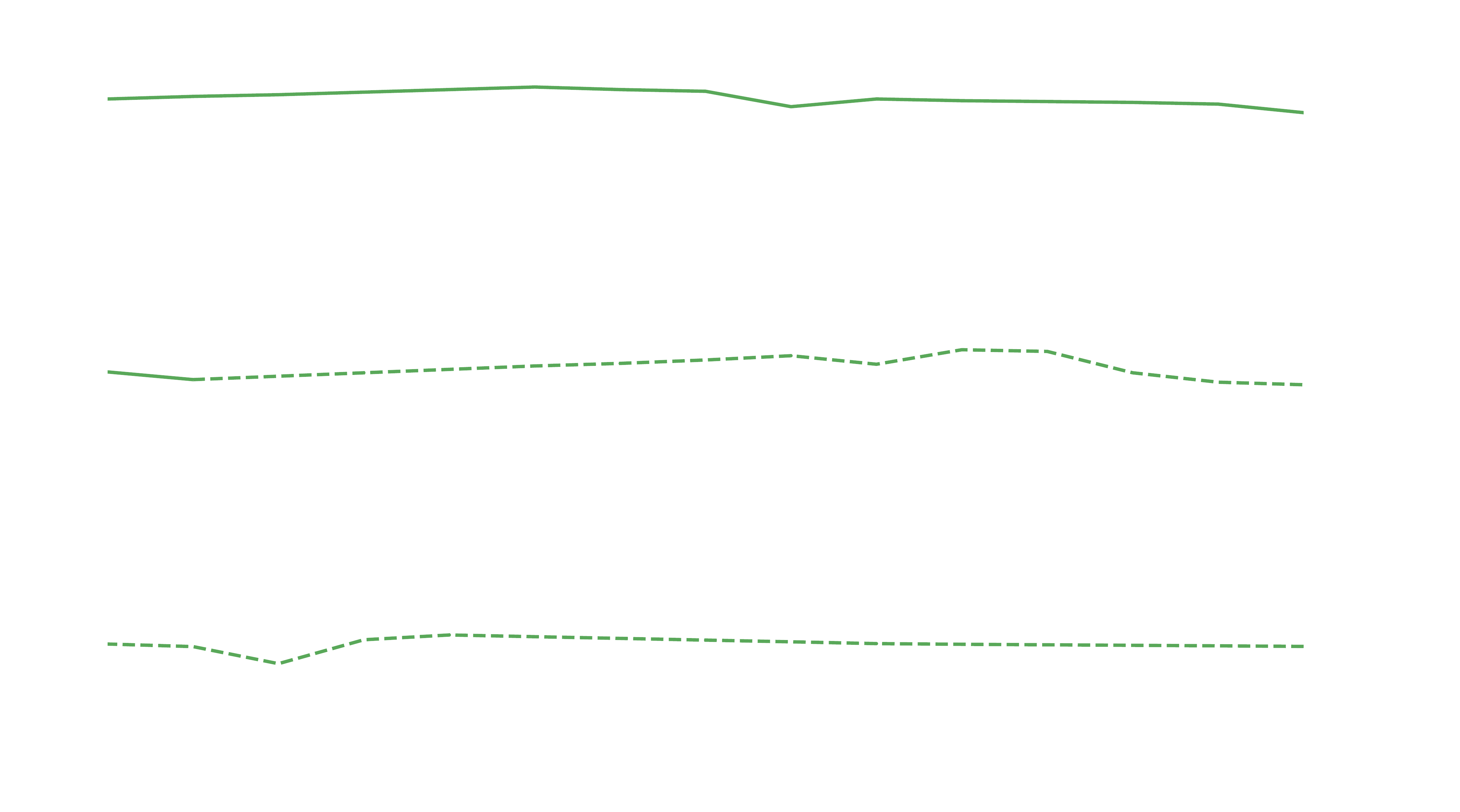}%
    \end{ocg}%
    \hspace{-0.9\textwidth}%
    \begin{ocg}{fig:outer_off}{fig:outer_off}{0}%
    \end{ocg}%
    \begin{ocg}{fig:outer_on}{fig:outer_on}{1}%
    \includegraphics[width=0.9\textwidth]{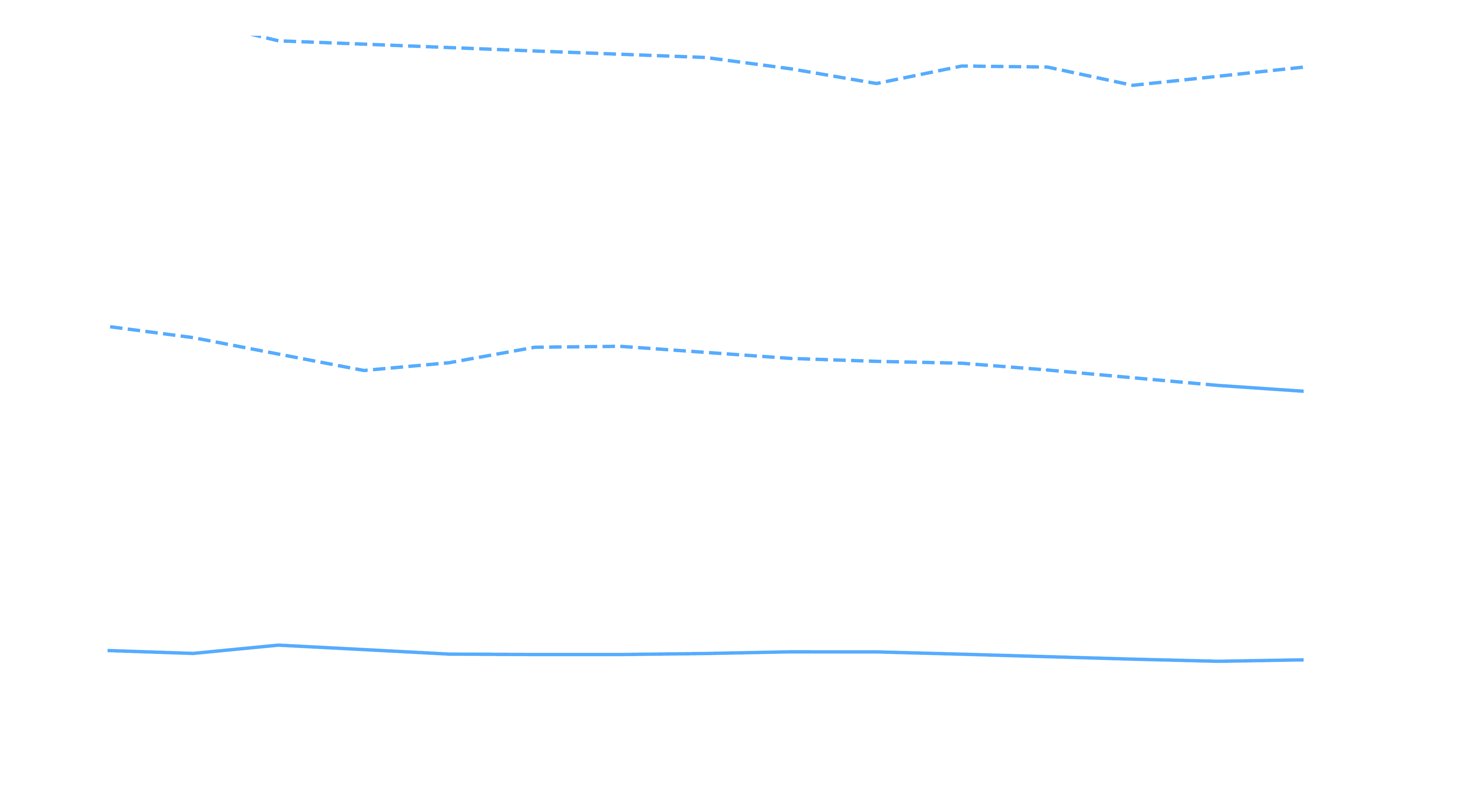}%
    \end{ocg}%
    \hspace{-0.9\textwidth}%
    \begin{ocg}{fig:contour_off}{fig:contour_off}{0}%
    \end{ocg}%
    \begin{ocg}{fig:contour_on}{fig:contour_on}{1}%
    \includegraphics[width=0.9\textwidth]{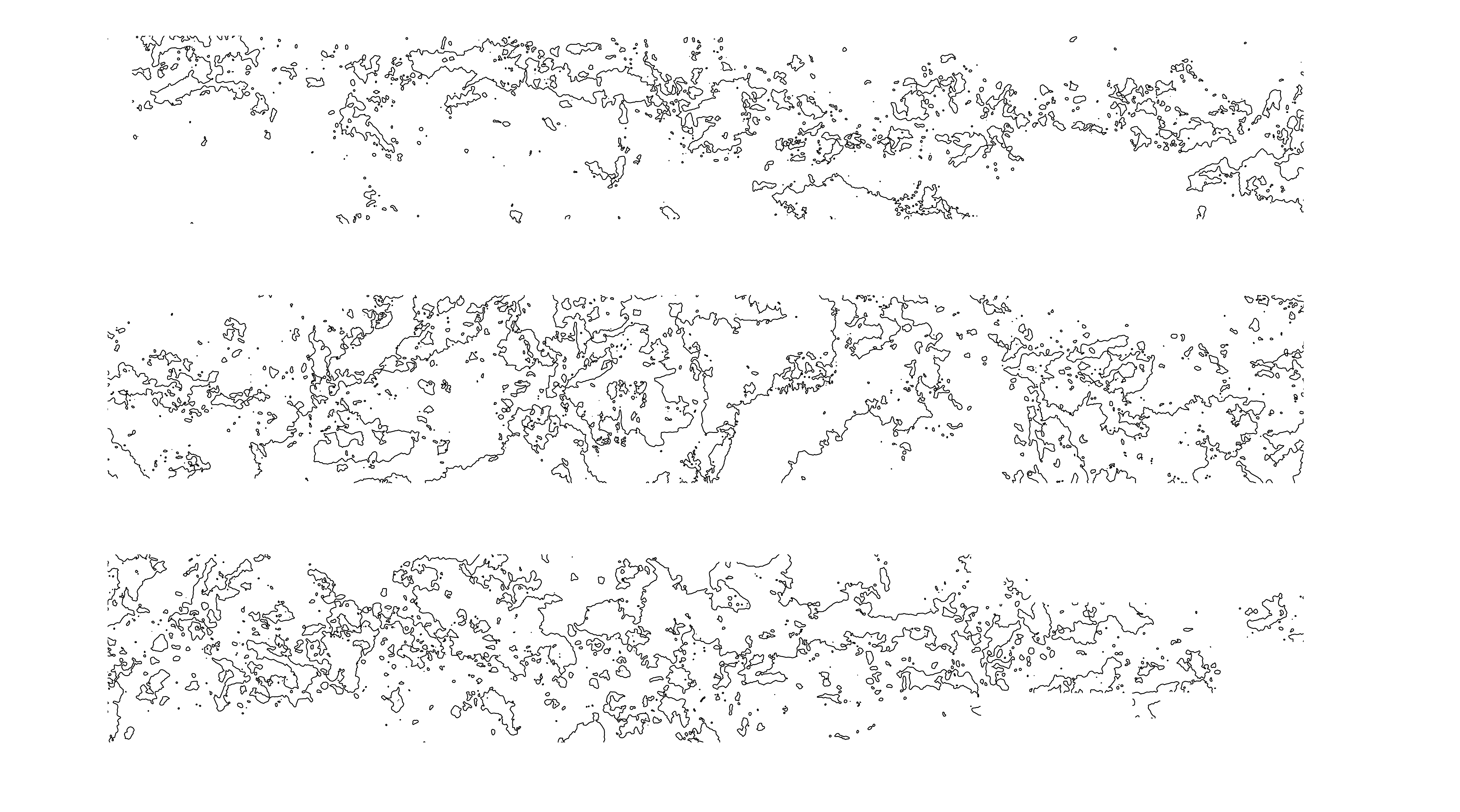}%
    \end{ocg}%
    \hspace{-0.9\textwidth}%
    \begin{ocg}{fig:fit_plane_off}{fig:fit_plane_off}{0}%
    \end{ocg}%
    \begin{ocg}{fig:fit_plane_on}{fig:fit_plane_on}{1}%
    \includegraphics[width=0.9\textwidth]{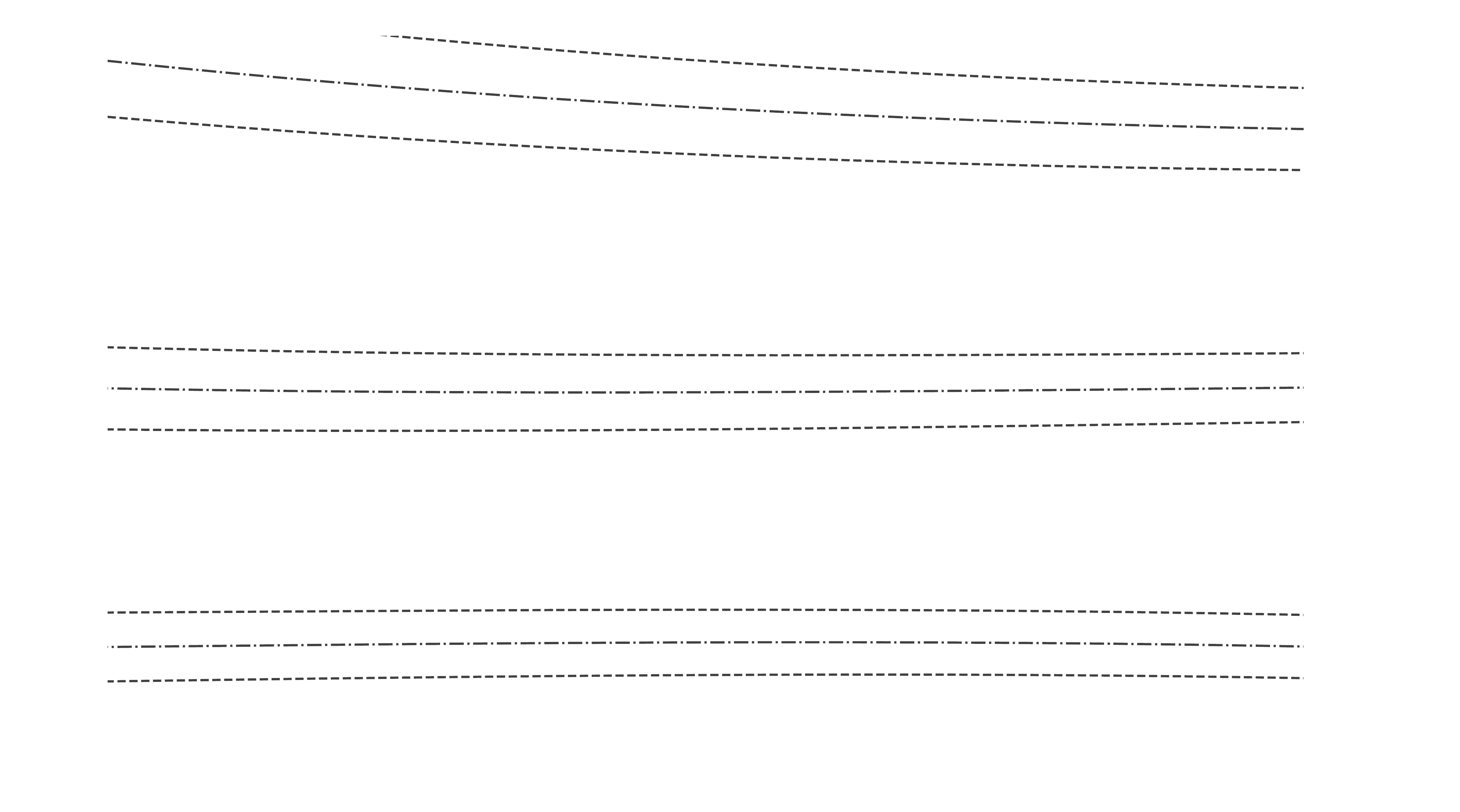}%
    \end{ocg}\\%
    \includegraphics[width=0.9\textwidth]{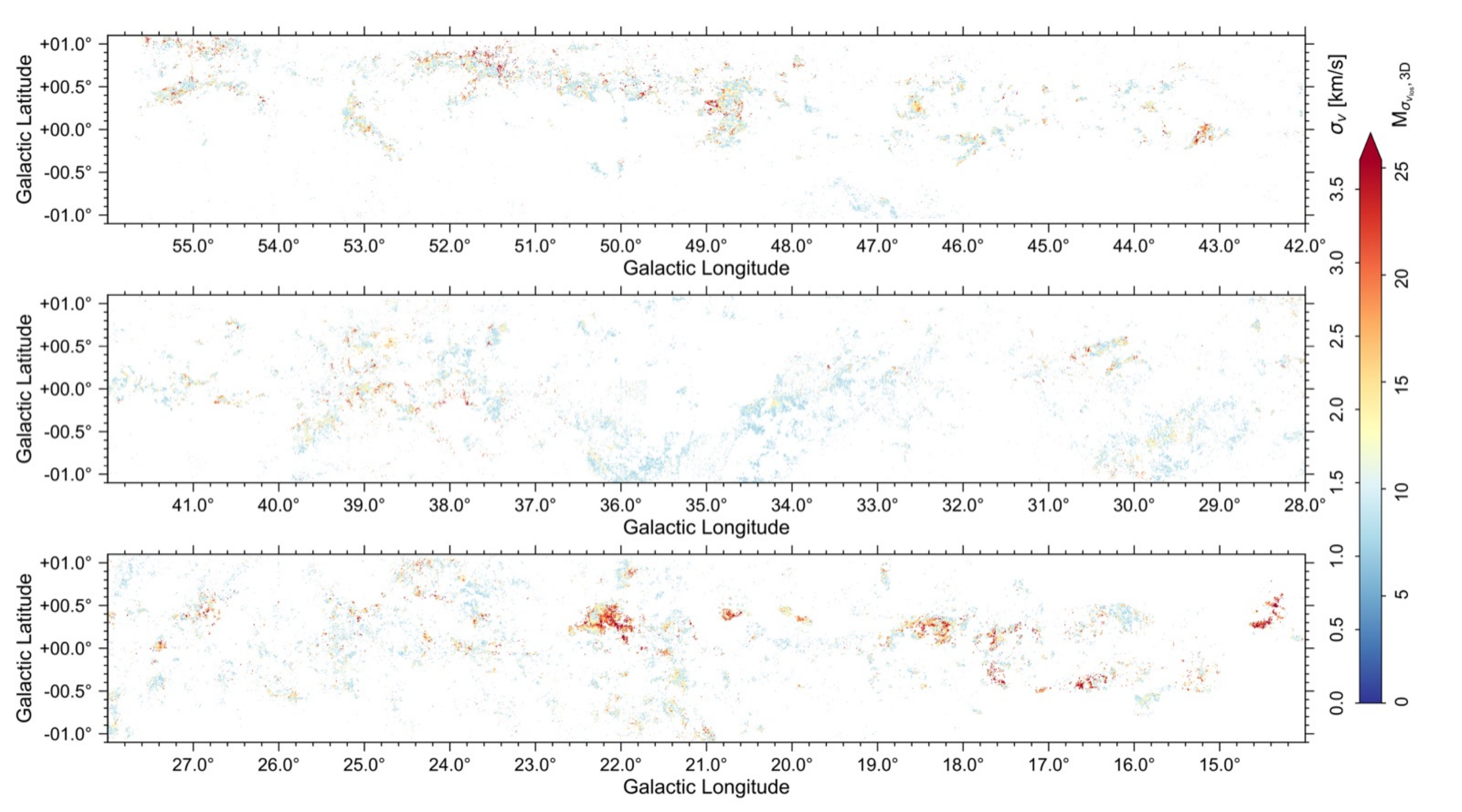}%
    \hspace{-0.9\textwidth}%
    \begin{ocg}{fig:far_off}{fig:far_off}{0}%
    \end{ocg}%
    \begin{ocg}{fig:far_on}{fig:far_on}{1}%
    \includegraphics[width=0.9\textwidth]{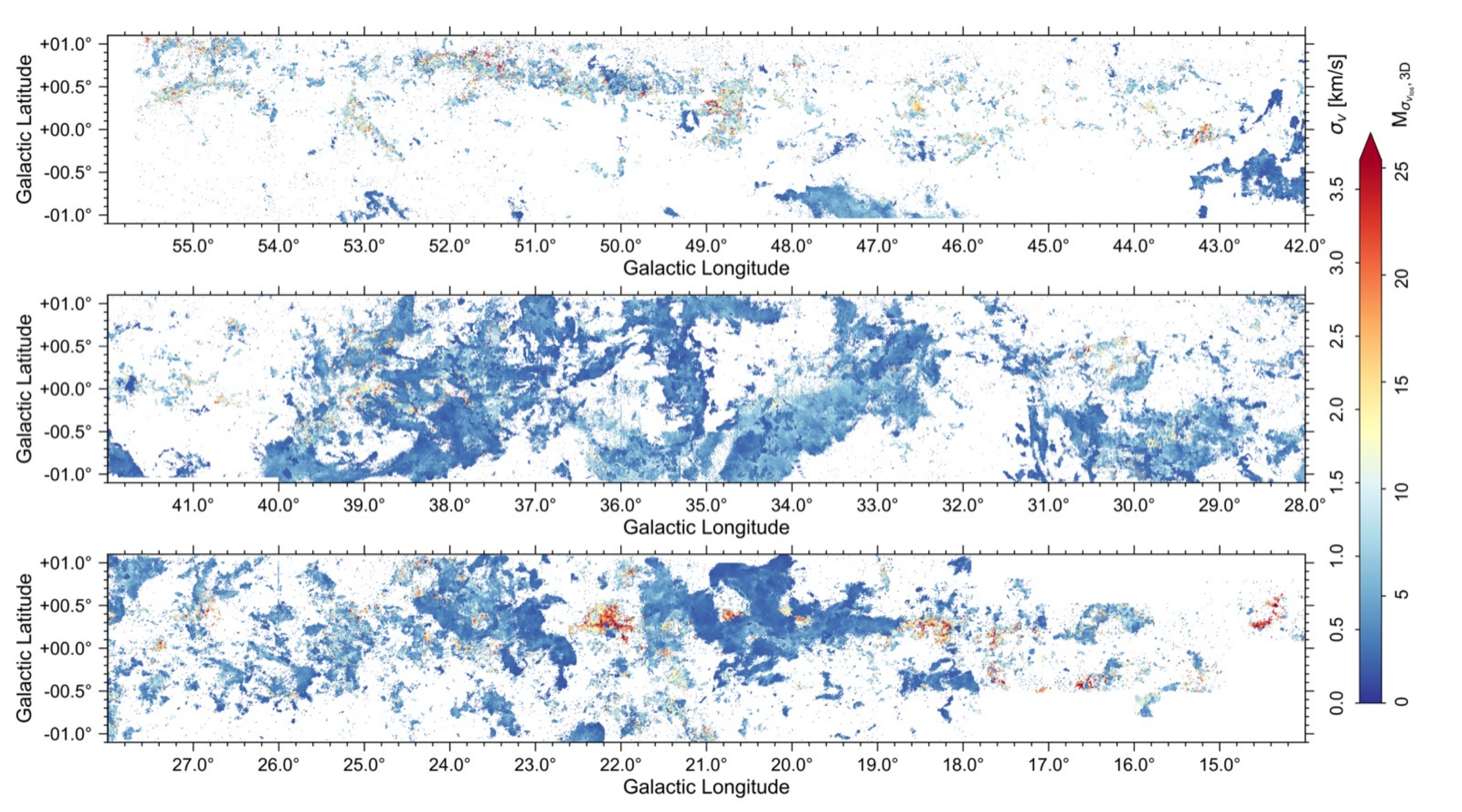}%
    \end{ocg}%
    \hspace{-0.9\textwidth}%
    \begin{ocg}{fig:perseus_off}{fig:perseus_off}{0}%
    \end{ocg}%
    \begin{ocg}{fig:perseus_on}{fig:perseus_on}{1}%
    \includegraphics[width=0.9\textwidth]{grs_mosaic_veldisp_layer_perseus.pdf}%
    \end{ocg}%
    \hspace{-0.9\textwidth}%
    \begin{ocg}{fig:outer_off}{fig:outer_off}{0}%
    \end{ocg}%
    \begin{ocg}{fig:outer_on}{fig:outer_on}{1}%
    \includegraphics[width=0.9\textwidth]{grs_mosaic_veldisp_layer_outer.pdf}%
    \end{ocg}%
    \hspace{-0.9\textwidth}%
    \begin{ocg}{fig:contour_off}{fig:contour_off}{0}%
    \end{ocg}%
    \begin{ocg}{fig:contour_on}{fig:contour_on}{1}%
    \includegraphics[width=0.9\textwidth]{grs_mosaic_veldisp_layer_contour.pdf}%
    \end{ocg}%
    \hspace{-0.9\textwidth}%
    \begin{ocg}{fig:fit_plane_off}{fig:fit_plane_off}{0}%
    \end{ocg}%
    \begin{ocg}{fig:fit_plane_on}{fig:fit_plane_on}{1}%
    \includegraphics[width=0.9\textwidth]{grs_mosaic_veldisp_layer_fit_plane.pdf}%
    \end{ocg}%
    \caption{\emph{Top}: Zeroth moment map of the decomposed data set integrated within $-5 < \vlsr < 20$~\kms. 
    \emph{Bottom}: Velocity dispersion values for the narrowest fit component within the same $\vlsr$ range as used in the top panel.
    The second scaling on the colourbar indicates corresponding estimates for upper limits of the turbulent Mach number.
    In both panels we show: the position of the Perseus and Outer spiral arms as inferred by \citet{Reid2019} (green and blue lines, respectively, cf. \fig\ref{fig:mach_vlsr}); black contours indicating a $\wco$ value of 0.5~K$\,$\kms; a fit to the positions with $\veldisp$ values $> 1$~\kms\ (dash-dotted black line); and the estimated FWHM extent of the molecular gas disk on the far side of the Milky Way (dashed black lines). 
    See \sect$\,\ref{sec:nearfar}$ for more details.
    When displayed in Adobe
    Acrobat, it is possible to switch to the \ToggleLayer{fig:original_on,fig:original_off}{\protect\cdbox{channel map of the original data set}}, show only the 
    \ToggleLayer{fig:far_on,fig:far_off}{\protect\cdbox{positions with $\veldisp$ values $> 1$~\kms}}, hide
    the \ToggleLayer{fig:contour_on,fig:contour_off}{\protect\cdbox{contours}}, hide the \ToggleLayer{fig:fit_plane_on,fig:fit_plane_off}{\protect\cdbox{fit to the positions with $\veldisp$ values $> 1$~\kms\ and estimated FWHM extent of the gas disk}},
    and hide the positions of the \ToggleLayer{fig:perseus_on,fig:perseus_off}{\protect\cdbox{Perseus arm}} and \ToggleLayer{fig:outer_on,fig:outer_off}{\protect\cdbox{Outer arm}}.
    }
  \label{fig:Mach_map}
\end{figure*}

In Sects.$\,$\ref{sec:kinematics} to \ref{sec:larson} we discussed what we would expect for the origin, extent, and shape of velocity components with $\vlsr$ values of $-5\ \text{to}\ 20$~\kms.
Now we will demonstrate how this applies to our decomposition results and how the $\veldisp$ values of the fit components can give us an indication of whether the emission line originates from regions close to the sun or farther away.

The upper part of \fig\ref{fig:Mach_map} shows the moment masked version of the decomposed GRS data set integrated within a velocity range of $-5 < \vlsr < 20$~\kms\ (if fit components extended beyond this range we only included the emission within these limits).
As already discussed in the previous sections, this channel map contains a combination of emission features that originate either from regions very close to the sun or as far away as $15$~kpc.
We also show the positions of the Perseus and Outer spiral arms as inferred by \citet{Reid2019}, which indicates where we would expect emission from the far side of the disk to appear in the channel map.
In the Galactic longitude range from about 17 to 43$\degr$, \fig\ref{fig:Mach_map} is dominated by emission from the Aquila Rift cloud (see \fig\ref{fig:grs-mosaic_pv+spiral_arms} and the schematic map in Fig.~9 of \citealt{Dame1985_Aquilarift}), which is located at a distance to the sun of about 250 to 500~pc \citep{Aquilarift_dist, Zucker2019distances}.
Due to this close distance, emission from the Aquila Rift is spread out and extends over the full survey coverage in Galactic latitude.

With the predictions from Sects.$\,$\ref{sec:kinematics} to \ref{sec:larson} in hand, we can now try to understand the spatial distribution of the observed velocity dispersions. 
The \emph{bottom} part of \fig\ref{fig:Mach_map} shows for each line of sight the $\veldisp$ value and corresponding upper limits of the turbulent Mach number of the narrowest fit components (with centroid position values within $-5 < \vlsr < 20$~\kms). 
The region dominated by the Aquila Rift shows overall much lower $\veldisp$ values for the Mach number.
The distribution of all $\veldisp$ values contained in the \emph{bottom} part of \fig\ref{fig:Mach_map} peaks at a value of $0.35$~\kms, matching the low $\veldisp$ values we would expect for the physical extent of the GRS beam at $250\ \text{to}\ 500$~pc derived from the size-linewidth relation in \sect\ref{sec:larson}.
Toward higher Galactic longitude values ($\ell \gtrsim 40\degr$), where confusion with emission from the Aquila Rift cloud is expected to be less severe, we can see a strip of increased $\veldisp$ values that seems confined in Galactic latitude to $\sim 1\degr$.
If these velocity components also originate from regions close to the sun, these must be regions with increased non-thermal motions.
However interestingly, fit components with higher $\veldisp$ values are less spread out in Galactic latitude than components with lower velocity dispersion values, which agrees with what we would expect from emission lines coming from regions farther away.
We thus speculate that this emission with high $\veldisp$ values is coming mostly from the Perseus and Outer arm on the far side of the Galactic disk (cf. \fig\ref{fig:mach_vlsr}).

Based on our arguments from a size-linewidth relationship we expect the emission lines originating from far distances to show increased $\veldisp$ values $> 1$~\kms\ (cf. Table~\ref{tbl:veldisp_expected}). 
Using this prediction as a threshold, we fitted a polynomial of the third order to all $\veldisp$ values $> 1$~\kms\ in the lower part of \fig\ref{fig:Mach_map}, which is indicated with dash-dotted black line.
For the polynomial fit we first calculated the average Galactic latitude position per Galactic longitude bin for all $\veldisp$ values $> 1$~\kms.
We also overplot the expected scale height of the molecular gas disk with dashed black lines, which we calculated for the average distance to emission lines with $\vlsr$ values of $-5\ \text{to}\ 20$~\kms\ on the far side (indicated with a dash-dotted black line in \fig\ref{fig:mach_vlsr}).
This estimated range for the scale height of the molecular gas disk well matches the height of the strip with increased $\veldisp$ values.
Assuming that the indicated strip of high $\veldisp$ values indeed corresponds to the Galactic midplane at the far distance, its position at positive Galactic latitude values would also be consistent with the shift expected from a contribution of the warp of the Galactic disk toward positive heights in the first quadrant (\sect\ref{sec:midplane}).
If our conjecture is true, this would point to an increase in the warp of the Galactic disk at Galactic longitude values $\gtrsim 42\degr$ at distances of $\sim 10\ \text{to}\ 12$~kpc (cf. \fig\ref{fig:mach_vlsr}).
Moreover, it would suggest that the majority of the increased $\veldisp$ values in \fig\ref{fig:Mach_map} are caused by effects of resolution, beam averaging, or turbulent motion on the scale of clouds (macroturbulence) rather than being introduced by highly turbulent subregions within nearby local clouds (microturbulence).

The Gaussian decomposition results can thus be useful in disentangling near and far emission for characterising and constraining Galactic structure.
In a follow-up work (Riener et al., in prep.) we will concentrate on establishing distances to the Gaussian fit components, for which the considerations from this section can serve as useful prior information on solving the kinematic distance ambiguity.


\section{Summary}

In this work we present Gaussian decomposition results obtained with \gausspyplus\ \citep{Riener2019} for the entire GRS data set at its full spatial and spectral resolution.
In total, we fitted $\sim 4.6$ million Gaussian components to the $\sim 2.3$ million $\co{13}{}$ emission line spectra of the GRS.
Especially spectra from lines of sight near the Galactic midplane showed great complexity, requiring 10 or more fit components for a good decomposition. 

The decomposition recovers $87.5\%$ of the flux contained in the GRS ($92.1\%$ of emission with a S/N ratio $> 3$).
Most of the non-recovered flux is due to diffuse or weak $\co{13}{}$ emission that could not be identified in the decomposition due to varying noise properties and our use of a single S/N threshold for the entire data set.

Assuming a uniform gas temperature of 18~K, we determined upper limits for the turbulent Mach number. 
We estimate from the velocity dispersion of our fit components with S/N ratio $>3$ that about $28.6\%$ are associated with turbulent Mach number values $> 10$.
We see a clear trend of higher velocity dispersion values for fit components with higher $\vlsr$ values, which is likely due to the effect of beam averaging of emission lines originating at larger distances to the sun.

We studied the distribution of velocity dispersion values along the Galactic coordinates and find that large velocity dispersions increase toward the Galactic midplane, likely due to the concentration of star forming regions in the Galactic plane.
We also find an increase in the $\veldisp$ values toward the inner Galaxy, which we speculate could be due to the influence of the Galactic bar.

The integrated intensities of $^{13}$CO correlate with the number of Gaussian fit components along the line of sight, indicating that larger integrated intensities are associated with more complexity in the spectra. 
We also compared the number of fitted components to H$_{2}$ surface density values inferred from dust emission by \citet{Marsh2017-ppmap} and find a similar trend in that higher $\Sigma_{\text{H}_{2}}$ values are associated with more fit components in the corresponding $\co{13}{}$ spectra, indicating that gas and dust emission is originating from the same structures along the line of sight.

We also demonstrated how the decomposition results can aid in resolving confusion from kinematic distance ambiguities. 
We use arguments based on Galactic structure and the Galactic rotation curve to disentangle emission from the nearby Aquila Rift molecular cloud and gas emission at distances $\gtrsim 8.5$~kpc that is likely associated with the Perseus and Outer spiral arms.
Our fitting results thus may be useful as prior information in determining kinematic distances.

In a forthcoming work we will present distance estimates to all the fit components presented in this paper, which will enable us to better associate emission peaks to locations in the Galaxy.
This will allow us to discuss the Galactic distribution of the gas emission and to infer how physical quantities vary with Galactocentric distance.

\begin{acknowledgements}
This project received funding from the European Union’s Horizon 2020 research and innovation program under grant agreement No 639459 (PROMISE). 

This publication makes use of molecular line data from the Boston University-FCRAO Galactic Ring Survey (GRS). 
The GRS is a joint project of Boston University and Five College Radio Astronomy Observatory, funded by the National Science Foundation under grants AST-9800334, AST-0098562, \& AST-0100793.
HB acknowledges support from the European Research Council under the European Community's Horizon 2020 framework program (2014-2020) via the ERC Consolidator Grant 'From Cloud to Star Formation (CSF)' (project number 648505). HB also acknowledges support from the Deutsche Forschungsgemeinschaft in the Collaborative Research Center (SFB 881) 'The Milky Way System' (subproject B1).
We would like to thank the MWISP collaboration for generously sharing the data of the G24 region with us.
We would also like to thank the referee, Mark Heyer, for constructive and useful suggestions that helped to improve the paper.
      \\\textbf{Code bibliography}:
      This research made use of \textsc{matplotlib} \citep{Hunter2007}, a suite of open-source python modules that provides a framework for creating scientific plots; \textsc{astropy}, a community-developed core Python package for Astronomy \citep{astropy}; \textsc{aplpy}, an open-source plotting package for Python \citep{APLpy}; and \textsc{seaborn} \citep{seaborn}.
\end{acknowledgements}

\bibliographystyle{aa} 
\bibliography{bibliography} 

\begin{appendix} 


\section{Chosen parameters, data preparation and decomposition runs for the GRS survey}
\label{app:grs-decomp}

In this appendix we describe the settings we used for the decomposition runs with \gausspyplus.
Where not specified otherwise, we used the default settings of \gausspyplus\ as described in \citet{Riener2019}.

\subsection{Preparatory steps}
\label{app:grs-preparation}

The cubes of the original GRS data set\footnote{\url{https://www.bu.edu/galacticring/new_data.html}} are centred on integer Galactic longitude values, with each cube overlapping by one degree with the following cube.
For the mosaicking we only used the cubes centred on odd integer values in Galactic longitude. 
Since the mosaicked cube of the entire GRS data set was too large for a decomposition with \gausspyplus, we split it again along the Galactic longitude axis into 23 individual, non-overlapping subcubes.\footnote{This was driven by restrictions in available computing memory.
For big spectral cubes we found it beneficial to split the data set into smaller subcubes for the \gausspyplus\ decomposition runs, but this will depend on the computing infrastructure available to the user.}
We performed all preparatory and decomposition steps with \gausspyplus\ on these individual subcubes; this also included the spatially coherent refitting phases, which means that the decomposition results might show discontinuities at the borders of the individual subcubes. 
We checked for such an effect on the map showing the number of fit components (\fig\ref{fig:component_map}) but did not find any obvious problems.

Especially close to the borders of the coverage in Galactic latitude and longitude, the GRS data set contains spectra with instrumental artefacts, such as strongly amplified noise fluctuations.
We thus masked out all spectra with extremely high noise values ($\rmsTa > 0.75\,$K; see \sect\ref{sec:noise-values} for more details).

The \gausspyplus\ package includes preparatory steps for the decomposition, such as the identification of regions in the spectrum estimated to contain signal and the automated masking of negative features that are likely noise spikes or artefacts (see Sect.~3.1 in \citealt{Riener2019} for more details).
We found that about $1.6\%$ of the spectra from the GRS data set contained significant negative features in the spectrum (see \sect$\,\ref{sec:recovered_flux}$ for more details).

\subsection{Choice of the \gausspy\ smoothing parameters}
\label{app:grs-training}

The \gausspy\ algorithm needs to be supplied with a training set consisting of a couple of hundred well-fit spectra, so that it can infer optimal smoothing parameters for the respective data set via its in-built supervised machine learning method (see \citealt{Lindner2015} for more details). 
The \gausspy\ algorithm uses higher derivatives of a spectrum to automatically decide on initial guesses for the number and shape of fit components. 
To suppress negative impacts of noise fluctuations, the spectrum needs to be smoothed before derivatives are taken.
For the decomposition of HI and CO spectra, a two-phase approach using two different smoothing parameters $\alpha_{1}$ and $\alpha_{2}$ (corresponding to the size of two different Gaussian smoothing kernels) was found to be beneficial \citep{Lindner2015, Riener2019}.
In the training step of \gausspy, the algorithm essentially iterates in a controlled manner through different values for the two smoothing parameters and compares the resulting decompositions of the training set with the user-provided corresponding best fit solutions for the spectra.
The \gausspy\ algorithm evaluates the accuracy of the decomposition results via the F$_{1}$ score:

\begin{equation}
	F_{1} = 2\cdot\dfrac{\text{precision}\cdot\text{recall}}{\text{precision} + \text{recall}},
\end{equation}

\noindent where precision is defined as the comparison of the fraction of correctly fitted components to the number of total fitted components and recall is defined as the fraction of components from the provided best fit solutions that had matching counterparts in the decomposition with the tested smoothing parameters.

The \gausspyplus\ package includes a routine that automatically creates a training set of high-quality decompositions that lead to values for the smoothing parameters that are close to the optimal ones \citep[see App.~B.5 in][]{Riener2019}.
For the GRS data set, we used the default settings of \gausspyplus\ to create nine different training sets each containing $500$ spectra.
The spectra for each training set were randomly sampled from the entire  GRS coverage and were then automatically decomposed with \gausspyplus\ (see \sect$\,3.1.4$ in \citealt{Riener2019} for more details about the decomposition method that is used to create training sets).

\begin{table}
    \caption{Smoothing parameters (in units of spectral channels) obtained for the \gausspyplus\ training sets.}
    \centering
    \small
    \renewcommand{\arraystretch}{1.2}
    \begin{tabular}{cccc}
    \hline\hline       
    Sample & $\alpha_{1}$ & $\alpha_{2}$ & F$_{1}$ score \\
    \hline       
    1 & 1.75 & 4.02 & 74.2$\%$ \\
    2 & 1.89 & 4.08 & 75.4$\%$ \\
    3 & 1.95 & 4.15 & 76.5$\%$ \\
    4 & 2.01 & 3.98 & 76.4$\%$ \\
    5 & 1.97 & 4.22 & 73.4$\%$ \\
    6 & 2.1 & 4.72 & 74.9$\%$ \\
    7 & 2.12 & 4.36 & 72.6$\%$ \\
    8 & 1.77 & 3.98 & 74.1$\%$ \\
    9 & 1.98 & 4.63 & 74.9$\%$ \\
    \hline
    \end{tabular}
    \label{tbl:trainingset_gpy}
\end{table}

In Table~\ref{tbl:trainingset_gpy} we list the value for the smoothing parameters $\alpha_{1}$ and $\alpha_{2}$ and the corresponding F$_{1}$ score we obtained with the machine learning functionality of \gausspy\ for these nine training sets.
In general, the resulting inferred smoothing parameter values are similar and compare well to each other.
We expect small to moderate deviations between the samples such as they are present in Table~\ref{tbl:trainingset_gpy}, since we randomly chose the spectra for the trainings sets from the entire GRS coverage, which contains spectra with significantly different noise values (\fig\ref{fig:noise_map}).
As discussed in App.~B.5 of \citet{Riener2019}, such small deviations of the smoothing parameters only have a limited impact on the decomposition results.
Moreover, with $\gausspyplus$ the fitting is not that dependent anymore on the exact values of the smoothing parameters, since it includes an improved fitting routine that aims to improve decompositions obtained with \gausspy\ that did not yield a good fit.

The training set decomposition method in \gausspyplus\ by default only includes fit solutions that have a $\chired$ value below 1.2.
As discussed by \citet{Andrae2010}, a threshold based on a fixed value of $\chired$ can be problematic, since for non-linear functions such as Gaussian fit components the degrees of freedom cannot be exactly determined and may vary substantially.
To check whether this $\chired$ threshold might have biased the training samples produced with \gausspyplus, we also tried a different approach to create training sets, by using the semi-automated spectral line fitting package \scousepy\ \citep{Henshaw2016, Henshaw2019}.

In \scousepy, the data set is first divided into spectral averaging areas (SAAs). 
All spectra contained in an individual SAA are then averaged and the user manually fits the resulting spectrum by deciding on the number of fit components and their shape.
The fit results from the SAA then help to inform the automated decomposition of the individual spectra contained in the SAA, which moreover leads to spatial coherence between the fit results of neighbouring spectra.

To create training sets with \scousepy, we split the mosaicked GRS data set along the Galactic longitude axis into 23 individual subcubes and randomly placed nine spectral averaging areas (SAAs) with $3\times3$ pixels in each subcube for a total of $207$ SAAs and $1863$ individual spectra.
We then proceeded with the suggested workflow for \scousepy:
First we manually fitted the averaged spectrum of each SAA, which informed the automated decomposition step for each individual spectrum contained in the SAA.
Then we visually inspected the best fit solutions for all individual spectra and refit them manually if the best fit results were not satisfactory, that means if they showed unfit peaks, strongly blended components or fit components with very broad line widths, or if the fit solution was not spatially coherent with its neighbours.

For the selection of good-quality decompositions for the training set, we checked the final \scousepy\ best fit results of all $1863$ decomposed spectra with their corresponding residuals again by eye, which left us with a working sample of $1639$ individual spectra. 
We then created three training sets, each of which contained $500$ randomly chosen spectra and decomposition results from our final selection (we drew the spectra without replacement, which means an individual spectrum could only appear in one training set).

\begin{table}
    \caption{Smoothing parameters (in units of spectral channels) obtained for training set decompositions with \scousepy\ and \gausspyplus.}
    \centering
    \small
    \renewcommand{\arraystretch}{1.2}
    \begin{tabular}{ccccccc}
    \hline\hline             
    & \multicolumn{3}{c}{\scousepy} & \multicolumn{3}{c}{\gausspyplus} \\             
    Sample & $\alpha_{1}$ & $\alpha_{2}$ & F$_{1}$ score & $\alpha_{1}$ & $\alpha_{2}$ & F$_{1}$ score \\
    \hline  
    1 & 2.02 & 4.48 & 73.9$\%$ & 2.38 & 4.28 & 77.9$\%$ \\
    2 & 2.0 & 4.08 & 73.9$\%$ & 2.12 & 4.17 & 75.9$\%$ \\
    3 & 2.14 & 4.8 & 73.6$\%$ & 1.41 & 4.49 & 71.4$\%$ \\
    \hline
    \end{tabular}
    \label{tbl:trainingset_scousepy}
\end{table}

Table~\ref{tbl:trainingset_scousepy} shows the resulting smoothing parameter values for these three training sets as determined by the machine learning routine of \gausspy.
A comparison with Table~\ref{tbl:trainingset_gpy} demonstrates that the inferred values for $\alpha_{1}$ and $\alpha_{2}$ cover the same range of values.
Table~\ref{tbl:trainingset_scousepy} also shows the results obtained if we instead used the training set decomposition technique of \gausspyplus\ to produce best fit solutions for \gausspy\ for the same three training sets.
The obtained values for the smoothing parameters are again similar to the values obtained with \scousepy\ and the values obtained for the \gausspyplus\ training sets in Table~\ref{tbl:trainingset_gpy}.
The larger spread of the obtained values for $\alpha_{1}$ is due to the presence of emission lines with low S/N ratio and non-Gaussian shapes in the random selection of spectra in the \scousepy\ training sets. 
Spectra containing such challenging spectral features can lead to increased $\chired$ values for the fit solution, which would usually cause their exclusion from the \gausspyplus\ training set selection. 
The similarity of the obtained smoothing parameter values is thus a reassuring confirmation that both \gausspyplus\ and \scousepy\ produce good training set decompositions.

Overall, the training sets created with \gausspyplus\ and \scousepy\ yielded similar values for the smoothing parameter values.
We thus decided to use the median values for $\alpha_{1}$ and $\alpha_{2}$ obtained from the nine training sets created with \gausspyplus\ as our chosen parameters for the decomposition of the GRS data set.
This yielded values of $\alpha_{1} = 1.97$ and $\alpha_{2} = 4.15$, which also compare well with the values inferred from the three training sets listed in Table~\ref{tbl:trainingset_scousepy}.
Given the highly non-uniform noise coverage of the GRS (see \fig\ref{fig:noise_map}), it is clear that our chosen values for the smoothing parameters will not be the optimal ones for all regions of the GRS data set. 
For example, \citet{Riener2019} inferred higher smoothing parameters of $\alpha_{1} = 2.89$ and $\alpha_{2} = 6.65$ for a small GRS test field that showed a strong noise gradient.
However, as discussed in \citealt{Riener2019}, the improved fitting and spatially coherent refitting routines of \gausspyplus\ should be able to salvage or mitigate negative effects introduced by the choice of non-optimal smoothing parameters.

\subsection{Decomposition parameters and steps}
\label{app:grs-decomposition}

The \gausspyplus\ decomposition proceeds in three stages (see \citealt{Riener2019} for more details).
In Stage~1, \gausspyplus\ fits each spectrum individually, with the AICc serving as decision criterion between different fit solutions for the same spectrum.
In Stage~2, fit solutions from Stage~1 can be flagged based on different user-selected criteria.
Neighbouring fit solutions are then used for refitting attempts of the flagged spectra, which already introduces local spatial coherence between the fit solutions.
Finally, in Stage~3 all fit solutions are checked for spatial coherence of the centroid position values of the fit components, with the aim of introducing spatial coherence more globally.
Neighbouring fit solutions again serve as refit templates for decomposition results that have deviating centroid position values.

By default \gausspyplus\ does not include flagged spectra as possible refit solutions in Stage~1. 
This was problematic for the GRS data set, as with our chosen flagging criteria there were regions near the Galactic midplane where almost all spectra were flagged and thus no refit solutions were available. 
We thus introduced the new parameter \texttt{use\_all\_neighbors} to \gausspyplus, with which users can allow flagged spectra to serve as templates in case no unflagged spectra are available or refit attempts using only the unflagged spectra were not successful.
In case flagged neighbouring spectra are used as refit templates, the flagged spectra are ranked according to their total flag values $\flag{tot}{}$ and the fit solution with the lowest $\flag{tot}{}$ value is used first.
Otherwise, we used the default settings of \gausspyplus\ as described in \citet{Riener2019}, with the exception of relaxing the $\Delta\mu_{max}$\footnote{See Sect.~3.3.1 in \citet{Riener2019} for more details.} parameter from its default default value of 2 to 4 channels in phase~2 of the spatially coherent refitting, as it was done for the decomposition of the GRS test field in \citet{Riener2019}. 
We also increased the minimum weight threshold $\pazocal{W}_{\Min}$ from its default value of $\sfrac{3}{6}$ to $\sfrac{4}{6}$, which means that in Stage~3 we only had two iterations with subsequent weight thresholds of $\pazocal{W} = \sfrac{5}{6}$ and $\pazocal{W} = \sfrac{4}{6}$ (see Sect.~3.3.2 in \citealt{Riener2019} for more details).
We tested the decomposition for one of the most complex regions in the GRS (with 10 or more emission peaks per spectrum) and found that a $\pazocal{W}_{\Min}$ value of $0.5$ was less beneficial, as it sometimes resulted in overly complex best fit solutions, especially for heavily blended structures.

\begin{table}
    \caption{Percentages of refitted spectra, added and subtracted fit components, and flagged spectra after each stage of \gausspyplus.
    The percentage is calculated relative to the total number of spectra in the GRS.}
    \centering
    \small
    \renewcommand{\arraystretch}{1.2}
    \begin{tabular}{cccc}
    \hline\hline             
    & Stage 1 & Stage 2 & Stage 3 \\
    \hline  
N$_{\mathrm{refit}}$ & -- & 58.9\% & 6.2\%\\
N$_{\mathrm{comp\,(+)}}$ & -- & 2.9\% & 1.1\%\\
N$_{\mathrm{comp\,(-)}}$ & -- & 1.8\% & 0.6\%\\
$\flag{tot}{}$ & $37.28\%$ & $31.67\%$ & $32.68\%$ \\
$\flag{blended}{}$ & $4.12\%$ & $1.86\%$ & $2.32\%$ \\
$\flag{neg.\,res.\,peak}{}$ & $0.13\%$ & $0.09\%$ & $0.10\%$ \\
$\pazocal{F}_{\Theta}$ & $17.39\%$ & $14.98\%$ & $15.70\%$ \\
$\pazocal{F}_{\Theta  > 50}$ & $8.05\%$ & $6.95\%$ & $7.11\%$ \\
$\flag{residual}{}$ & $18.65\%$ & $15.55\%$ & $15.65\%$ \\
$\pazocal{F}_{N_{\mathrm{comp}}}$ & $4.73\%$ & $3.69\%$ & $3.74\%$ \\
    \hline
    \end{tabular}
    \label{tbl:decomp_results}
\end{table}

Table~\ref{tbl:decomp_results} gives an overview of the percentage of spectra refitted in each stage and the percentage of spectra flagged after each stage.
The N$_{\mathrm{refit}}$ parameter gives the percentage of spectra that obtained a new best fit solution in the spatially coherent refitting phases.
In Stage~2, almost $59\%$ of the spectra were successfully refit based on a neighbouring fit solution, whereas in Stage~3 only about $6\%$ of the spectra were refit based on the neighbouring centroid position values. 
Based on this high fraction of refitted spectra, it is reasonable to assume that Stage~2 of \gausspyplus\ already managed to introduce a large amount of spatial coherence.
The N$_{\mathrm{comp\,(+)}}$ and N$_{\mathrm{comp\,(-)}}$ parameters give the percentage of added and removed fit components from the total number of fit components used for the entire decomposition in Stages~2 and 3 of \gausspyplus. 
Similar as for the decomposition of the GRS test field in \citet{Riener2019}, we find that the spatially coherent refitting stages tend to add more components to the fit solution.
This is expected, as the fit solutions in Stage~1 are guided by the AICc, which aims at a good trade-off between the number of fit components and the resulting goodness of fit of the model (see Sects.~3.2.1 and 3.2.3 in \citealt{Riener2019} for more information).
Stages~2 and 3 are designed to reduce flagged spectra based on flags that can be selected by the user.
For the decomposition of the GRS we use the default flags of \gausspyplus, which flag spectra with: blended fit components ($\flag{blended}{}$); negative residual peaks ($\flag{neg.\,res.\,peak}{}$); broad fit components ($\pazocal{F}_{\Theta}$); non-normally distributed residual values ($\flag{residual}{}$); number of fit components not compatible with neighbouring fits ($\pazocal{F}_{N_{\mathrm{comp}}}$). 
The $\flag{tot}{}$ value gives the percentage of spectra that were flagged by any of the flagging criteria listed above. 
The $\pazocal{F}_{\Theta}$ criterion only flags spectra that contain fit components that are broad compared to other fit components in the spectrum or in fit solutions of directly neighbouring spectra. 
However, it is also interesting to see whether the decomposition managed to reduce the fraction of fit components with very high absolute values for the linewidth. 
We thus also give the percentage of spectra that contain fit components with FWHM values that are higher than 50 spectral channels or about $10.5$~\kms\ ($\pazocal{F}_{\Theta  > 50}$), even though this was not a flagging criterion used in the \gausspyplus\ decomposition. 
We chose this upper limit for the FWHM as it is close to the maximum line widths of $9.8$ and $8.3$~\kms\ that \citet{Rathborne2009} found in their catalogue of GRS clouds and clumps, respectively.
Table~\ref{tbl:decomp_results} shows that in Stage~2 we are able to significantly reduce the percentage of flagged spectra for all flagging criteria.
Since in Stage~3 we are more concerned with enforcing spatial coherence based on the centroid velocity values of neighbouring fit components, the number of spectra flagged as having blended and broad fit components increases again slightly.

\subsection{Additional remarks on the decomposition results}
\label{app:add_remarks}

Some of the original GRS subcubes we used to produce a big mosaic of the entire data set (\sect~\ref{app:grs-preparation}) overlapped with each other. 
This overlap caused an averaging of the columns at the edges of the respective subcubes at even integer values in Galactic longitude ($\ell = 16, 18, \dots, 54\degr$).
This averaging produced lower noise values and higher S/N ratios of peaks in the spectrum, which led to a large number of fit components. 
We include these averaged columns in the best fit solutions presented in Table~\ref{tbl:decomp_results}, but we note that these fit solutions should be treated with caution.

There is an instrumental artefact present at $\vlsr$ values of $63.27\ \text{to}\ 66.09$~\kms and ranges in Galactic coordinates of $32.2\degr < \ell < 32.7\degr$ and $-1.1\degr < b < -0.7\degr$, which causes very narrow positive and negative high amplitude spikes, whose effect is especially visible in Figs.~\ref{fig:grs-mosaic_pv+spiral_arms} and \ref{fig:histogram_2d_centroid_vel-vel_disp}.
The final decomposition results still contain the fit components of the positive spikes for this region.


\section{Quality assurance metrics}
\label{sec:quality}

In this appendix we present different quality assurance metrics for our decomposition results. 
We discuss the distributions of our calculated noise values, goodness of fit values, and measures for the distribution of the residual values.

\subsection{Noise values}
\label{sec:noise-values}

\begin{figure*}
  \centering
  \includegraphics[width=0.9\textwidth]{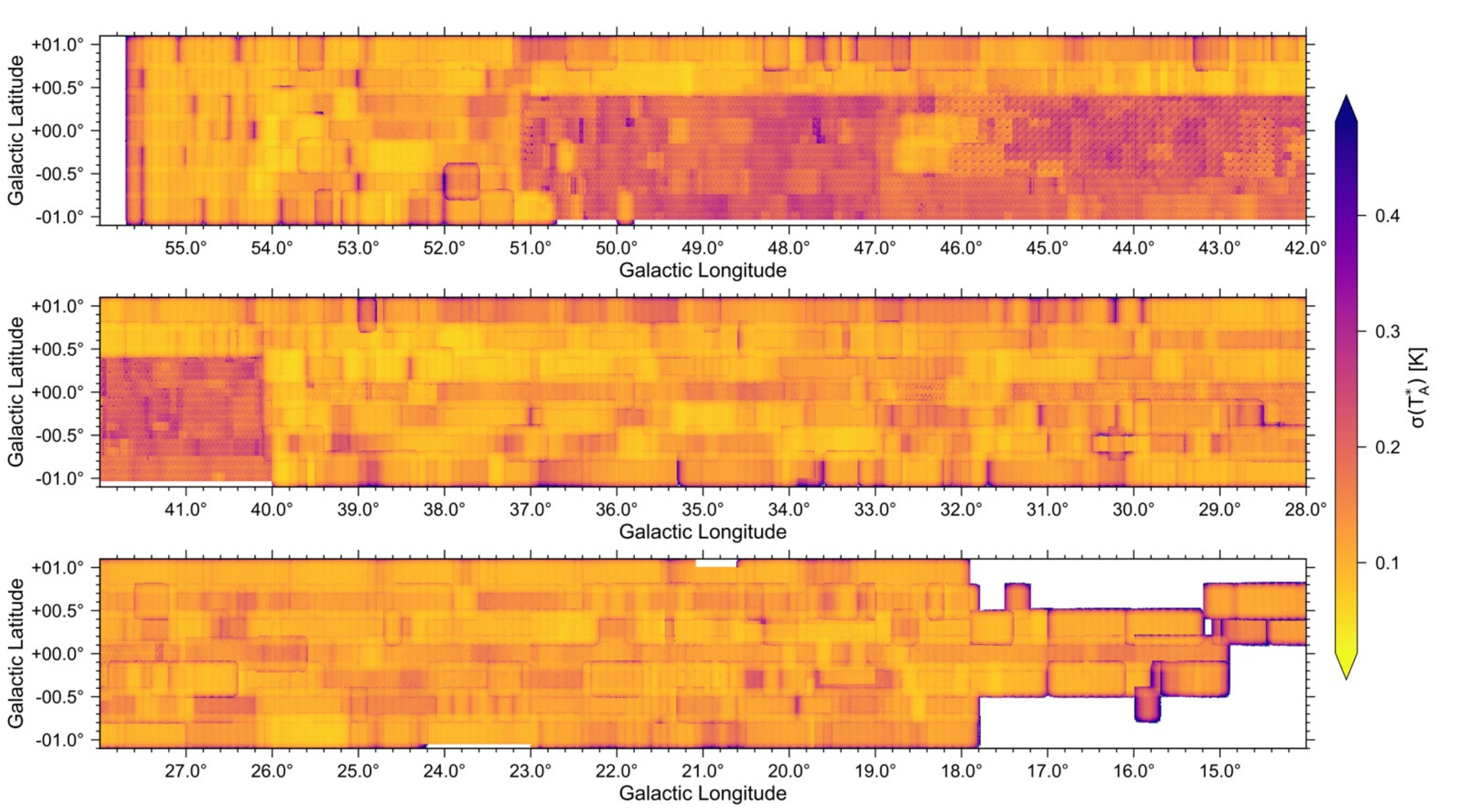}
  \caption{Map of noise values.}
  \label{fig:noise_map}
\end{figure*}

Vital parameters of the \gausspyplus\ algorithm are based on S/N thresholds, which is why a reliable noise estimate is an essential basis for obtaining good decomposition results \citep{Riener2019}.
Figure~\ref{fig:noise_map} shows the noise map of the entire coverage of the GRS data set that was obtained with the automated noise estimation routine of \gausspyplus\ (see Sect.~3.1.1 in \citealt{Riener2019} for more details on the noise estimation).
A comparison to the noise map published in the GRS overview paper (Fig.~7 from \citealt{Jackson2006}) shows that our noise map reproduces the overall large-scale patterns; however the individual noise values on a line of sight scale are more accurate and can show significant differences  (see Sect.~5.2 in \citealt{Riener2019} for more details).

\begin{figure}
    \centering
    \includegraphics[width=\columnwidth]{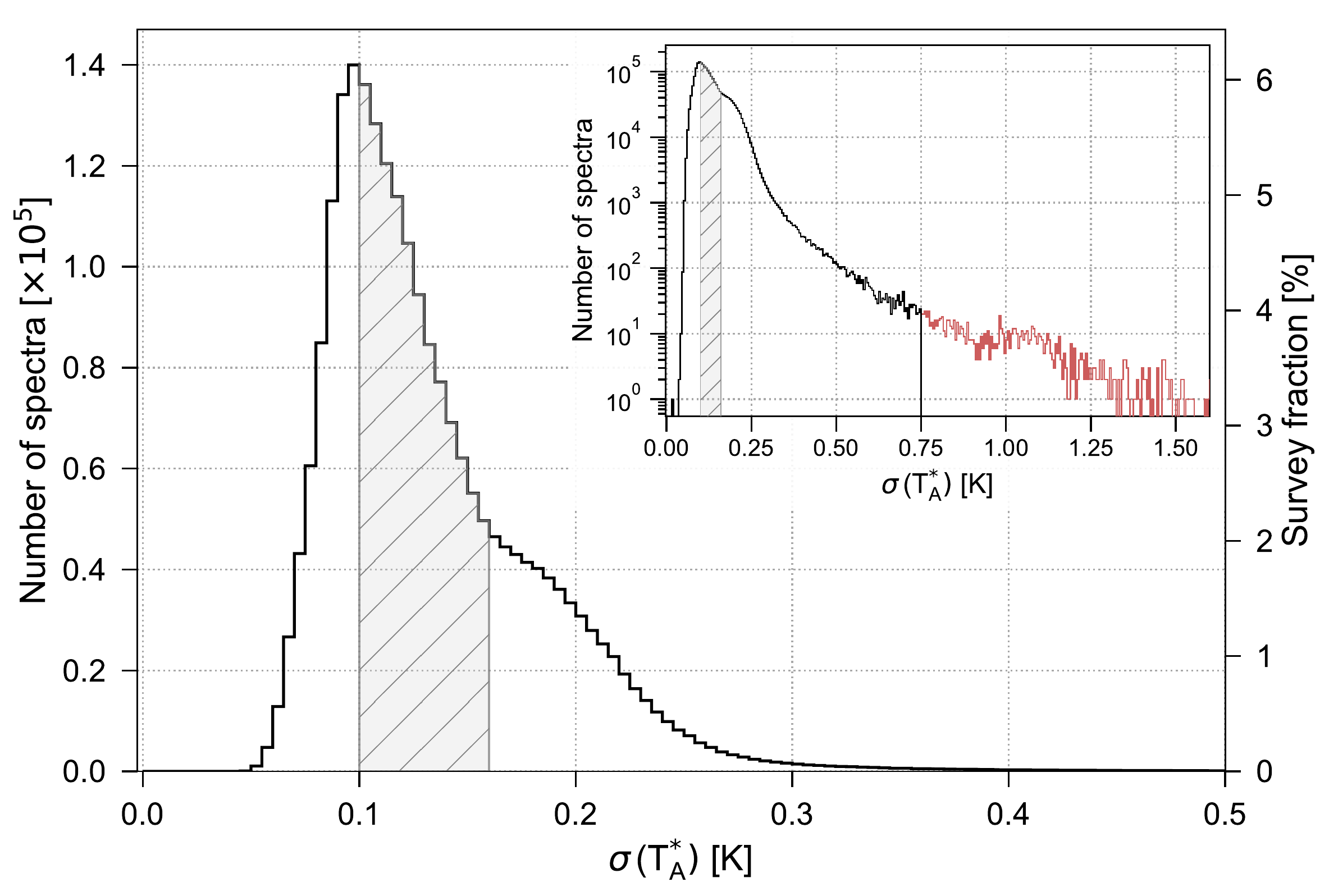}
    \caption{Histogram of determined rms noise (given in antenna temperature values) for all spectra in the GRS.
    The inset shows the same distribution on a logarithmic scale.
    The red line in the inset indicates masked out spectra with high noise values.
    The grey-shaded area marks the IQR, ranging from about $0.1 - 0.16$~K.
    The bin width is $0.005$~K.}
    \label{fig:histogram_noise}
\end{figure}

We show a histogram of the $\rms$ values (given in antenna temperatures) estimated by \gausspyplus\ in \fig\ref{fig:histogram_noise}.
The noise distribution peaks at a value of $\sigma(T_{A}^{*}) = 0.098$~K and shows a clear second bump after the IQR indicated with the hatched grey area. 
This bimodal noise distribution was already discussed in \citet{Jackson2006} and is explained by different observing modes used in the GRS survey (cf. their Fig.~8 but note that they do not include positions of $\ell < 18\degr$ and  $\vert b\vert > 1\degr$).
We decided to mask out 1188 spectra with $\rmsTa$ values $> 0.75$~K, which corresponds to the top $0.05\%$ of the noise distribution.
We found that such high $\rmsTa$ values can be indicative of instrumental artefacts.
The inset in \fig\ref{fig:histogram_noise} shows the noise distribution plotted on a logarithmic scale, and the $\rmsTa$ values of the masked spectra with high noise values are indicated in red.

\subsection{Goodness of fit statistics}
\label{sec:goodness-of-fit}

\begin{figure}
    \centering
    \includegraphics[width=\columnwidth]{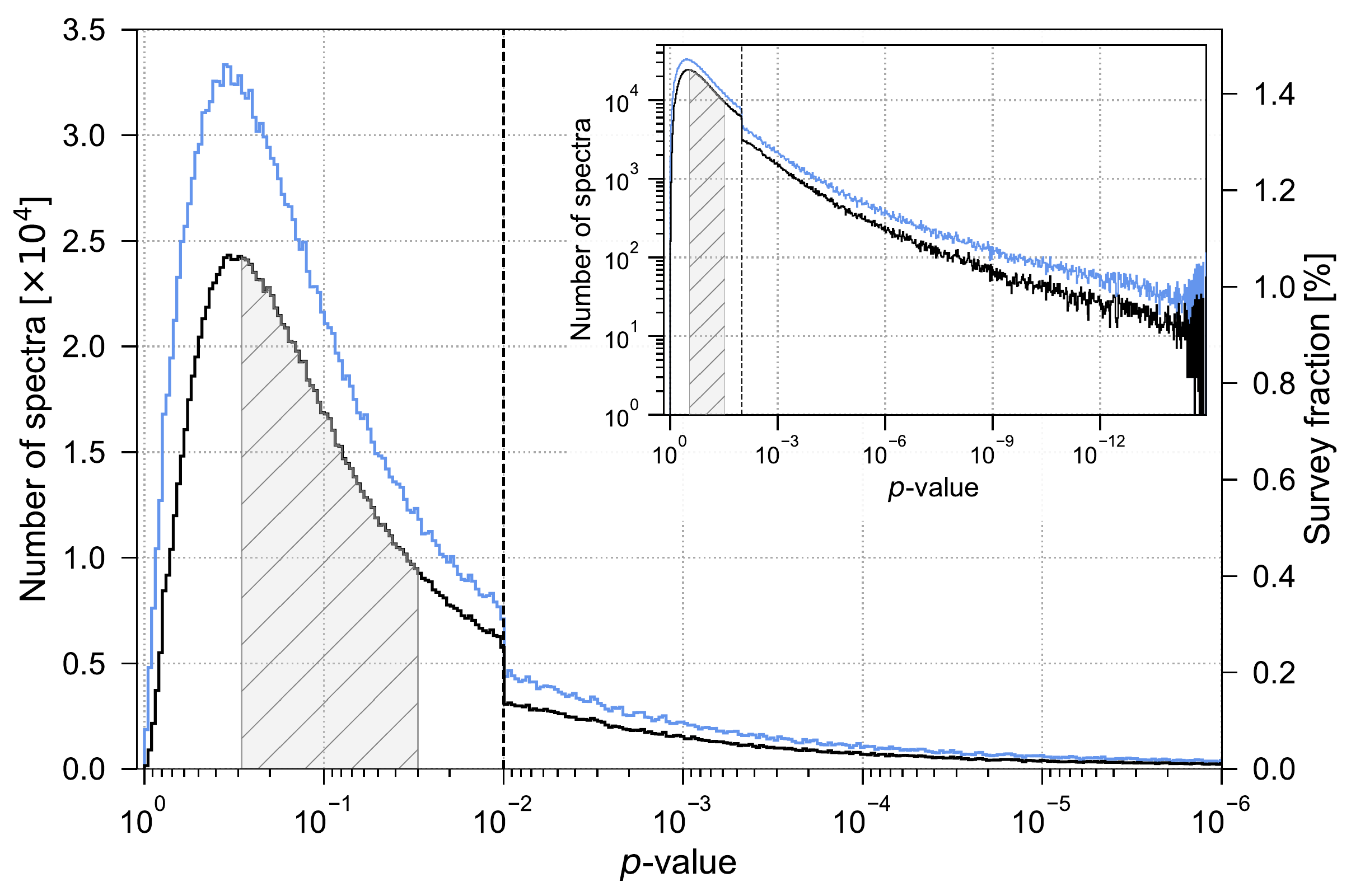}
    \caption{Histogram of $p$-values indicating normally distributed residual values for all GRS spectra (blue) and GRS spectra that have at least one fit component (black).
    The inset shows the same distribution on a logarithmic scale.
    The dashed vertical line marks the default $p$-value limit, below which \gausspyplus\ tries to refit spectra if possible.
    The grey-shaded area marks the IQR for the distribution of spectra with fitted components, ranging from about $0.03 - 0.3$.
    The bin width is $0.02$~dex.}
    \label{fig:histogram_pvalue}
\end{figure}

One of the goodness-of-fit estimates employed by \gausspyplus\ is a check of whether the normalised residuals of the spectra show a normal distribution and are thus consistent with Gaussian or white noise (see Sect.~3.2.1 in \citealt{Riener2019} for more details).
This check yields a $p$-value for the null hypothesis that the residuals are consistent with white noise.
Figure~\ref{fig:histogram_pvalue} shows the distribution of $p$-values for all spectra (blue line) and spectra that were fit by at least one component in the final decomposition results (black line). 
The inset gives the same distribution for a logarithmically scaled ordinate.
In the default settings of \gausspyplus, a $p$-value of $< 1\%$ serves as an indication that the residual contains features inconsistent with Gaussian noise, which is used to initiate refit attempts or decide between alternative fit solutions.
This threshold for the $p$-value is indicated with the vertical dashed line and we can see a clear jump in the distribution at this value.

Another goodness-of-fit statistic that is often used to report the quality of fit results is the $\chired$ value that is defined as 

\begin{equation}
	\chired = \frac{1}{N - k} \sum^{N}_{i=1} \frac{\left(y_{i} - Y_{i}\right)^{2}}{\sigma_{\mathrm{rms}}^{2}},
\end{equation}

\noindent where in the case of the Gaussian decomposition $N$ corresponds to the number of spectral channels, $k$ corresponds to the number of free parameters used in the fit solution and $y_{i}$ and $Y_{i}$ are the data and fit value at channel position $i$.
Since for non-linear models the $\chired$ estimate can suffer from large uncertainties \citep{Andrae2010} it is not the best goodness-of-fit metric for the Gaussian decomposition results. As discussed in \citet{Riener2019}, another problem of $\chired$ estimates is that the inclusion of a large number of spectral channels containing only noise can mask bad fit results.
Given these caveats, our reported $\chired$ values for our fit results should be taken with caution.
If \gausspyplus\ identified signal intervals in a spectrum we only include the spectral channels identified in those intervals for the $\chired$ estimate, which increases its ability to identify potentially incorrect or insufficient fit results.

\begin{figure}
    \centering
    \includegraphics[width=\columnwidth]{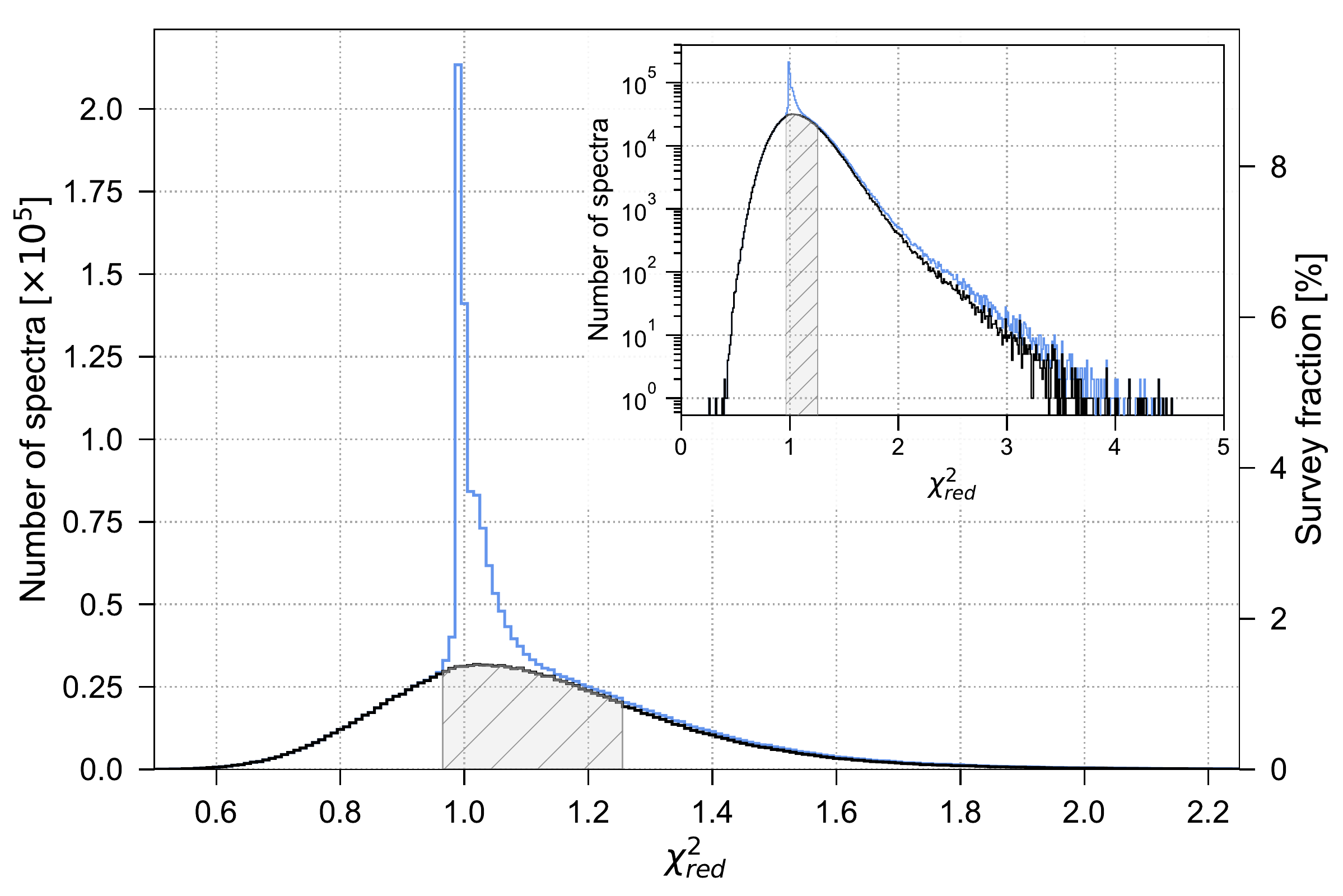}
    \caption{Histogram of $\chired$ values for all GRS spectra (blue) and GRS spectra with at least one fit component (black).
    The inset shows the same distribution on a logarithmic scale.
    The grey-shaded area marks the IQR for the distribution of spectra with fitted components, ranging from about $0.96 - 1.25$.
    The bin width is $0.01$.}
    \label{fig:histogram_rchi2}
\end{figure}

The distribution of the $\chired$ values for the spectra from the GRS is shown in \fig\ref{fig:histogram_rchi2}.
The distribution marked with the blue line contains all spectra of the GRS data set, irrespective of whether they were fit in the decomposition.
The black line shows the distribution of the $\chired$ values for spectra that have at least one fitted Gaussian component. 
Both of the distributions peak at a value of 1.
Spectra with no fitted Gaussian components are expected to have a $\chired$ value close to 1 if they only contain noise and the $\rms$ value was estimated correctly.
Increased $\chired$ values for these unfitted spectra can thus indicate spectra with valid signal that could not be fit, an incorrect noise estimation, or artefacts in the spectrum such as insufficient baseline subtraction.

About $35\%$ of the spectra with fitted Gaussian components have $\chired$ values below $1$; about $0.4\%$ of the spectra with fitted Gaussian components have $\chired$ values above $2$.
Since we aimed to mostly include channels that contain signal in the calculation of the goodness of fit criterion, for most of the spectra low $\chired$ values do very likely not indicate an overfitting of the data, but rather reflect a smaller amount of spectral channels that were used for the $\chired$ calculation.

\begin{figure*}
  \centering
  \includegraphics[width=0.9\textwidth]{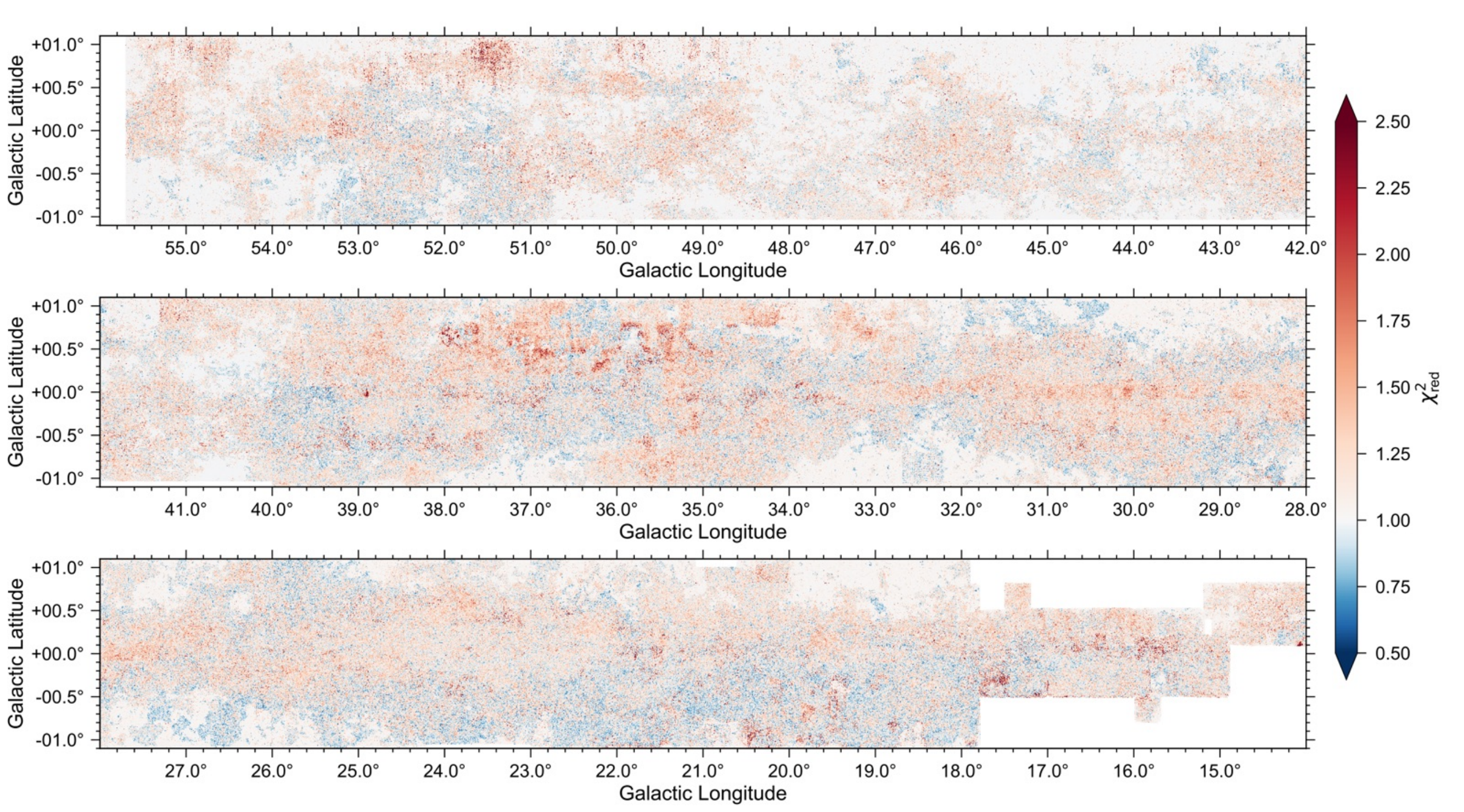}
  \caption{Map of $\chired$ values.}
  \label{fig:rchi2_map}
\end{figure*}

Figure~\ref{fig:rchi2_map} shows a map of the $\chired$ values for the entire GRS data set. 
This map does not show any obvious correlations with the maps showing the normalised residuals (\fig\ref{fig:zero_mom}), noise values (\fig\ref{fig:noise_map}), and number of fit components (\fig\ref{fig:component_map}).
However, we can see regions of increased $\chired$ values (e.g. around $\ell\sim 51.5\degr$ and $b\sim 1\degr$), which could indicate problems with the decomposition, such as missed or incorrectly fit signal peaks, or problems in the original data, such as insufficient baseline corrections.


\section{Effects of optical depth on the decomposition results}
\label{sec:optical_depth}

Interpretation of the Gaussian decomposition data can potentially be affected by optical depth effects; we examine in this section how severe such effects may be. 
Moderate optical depth effects can lead to a broadening of the emission lines, which affects the line widths (and further, estimated Mach numbers; e.g. \citealt{Hacar2016opacity}).
Increased optical depth effects can cause flat-topped or self-absorbed emission lines, which can yield two-component fits that are unphysical.
The version of \gausspyplus\ used in this work does not account for any possible optical depth effects.
If an emission line is affected by self-absorption, \gausspyplus\ will likely fit it with multiple Gaussian components. 
Our aim here is to establish in which regions, or for which spectra in the GRS, optical depth effect may affect the decomposition result. 
We will start by recalling previous results about optical depth effects, depletion, and freeze out on the $\co{13}{}\transition{1}{0}$ transition. 
We will then present two methods of calculating the optical depth $\opacity{13}$ of $\co{13}{}\transition{1}{0}$.
Subsequently, we will apply these methods to observations from the MWISP survey targeting a dense molecular cloud and establish how severe the optical depth effects are and how we can apply these results to the GRS decomposition. 

\subsection{Depletion, freeze-out, and self-absorption of $^{13}$CO}
\label{sec:tau_literature}

Evidence for depletion or freeze-out of CO isotopologues is already well-documented for IRDCs and clumps contained in the GRS data set \citep[e.g.][]{Hernandez2011irdcs, Pitann2013irdcs, Giannetti2014atlasgal, Pon2016irdcs, Barnes2018-irdcs}, albeit for observations at spatial and spectral resolutions that are typically at least a factor of two better than for the GRS.
For clumps, the depletions are usually most severe for their densest and coldest parts \citep{Hernandez2011irdcs, Giannetti2014atlasgal}.
In the GRS data set, such small regions are only resolved for gas emission that is located within $\sim 3$~kpc of the sun, which is only a small fraction of the emission contained in the GRS.
We thus conclude that freeze-out and depletion effects will not be a problem at the scales probed by the GRS.

Optical depth values have already been calculated for the GRS, albeit only on scales of molecular clouds and for spatially and spectrally smoothed versions of the data set.
For their catalogue of molecular clouds in the GRS, \citet{Rathborne2009} found the $^{13}$CO emission to be optically thin with average and maximum $\opacity{13}$ values of $0.13$ and $0.5$, respectively.
\citet{Roman-Duval2010} repeated the $\opacity{13}$ estimation for 583 molecular clouds from \citet{Rathborne2009}, but performed the calculation on a voxel-by-voxel basis first before averaging; this resulted in a higher mean optical depth value of $1.46$, indicating that the GRS data set includes optically thick emission.
\citet{Rigby2016} also found indications for self-absorption in the GRS data set, by comparing it to $\co{13}{}\transition{3}{2}$ and $\co{}{18}\transition{3}{2}$ transitions of the CHIMPS survey (that has a spatial resolution about 3 times better than GRS).

However, there are also counterexamples that indicate that $\co{13}{}\transition{1}{0}$ stays mostly optically thin even in dense regions. 
For example, \citet{Beuther2007_irdcs} studied a sample of 43 IRDCs and found that the peaks in $\co{13}{}\transition{1}{0}$ correspond very well with peaks of the high density tracer $\text{H}^{13}\text{CO}^{+}\transition{1}{0}$ even for regions in the spectrum where $\co{12}{}\transition{2}{1}$ seems to get optically thick.
Since these were IRAM-30m observations at about half the spatial resolution of GRS, one would conclude that for a larger physical beam such effects should be even more reduced due to beam averaging.

\subsection{Calculation of optical depth values}
\label{sec:tau_calc}

Here we review two common ways of estimating the optical depth of $\co{13}{}$, which we will later apply in \sect\ref{sec:tau_g24} to assess the importance of opacity effects on the decomposition results. 
The first method for calculating $\opacity{13}$ requires information about the excitation temperature $\Tex$, obtained from $\co{12}{}\transition{1}{0}$ emission that is assumed to be optically thick.
The second method for calculating $\opacity{13}$ is based on the relative abundance of $\co{13}{}$ to another isotopologue that traces higher column densities and is assumed to be optically thin; in our case this is the $\co{}{18}\transition{1}{0}$ emission line.
This method is thus best applicable to dense regions of molecular clouds, such as IRDCs, filaments, clumps, and cores.

\subsubsection{Method 1: Estimating $\opacity{13}$ with additional $\co{12}{}\transition{1}{0}$ observations}
\label{sec:tau_calc_m1}

Under the condition of local thermodynamic equilibrium and the assumption that the $\co{12}{}\transition{1}{0}$ line is optically thick, the excitation temperature $\Tex$ can be determined as \citep{Wilson+Rohlfs}:

\begin{equation}
    \Tex = T_{0}^{12} \cdot \left(
        \text{ln}\left[
        1 + \dfrac{T_{0}^{12}}{\Tb^{12} + 0.82} \right]
        \right)^{-1},
    \label{eq:tex}
\end{equation}

\noindent where $\Tb^{12}$ is the maximum main beam brightness temperature of the $\co{12}{}\transition{1}{0}$ line and $T_{0}^{12} = h \nu^{12} k_{\mathrm{B}} \sim 5.5$~K at the rest frequency of $\co{12}{}$ ($\nu^{12} = 115.271$~GHz).
Assuming that $\Tex$ is the same for the $\co{12}{}\transition{1}{0}$ and $\co{13}{}\transition{1}{0}$ transitions, the optical depth for the $\co{13}{}\transition{1}{0}$ line is then given by \citep{Wilson+Rohlfs}:

\begin{equation}
    \opacity{13} = -\text{ln}
    \left[
    	1 - \dfrac{\Tb^{13}}{T_{0}^{13}} \left(
        	\dfrac{1}{e^{T_{0}^{13} / \Tex} - 1} - 0.16\right)^{-1}
    \right],
    \label{eq:tau}
\end{equation}

\noindent where $\Tb^{13}$ is the maximum main beam brightness temperature of the $\co{13}{}\transition{1}{0}$ line and $T_{0}^{13} = h \nu^{13} k_{\mathrm{B}} \sim 5.3$~K at the rest frequency of $\co{13}{}$ ($\nu^{13} = 110.201$~GHz).

\begin{figure}
    \centering
    \includegraphics[width=\columnwidth]{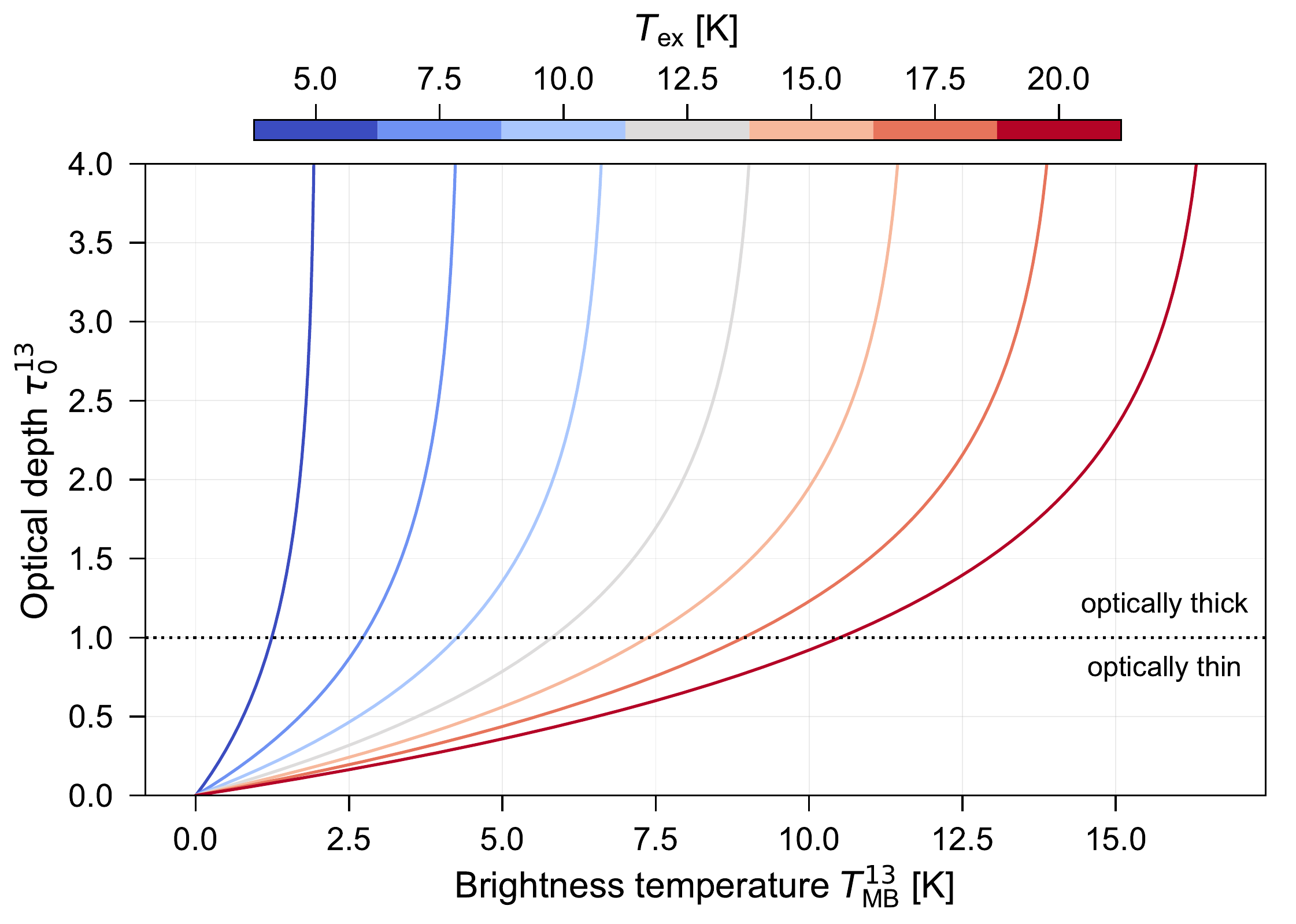}
    \caption{Optical depth of $^{13}$CO as a function of main beam brightness temperature for different excitation temperature ($\Tex$) values.
    The dotted horizontal line marks the threshold between the regimes where $^{13}$CO is optically thin or thick.}
    \label{fig:optical-depth}
\end{figure}

\begin{table}
    \caption{Threshold between optically thin and thick emission for given excitation temperature values.}
    \centering
    \small
    \renewcommand{\arraystretch}{1.3}
    \begin{tabular}{ccc}
    \hline\hline
    $\Tex$ [K] & T$^{13}_{\text{B,~crit.}}$ [K] & T$^{13}_{\text{B,~GRS}}$ > T$^{13}_{\text{B,~crit.}}$ \\
    \hline
    5.0 & 1.2 & 41.96\% \\
    7.5 & 2.7 & 8.55\% \\
    10.0 & 4.3 & 1.84\% \\
    12.5 & 5.8 & 0.47\% \\
    15.0 & 7.4 & 0.16\% \\
    17.5 & 8.9 & 0.06\% \\
    20.0 & 10.5 & 0.02\% \\
    \hline
    \end{tabular}
    \label{tbl:optical_depth}
\end{table}

Figure~\ref{fig:optical-depth} shows how $\opacity{13}$ increases as a function of $\Tb^{13}$ for different fixed values of $\Tex$. 
Table~\ref{tbl:optical_depth} gives the $\Tb^{13}$ values for which we expect the $\co{13}{}$ emission to get optically thick for a given $\Tex$ value.
The table also shows the percentage of fit components from the GRS decomposition that have amplitude values greater than this $\Tb^{13}$ value.
For example, for an assumed fixed value of $\Tex = 10$~K, about $1.8$\% of all Gaussian fit components in the GRS survey would pass the threshold between optically thin and optically thick gas.
A comparison with \fig\ref{fig:histogram_intensity} shows that optical depth effects are important for our decomposition results if the excitation temperature values are $\lesssim 10$~K.

\subsubsection{Method 2: Estimating $\opacity{13}$ with additional $\co{}{18}\transition{1}{0}$ observations}
\label{sec:tau_calc_m2}

In case additional $\co{}{18}\transition{1}{0}$ observations are available, $\opacity{13}$ can be determined based on the expected relative abundance of $\co{13}{}$ to $\co{}{18}$.
Under the assumption that the $\co{}{18}$ emission is optically thin, a comparison of the integrated emission $\wco$ of $\co{13}{}$ and $\co{}{18}$ can be put in direct relation to their opacities $\opacity{13}$:

\begin{equation}
    \dfrac{W_{\co{13}{16}}}{W_{\co{12}{18}}} = 
    \dfrac{1 - e^{-\opacity{13}}}{1 - e^{-\opacity{18}}}.
    \label{eq:wco_tau}
\end{equation}

\noindent We can use information about the isotopic ratios of carbon and oxygen to rewrite $\opacity{18}$ in terms of $\opacity{13}$:

\begin{equation}
    \opacity{18} = 
    \dfrac{[^{12}\text{C}]/[^{13}\text{C}]}{[^{16}\text{O}]/[^{18}\text{O}]} \cdot \opacity{13}.
    \label{eq:relative_abundance}
\end{equation}

\noindent \citet{Giannetti2014atlasgal} determined that these isotope ratios vary with Galactocentric distance R$_{\text{gal}}$ as $[^{12}\text{C}]/[^{13}\text{C}] = 6.1\times \text{R}_{\text{gal}}~\text{[kpc]} + 14.3$ and $[^{16}\text{O}]/[^{18}\text{O}] = 58.8\times \text{R}_{\text{gal}}~\text{[kpc]} + 37.1$.
Substituting Eq.~\ref{eq:relative_abundance} in Eq.~\ref{eq:wco_tau} thus allows to solve for $\opacity{13}$.

\subsection{Optical depth effects in the G24 region}
\label{sec:tau_g24}

In this section we study the effects of optical depth on the decomposition with the help of a dense, elongated giant molecular cloud.
The cloud, G24, has been identified by \citet{Wang2015gmfs} from the HiGAL and GRS data sets. 
G24 is located at the Galactic coordinates $\ell=24\degr, b=+0.48\degr$ and spans $\vlsr$ values of $93 \text{--} 99$~\kms\ \citep{Wang2015gmfs}.
The total mass of G24 has been estimated to be about $10^5$~\msun, the length about 100 pc, and the distance about 5.8~kpc \citep[][]{Zucker2018-gmfs, Zhang2019-gmfs}. 
The cloud is star-forming and its star formation rate and efficiency are typical for similar clouds in the Milky Way \citep{Zhang2019-gmfs}.
G24 overlaps with the molecular cloud GRSMC~G024.09+00.44 identified by \citep{Rathborne2009}, whose total mass was estimated to be $2.8 \times 10^{5}$~\msun\ \citep{RomanDuval2010gmcs}.

\begin{figure}
    \centering
    \includegraphics[width=0.85\columnwidth]{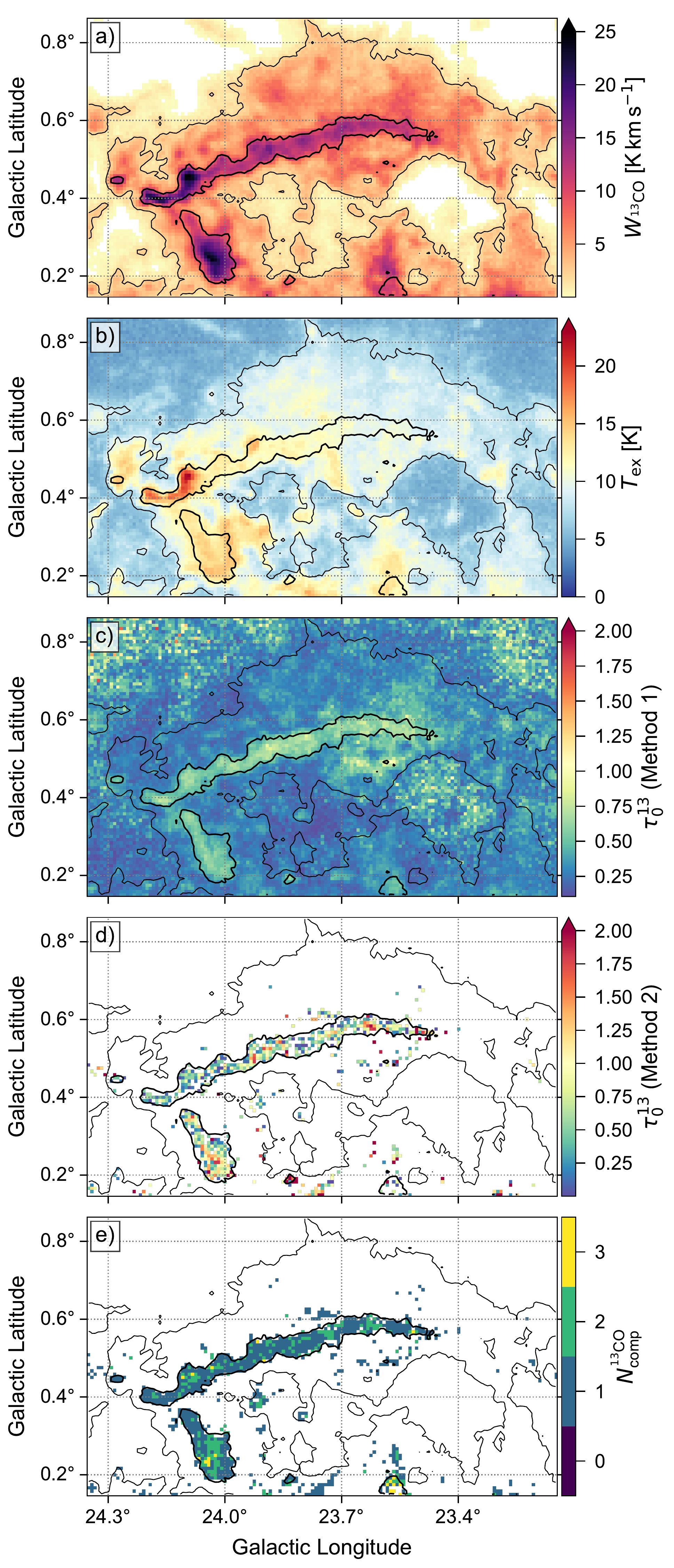}
    \caption{
    Maps for the region containing the elongated molecular cloud G24.
    \emph{a)}: Zeroth moment map of the $\co{13}{}$ emission.
    \emph{b)}: Map of determined excitation temperature values.
    \emph{c)}: Map of determined $\opacity{13}$ values using Method~1 (App.$\,$\ref{sec:tau_calc_m1}).
    \emph{d)}: Map of determined $\opacity{13}$ values using Method~2 (App.$\,$\ref{sec:tau_calc_m2}).
    \emph{e)}: Map showing number of $\co{13}{}$ fit components associated with a single $\co{}{18}$ fit component.
    The black contours indicate $W_{\co{13}{}}$ values of 2~\Kkms\ (thin line) and $W_{\co{}{18}}$ values of 1~\Kkms\ (thick line).
    All maps were obtained for the data points between $93 < \vlsr < 99$~\kms.
    }
    \label{fig:g24_tex}
\end{figure}

There are two reasons that make G24 a good example to test if optical depth effects affect our decomposition.
First, G24 is located in one of the most complex regions of the GRS data set, judging by the number of fitted components along the line of sight (\fig\ref{fig:component_map}). 
The decomposition is most challenging for such complex lines of sight. 
Second, G24 is among the most massive structures identified from GRS as distinct clouds, with a gas surface density of $\sim 105$~\msun$\,$pc$^{-2}$ \citep{Zhang2019-gmfs}.
As a result, if optical depth effects hamper the GRS decomposition in general, we should definitely see the effects in G24.

\subsubsection{Observations}

We used data of the G24 region from the MWISP project \citep{Su2019}, which is an ongoing CO survey for the northern Galactic plane ($\ell=-10\ \text{to}\ 250\degr$, $\vert b\vert\leq 5.2\degr$) using the PMO-13.7$\,$m single-dish telescope located at Delingha in China.
We obtained observations of $\co{12}{} \transition{1}{0}$, $\co{13}{} \transition{1}{0}$, and $\co{}{18} \transition{1}{0}$ covering a $2.5\degr \times 1\degr$ field centred at Galactic coordinates $\ell=23.75\degr$ and $b=0.5\degr$.
The angular resolution of the data is $54.8^{\prime\prime}$ with a pixel sampling of $30^{\prime\prime}$.
The covered velocity range is from -500 to + 500 \kms with a velocity resolution of $\sim$ 0.16~\kms and 0.17~\kms for the $\co{12}{}$ and $^{13}$CO/$\co{}{18}$ observations, respectively.
The data was already supplied in main beam temperatures.

Figure~\ref{fig:g24_tex}\emph{a} shows a zeroth moment map for the G24 region, which was obtained from the $\co{13}{}$ MWISP data integrated between $93 < \vlsr < 99$~\kms\ using the moment masking technique described in \citet{Dame2011}.
The very good correspondence of the $W_{\co{}{18}}$ contour defined on the corresponding $\co{}{18}$ data with the zeroth moment map of $\co{13}{}$ confirms that $\co{}{18}$ emission is coming from areas with increased $\co{13}{}$ emission and thus increased column density.
The $W_{\co{}{18}}$ contour shown in \fig\ref{fig:g24_tex}\emph{a} corresponds to a $W_{\co{13}{}}$ contour of $\sim 10\,$\Kkms.

\subsubsection{\gausspyplus\ decomposition}
\label{app:mwisp-decomp}

\begin{table}
    \caption{Smoothing parameters (in units of spectral channels) obtained for decompositions of the \gausspyplus\ training sets for the $\co{13}{}$ and $\co{}{18}$ MWISP data sets.}
    \centering
    \small
    \renewcommand{\arraystretch}{1.3}
    \begin{tabular}{cccc}
    \hline\hline             
    Isotopologue & $\alpha_{1}$ & $\alpha_{2}$ & F$_{1}$ score \\
    \hline  
    $\co{13}{}$ & 2.18 & 4.94 & 71.1$\%$ \\
    $\co{}{18}$ & 2.98 & 5.75 & 70.2$\%$ \\
    \hline
    \end{tabular}
    \label{tbl:trainingset_mwisp}
\end{table}

For the decomposition of the MWISP data set we reduced the velocity axis to a range of $-50\ \text{to}\ +150$~\kms, which correspond to 1207 and 1202 spectral channels for $\co{13}{}$ and $\co{}{18}$, respectively. 
Similar as in App.~\ref{app:grs-training}, we used the default settings of \gausspyplus\ to create training sets with 500 decomposed spectra for both isotopologues.
In Table~\ref{tbl:trainingset_mwisp} we list the determined smoothing parameters and corresponding F$_{1}$ score for both training sets.
The $\co{}{18}$ data has a much lower S/N ratio and noise properties can thus affect the line shape already significantly, so the data needed to be smoothed more to yield a good decomposition performance.
Apart from the different smoothing parameter values, we used the same \gausspyplus\ settings to prepare and decompose the MWISP $\co{13}{}$ and $\co{}{18}$ spectra as for the GRS data set (described in Apps.~\ref{app:grs-preparation} and \ref{app:grs-decomposition}).

\subsubsection{Comparison of the decomposition results between the GRS and MWISP}
\label{sec:comparison}

\begin{figure}
    \centering
    \includegraphics[width=\columnwidth]{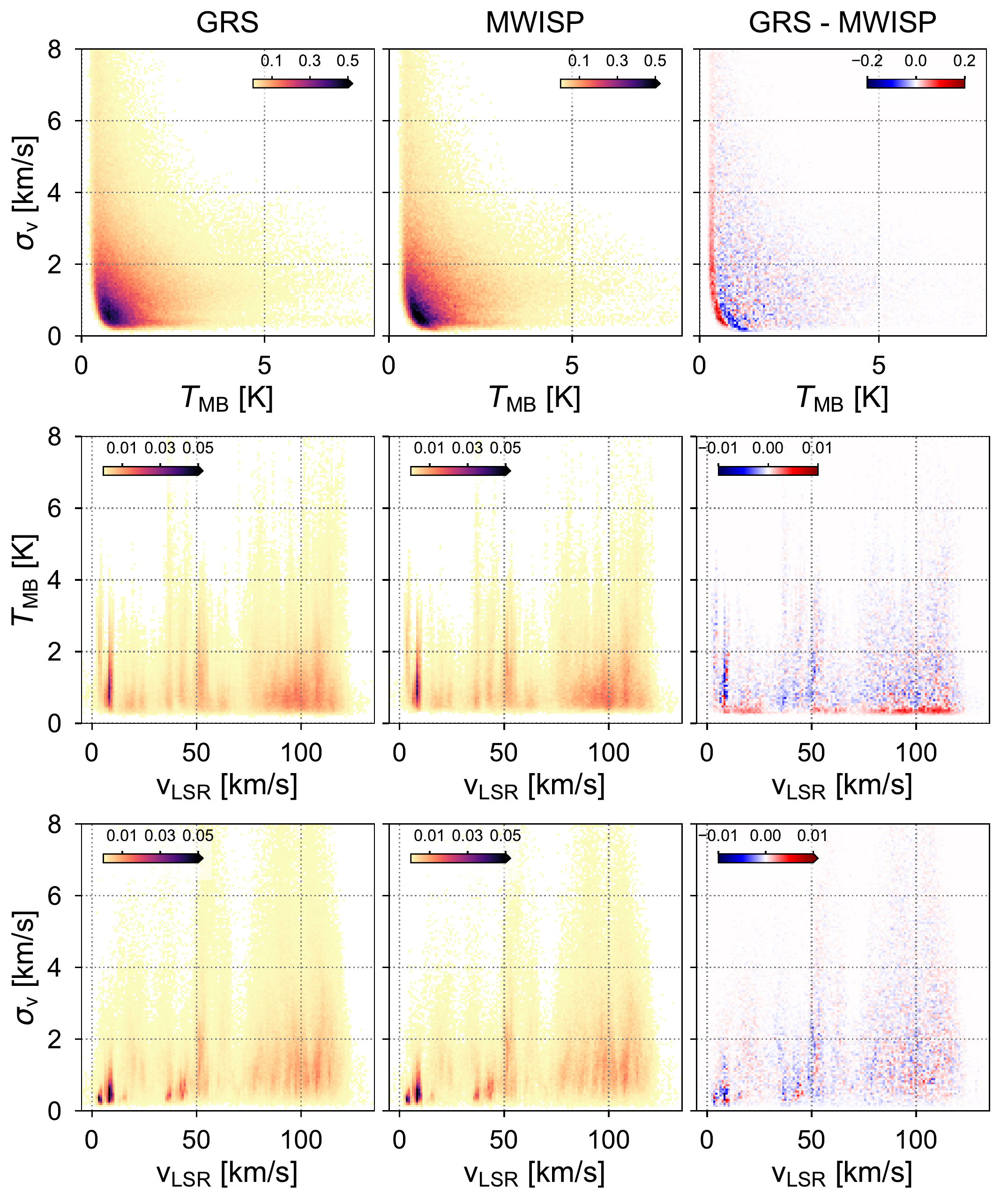} 
    \caption{Distributions of fit parameters for the decomposition of the GRS (\emph{left}), MWISP (\emph{middle}), and the difference between the two data sets (\emph{right}).
    The rows show normalised 2D histograms of: peak main beam brightness temperature and velocity dispersion values (\emph{top}), centroid velocity and peak main beam brightness temperature values (\emph{middle}), centroid velocity and velocity dispersion values (\emph{bottom}).
    Colourbars in all panels indicate the values of the normalised 2D distributions.  
    }
    \label{fig:comp_grs_mwisp}
\end{figure}

We now compare the \gausspyplus\ decomposition results obtained for the GRS and MWISP data sets. 
If the fit results for the two data sets are comparable, we can conclude that also our following results about the optical depth effects derived for the MWISP data will apply similarly to the GRS.
We cannot perform a straightforward spectrum-per-spectrum comparison, since the spectral channel widths and pixel scales of the data sets are not identical.
We thus opted for a simpler approach of comparing the fit parameter statistics.

For this comparison we used all fit components within $22.5\degr \lesssim \ell \lesssim 25\degr$, $0\degr \lesssim b \lesssim 1\degr$, and $-5 \lesssim \vlsr \lesssim 135\,\text{\kms}$.
Due to the different spatial resolutions and pixel scales this selection included $58\,635$ spectra with $271\,896$ fit components in the GRS data set and $32\,153$ spectra with $144\,555$ fit components in the MWISP data set.
Figure~\ref{fig:comp_grs_mwisp} shows normalised 2D histograms for all possible combinations of the Gaussian fit parameters for the GRS (\emph{left panels}) and MWISP (\emph{middle panels}) data sets.
The shapes of the 2D distribution of the GRS and MWISP match very well, which already demonstrates that the two decompositions yielded similar results.
To better quantify how similar the 2D distributions are, we show the difference between their distributions in the \emph{right panels}, where red and blue indicate higher values in the GRS and MWISP distributions, respectively.
The GRS data set includes regions with very low $\rms$ values, which led to fitted components with low $\Tb$ values that could not be detected in the MWISP data set.
Otherwise these distributions do not show any noticeable biases; there are individual small differences, but these can be explained by the variation in the resolution elements and the noise coverage.

We thus conclude that the GRS and MWISP decomposition results are sufficiently similar.
Thus, we expect that the following results about optical depth effects derived from the MWISP data can be applied to the GRS.

\subsubsection{Estimated $\opacity{13}$ values with Method 1}
\label{sec:tau_g24_m1}

\begin{figure}
    \centering
    \includegraphics[width=\columnwidth]{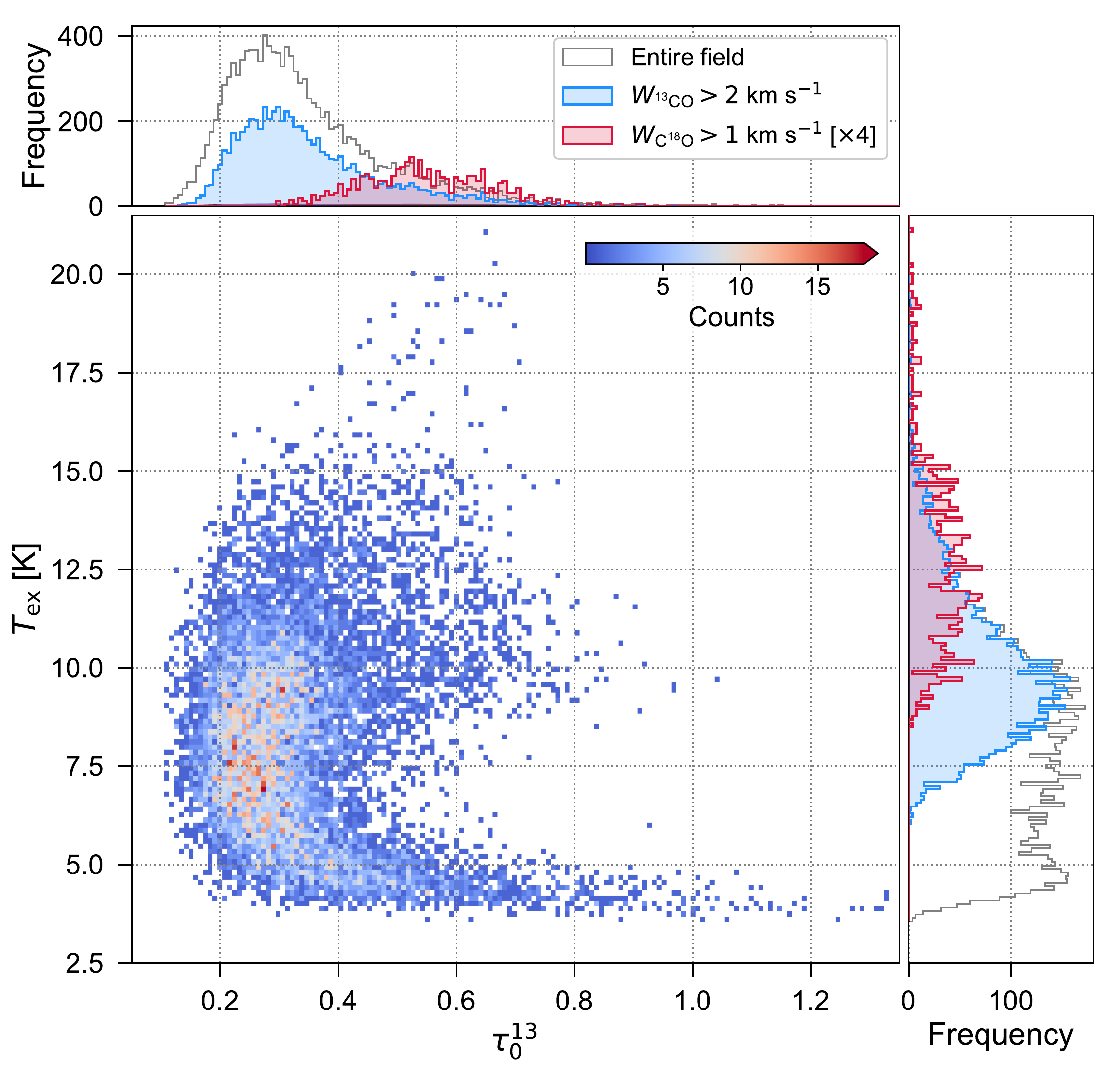}
    \caption{
    2D distribution of determined $\Tex$ and $\opacity{13}$ values shown in \fig\ref{fig:g24_tex}\emph{b} and \emph{c}.
    Marginal distributions on top and to the right are for the entire region shown in \fig\ref{fig:g24_tex} (unfilled histograms), and subsets of that region within the $W_{\co{13}{}} = 2$~K$\,$\kms\ contour (blue histograms) and within the $W_{\co{}{18}} = 1$~K$\,$\kms\ contour (red histograms; the counts are scaled by a factor 4 for better visibility).
    }
    \label{fig:g24_tau}
\end{figure}

With the $\co{12}{}$ and $\co{13}{}$ MWISP observations of the G24 region and Eqs.~\ref{eq:tex} and \ref{eq:tau} we can estimate values for $\Tex$ and $\opacity{13}$, respectively.
For the $\Tex$ and $\opacity{13}$ calculation we only considered data points within $93 < \vlsr < 99$~\kms.
We show the resulting maps for $\Tex$ and $\opacity{13}$ in \emph{panel b} and \emph{c} of \fig\ref{fig:g24_tex}, which already shows that the densest part of G24 is associated with higher $\opacity{13}$ values.
To better quantify how the optical depth varies with density and excitation temperature, we show 2D distributions of the $\Tex$ and $\opacity{13}$ values in \fig\ref{fig:g24_tau}.
Inside the contour defined by $W_{\co{13}{}} > 2$~\Kkms, the $\Tex$ values range from about $6$ to $23.5$~K and the $\opacity{13}$ values range from 0.12 to 1.04, with a median $\opacity{13}$ value of 0.32.
The area of G24 within a contour of $W_{\co{}{18}} > 1$~\Kkms\ is also associated with the highest $\Tex$ values (from about 8.5 to 23.5~K) as can be clearly seen in the respective marginal distribution shown in red in \fig\ref{fig:g24_tau}.
The corresponding $\opacity{13}$ values in this area range from 0.3 to 0.92, and the median $\opacity{13}$ value is 0.54.
While the densest regions of G24 definitely show increased values for $\opacity{13}$, the $\co{13}{}\transition{1}{0}$ transition is for most lines of sight still well below optical depth values of 1.
Based on the calculation of $\opacity{13}$ with Method 1, we conclude that optical depth effects do not play a significant role in the GRS data set.

\subsubsection{Estimated $\opacity{13}$ values with Method 2}
\label{sec:tau_g24_m2}

With the $\co{13}{}$ and $\co{18}{}$ MWISP observations of the G24 region and Eq.~\ref{eq:wco_tau} we can give an independent estimate for $\opacity{13}$.
We took values for the integrated intensity from our \gausspyplus\ decomposition of the MWISP $\co{13}{}$ and $\co{}{18}$ data sets (App.~\ref{app:mwisp-decomp})\footnote{In principle we could use moment analysis to get the integrated intensity, but blending with nearby emission features make this more uncertain}.
We selected all $\co{}{18}$ fit components in the region whose centroid positions were within $93 \lesssim \vlsr \lesssim 99$~\kms.
For each of these fitted $\co{}{18}$ components we associated $\co{13}{}$ fit components along the same line of sight whose centroid position was contained within the FWHM interval ($\vlsr \pm \text{FWHM}$) of the $\co{}{18}$ fit component. 
Thus a single $\co{}{18}$ fit component could be associated with multiple $\co{13}{}$ fit components.

For a calculated Galactocentric distance of G24 of about 3.8~kpc we expect isotope ratios of $[^{12}\text{C}]/[^{13}\text{C}] \sim 38$ and $[^{16}\text{O}]/[^{18}\text{O}] \sim 263$ \citep[based on the work of][]{Giannetti2014atlasgal}.
Using these isotope ratios to rewrite $\opacity{18}$ in terms of $\opacity{13}$ together with the integrated emission inferred from the $\co{13}{}$ and $\co{}{18}$ fit components, we can solve for $\opacity{13}$ in Eq.~\ref{eq:wco_tau}.
We show the resulting map of $\opacity{13}$ values estimated with Method~2 in \fig\ref{fig:g24_tex}\emph{d}.
The IQR of the $\opacity{13}$ values goes from 0.29 to 0.96, with a median value of 0.57.
So even though the median $\opacity{13}$ value is comparable to what was obtained with Method 1 for the same region, individual $\opacity{13}$ values can be much higher for Method~2.
However, the median S/N ratio for the fit components in $\co{}{18}$ is only $\sim 2.8$, which means that the uncertainty on their fit parameters due to impacts of the noise and thus also the uncertainty on the $\opacity{13}$ calculation will be increased.
These increased uncertainties might also explain the surprising lack of correlation we find between the two independent results for $\opacity{13}$ obtained for 471 positions with Method~1 and 2 (with a Pearson correlation coefficient of 0.1 and corresponding $p$-value of $1.2\%$).
Based on the $\opacity{13}$ calculation with Method~2, we thus conclude that optical depth effects can get important for the densest structures in the GRS, but still seem to not be an overwhelming issue even for these most problematic regions.

\subsection{Effects on the decomposition}
\label{sec:tau_g24_effects}

Our main concern is whether $\opacity{13}$ values are high enough to impact the fitting so that the underlying line gets fitted by multiple components instead of a single one.
We saw in the last two sections that we find increased optical depth values in the densest regions of G24 (\emph{panels c} and \emph{d} in \fig\ref{fig:g24_tex}), where we also detect the $\co{}{18}$ line.
For these dense regions, we can thus compare the position of the fit component(s) of the $\co{13}{}$ emission to the position of the fitted $\co{}{18}$ line.
If multiple fit components in $\co{13}{}$ are associated with a single $\co{}{18}$ fit component, it can serve as an indication that the $\co{13}{}$ emission might suffer from optical depth effects that could have led to the fitting of multiple components. 
We associate a fit component in $\co{13}{}$ with a fit component in $\co{}{18}$ if the centroid position of the former is contained within the FWHM interval of the latter.

We show a map of the number of $\co{13}{}$ fit components associated with a single $\co{}{18}$ fit component in \fig\ref{fig:g24_tex}\emph{e}.
Within the $W_{\co{}{18}} = 1$~\Kkms\ contour, that means the densest part of the G24 region, $71.8\%$ of the $\co{}{18}$ fit components are associated with a single $\co{13}{}$ component; $25.5\%$ of the $\co{}{18}$ components are associated with two $\co{13}{}$ components, and the remaining $2.8\%$ are associated with three $\co{13}{}$ components. 
Based on these results, we conclude that for about a third of the positions in the most dense region of G24 the decomposition may fit two or three components for what may be a single $\co{13}{}$ emission line, whose shape is non-Gaussian due to optical depth effects. 
Another possibility is that complex line structure is detected in $\co{13}{}$, but not in $\co{}{18}$, due to differences in sensitivity.

However, the low S/N ratio of the $\co{}{18}$ components makes this analysis rather uncertain.
We checked whether the association of a single $\co{}{18}$ component with multiple $\co{13}{}$ fit components could be due to broader $\veldisp$ values of the $\co{}{18}$ component.
The median $\veldisp$ value for $\co{}{18}$ components associated with one, two, or three $\co{13}{}$ fit components is $1.0$, $1.4$, and $2.4$~\kms, respectively, and the respective S/N ratio of the $\co{}{18}$ components is 3.2, 3.0, and 2.3.
It thus seems likely that the association of multiple $\co{13}{}$ fit components with a single $\co{}{18}$ component is at least partly affected by the low S/N ratio of the $\co{}{18}$ components.

Finally, we calculate $\co{13}{}$ column densities for the densest region in G24, which can serve as a guideline for when optical depth effects may affect the fitting results.
We can calculate $\co{13}{}$ column densities including a correction for optical depth effects via \citep{Wilson+Rohlfs}:
\begin{equation}
    N_{\text{CO}}^{13} \left[\text{cm}^{-2} \right] = 
    3\times 10^{14} \cdot \left( 1 - e^{T_{0}^{13} / \Tex}\right)^{-1} \cdot \dfrac{\opacity{13}}{1 - e^{-\opacity{13}}}\cdot W_{\co{13}{}},
    \label{eq:column_density}
\end{equation}

\noindent where we plug in the values for $W_{\co{13}{}}$, $\Tex$, and $\opacity{13}$ shown in \fig\ref{fig:g24_tex}\emph{a}--\emph{c}.
The region within the $W_{\co{}{18}} = 1$~\Kkms\ contour has a mean $N_{\text{CO}}^{13}$ value of $1.4\times10^{16}$~ cm$^{-2}$ (with minimum and maximum $N_{\text{CO}}^{13}$ values of $5.0\times10^{15}$ and $5.1\times10^{16}$~ cm$^{-2}$, respectively).
We compared the $\opacity{13}$ values calculated with Method~2 (\fig\ref{fig:g24_tex}$\,$\emph{d}) with the corresponding $N_{\text{CO}}^{13}$ values but did not find any correlation. 
However, based on the discussion in this section, we infer that optical depth effects might have an impact on the decomposition if $\co{13}{}$ column densities exceed values of $N_{\text{CO}}^{13} \sim 1\times 10^{16}$~cm$^{-2}$.

\end{appendix}

\end{document}